\documentclass[aps,pra,groupedaddress,nofootinbib,notitlepage,showpacs,floatfix,superscriptaddress,twocolumn,longbibliography]{revtex4-1}
\usepackage{color,graphicx,graphics,times,bbm,bm,amssymb,amsmath,amsfonts,amsthm,mathrsfs,xr}
\usepackage{subfigure}
\usepackage{hyperref}
\usepackage{placeins}
\hypersetup{
    colorlinks=true,       
    linkcolor=cyan,          
    citecolor=magenta,        
    filecolor=magenta,      
    urlcolor=cyan,           
    runcolor=cyan
}

\DeclareMathOperator*{\argmax}{arg\,max}
\DeclareMathOperator*{\argmin}{arg\,min}
\newcommand{\mc}[1]{\mathcal{#1}}

\newcommand{\bes} {\begin{subequations}}
\newcommand{\ees} {\end{subequations}}

\begin{document}
\title{Limitations of error corrected quantum annealing in improving the performance of Boltzmann machines}
\author{Richard Y. Li}
 \affiliation{Department of Chemistry, University of Southern California, Los Angeles, California 90089, USA}
 \affiliation{Center for Quantum Information Science \& Technology, University of Southern California, Los Angeles, California 90089, USA}
\author{Tameem Albash}
\affiliation{Department of Electrical \& Computer Engineering, University of New Mexico, Albuquerque, New Mexico 87131, USA}
\affiliation{Department of Physics and Astronomy, University of New Mexico, Albuquerque, New Mexico 87131, USA}
\affiliation{Center for Quantum Information and Control, University of New Mexico, Albuquerque, New Mexico 87131, USA}
\author{Daniel A. Lidar}
 \affiliation{Department of Chemistry, University of Southern California, Los Angeles, California 90089, USA}
 \affiliation{Center for Quantum Information Science \& Technology, University of Southern California, Los Angeles, California 90089, USA}
 \affiliation{Department of Electrical Engineering, University of Southern California, Los Angeles, California 90089, USA}
\affiliation{Department of Physics and Astronomy, University of Southern California, Los Angeles, California 90089, USA}

\date{\today}

\begin{abstract}
Boltzmann machines, a class of machine learning models, are the basis of several deep learning methods that have been successfully applied to both supervised and unsupervised machine learning tasks. These models assume that some given dataset is generated according to a Boltzmann distribution, and the goal of the training procedure is to learn the set of parameters that most closely match the input data distribution. Training such models is difficult due to the intractability of traditional sampling techniques, and proposals using quantum annealers for sampling hope to mitigate the cost associated with sampling. However, real physical devices will inevitably be coupled to the environment, and the strength of this coupling affects the effective temperature of the distributions from which a quantum annealer samples. To counteract this problem, error correction schemes that can effectively reduce the temperature are needed if there is to be some benefit in using quantum annealing for problems at a larger scale, where we might expect the effective temperature of the device to be too high. To this end, we have applied nested quantum annealing correction (NQAC) to do unsupervised learning with a small bars and stripes dataset, and to do supervised learning with a coarse-grained MNIST dataset, which consists of black-and-white images of hand-written integers. For both datasets we demonstrate improved training and a concomitant effective temperature reduction at higher noise levels relative to the unencoded case. 
We also find better performance overall with longer anneal times and offer an interpretation of the results based on a comparison to simulated quantum annealing and spin vector Monte Carlo. A counterintuitive aspect of our results is that the output distribution generally becomes less Gibbs-like with increasing nesting level and increasing anneal times, which shows that improved training performance can be achieved without equilibration to the target Gibbs distribution.
\end{abstract}
\maketitle

\section{Introduction}
The existence of commercially available quantum annealers of a non-trivial size  \cite{Harris:2010kx,Dwave,Bunyk:2014hb} along with the experimental verification of entanglement \cite{DWave-entanglement} and multi-qubit tunneling \cite{Boixo:2014yu} have ignited interest and a healthy debate concerning whether quantum annealing (QA) may provide an advantage in solving classically hard problems. QA can be considered a special case of adiabatic quantum computation (AQC)  (for a review see Ref.~\cite{Albash-Lidar:RMP}). In AQC, computation begins from an initial Hamiltonian, whose ground state is easy to prepare, and ends in a final Hamiltonian, whose ground states encodes the solution to the computational problem. In a closed-system setting with no coupling to the external environment, the adiabatic theorem guarantees that the system will remain in an instantaneous ground state, provided the interpolation from initial to final Hamiltonian is sufficiently slow, such that all non-adiabatic transitions are suppressed. The runtime to ensure this happens is related to the inverse of the minimum gap encountered during computation~\cite{Jansen:07,lidar:102106}. In this setting, errors arise only from non-adiabatic transitions and any control errors (e.g., in the annealing schedule or in the initial and final Hamiltonians). In QA, instead of running the computation once at a sufficiently long anneal time such that the adiabatic theorem is obeyed, one may choose to run the computation multiple times for a shorter anneal time, such that the time it takes to find the ground state with high probability is minimized~\cite{speedup}. 

Physical implementations of QA, however, will be coupled to the environment, which may introduce additional sources of errors, such as dephasing errors and thermal excitation errors~\cite{childs_robustness_2001,PhysRevA.74.052330,Albash:2015nx}. Theoretical and experimental studies have indicated that due to relatively strong coupling to a thermal bath, current quantum annealing devices operate in a quasi-static regime \cite{Amin:2015qf,Venuti:2015kq,Marshall:2017aa,Chancellor:2016pa,Albash:2017ab}. In this regime there is an initial phase of quasi-static evolution in which thermalization times are much shorter than the anneal time, and thus the system closely matches a Gibbs distribution of the instantaneous Hamiltonian. Towards the end of the anneal, thermalization times grow and eventually become longer than the anneal time, and the system enters a regime in which the dynamics are frozen. The states returned by a quantum annealer operating in this regime therefore more closely match a Gibbs distribution not of the final Hamiltonian, but of the Hamiltonian at the freezing point.

The fact that open-system QA prepares a Gibbs state may be a bug for optimization problems \cite{Albash:2017ab} but it could be a feature for sampling applications. Recently, there has been interest in using QA to sample from classical or quantum Gibbs distributions (see Ref.~\cite{Perdomo:17qml} for a review),
and there is interest in whether QA can prepare such distributions faster than using temperature annealing methods \cite{Venuti:2017aa}. One application where sampling plays an important role is the training of Boltzmann machines (BMs). These are a class of probabilistic energy-based graphical models which are the basis for powerful deep learning models that have been used for both supervised (with labels) and unsupervised (without labels) learning tasks \cite{Salakhutdinov:2012,Tang:2012}. 

As its name suggests, a Boltzmann machine assumes that the target dataset is generated by some underlying probability distribution that can be modeled by a Boltzmann or Gibbs distribution. Whereas the temperature does not play a large role for classical methods, as the values of model parameters can simply be scaled as needed, physical constraints on current (and future) quantum annealers  limit the range of model parameters that can be programmed. As such, an important parameter for using a physical quantum annealer is the effective temperature $T_\mathrm{eff}$, corresponding to the best-fit classical Gibbs temperature for the distribution output by the annealer. 
For example, $T_\mathrm{eff}$ is effectively infinite for the initial state of the quantum annealer, the uniform-superposition state, since the samples drawn from a quantum annealer at this point would be nearly random, and this would make training of a Boltzmann machine nearly impossible. 

Nested quantum annealing correction (NQAC)~\cite{vinci2015nested} is a form of quantum annealing correction (QAC)~\cite{PAL:13,PAL:14,MNAL:15,Mishra:2015} tailored for use on quantum annealing devices, including commercially available ones, that was developed to address some of these concerns. NQAC achieves error suppression by introducing an effective temperature reduction \cite{vinci2015nested,Vinci:2017ab,Matsuura:2018}, and previous work has shown that NQAC can be used to improve optimization performance and obtain more accurate estimates of the gradient in the training step of a BM \cite{Vinci:2017ab}. In this work we apply NQAC to an entire training procedure of fully-visible Boltzmann machines. We demonstrate an improvement for both supervised and unsupervised machine learning tasks using NQAC, explore the effects of increased anneal times, and make comparisons to spin-vector Monte Carlo (SVMC) \cite{SSSV} and simulated quantum annealing (SQA) \cite{Santoro} to probe the underlying physics of using a D-Wave (DW) quantum annealer as a sampler. 

The remainder of the paper is structured as follows. We provide some more technical background, both of Boltzmann machines and the NQAC construction in Sec.~\ref{sec:background}. In Sec.~\ref{sec:methods} we give an overview of the methods, and results are presented in Sec.~\ref{sec:results}. We conclude in Sec.~\ref{sec:conclusions}.

\section{Technical Background}
\label{sec:background}

A standard quantum annealing protocol defines the following time-dependent Hamiltonian:
\begin{align}
 H(s) = A(s)H_X + B(s)H_P, \quad s \in [0,1],
 \label{eq:annealing}
\end{align}
where $s = {t}/{t_f}$ is the dimensionless time ($t_f$ is the total anneal time), and $A(s)$ and $B(s)$ are the transverse field and longitudinal field annealing schedules, respectively~\cite{kadowaki_quantum_1998}. $H_X$, the initial (or ``driver'') Hamiltonian, is usually defined as $H_X = -\sum_i\sigma_i^x,$ such that the initial state of the system is in a uniform superposition over all input computational basis states (defined in the $\sigma^z$ basis). $H_P$, the problem Hamiltonian, is defined on a graph $\mathcal{G} = (\mathcal{V},\mathcal{E})$ composed of a set of $N = |\mathcal{V}|$ vertices and edges $\mathcal{E}$:

\begin{align}
 H_P = \sum_{i\in \mathcal{V}}h_i\sigma_i^z + \sum_{(i,j)\in \mathcal{E}}J_{ij}\sigma_i^z\sigma_j^z.
 \label{eq:H_P}
\end{align}
The local fields $\{h_i\}$ and the couplings $\{J_{ij}\}$ are used to represent the computational problem, and are programmable parameters in hardware implementations of quantum annealing. 

In the remainder of this section we present an overview of NQAC and Boltzmann machines. 

\subsection{NQAC}
NQAC is an implementation of a repetition code that can encode problems with arbitrary connectivity, allows for a variable code-size and also can be implemented on a generic quantum annealing device~\cite{vinci2015nested}. 

In general, to implement QAC, we encode the original (or ``logical'') quantum annealing Hamiltonian $H(s)$ in an ``encoded physical Hamiltonian'' $\bar{H}(s)$, using a repetition code \cite{PAL:13,PAL:14,Vinci:2015jt}:
\begin{align}
 \bar{H}(s) = A(s)H_X + B(s)\bar{H}_P, \quad s \in [0,1],\label{eq:encoded_H}
\end{align}
where $\bar{H}_P$ is the ``encoded physical problem Hamiltonian'' and all terms in $\bar{H}_P$ are defined over a set of physical qubits that is larger than the number of logical qubits in the original unencoded problem. 
The states of the logical problem Hamiltonian $H_P$ can then by recovered by properly decoding the states of $\bar{H}_P$. Encoding the driver Hamiltonian would make this a full stabilizer code \cite{Gottesman:1997ub} and would provide improved performance with error correction since it would enable the implementation of fully encoded adiabatic quantum computation, for which rigorous error suppression results have been proven~\cite{jordan2006error,Bookatz:2014uq,Jiang:2015kx,Marvian-Lidar:16,Marvian:2017aa,Lidar:2019ab}. However, unfortunately this is not possible with present implementations of quantum annealers; i.e., only $H_P$ is encoded in QAC.

\begin{figure*}
 \includegraphics[width=\linewidth]{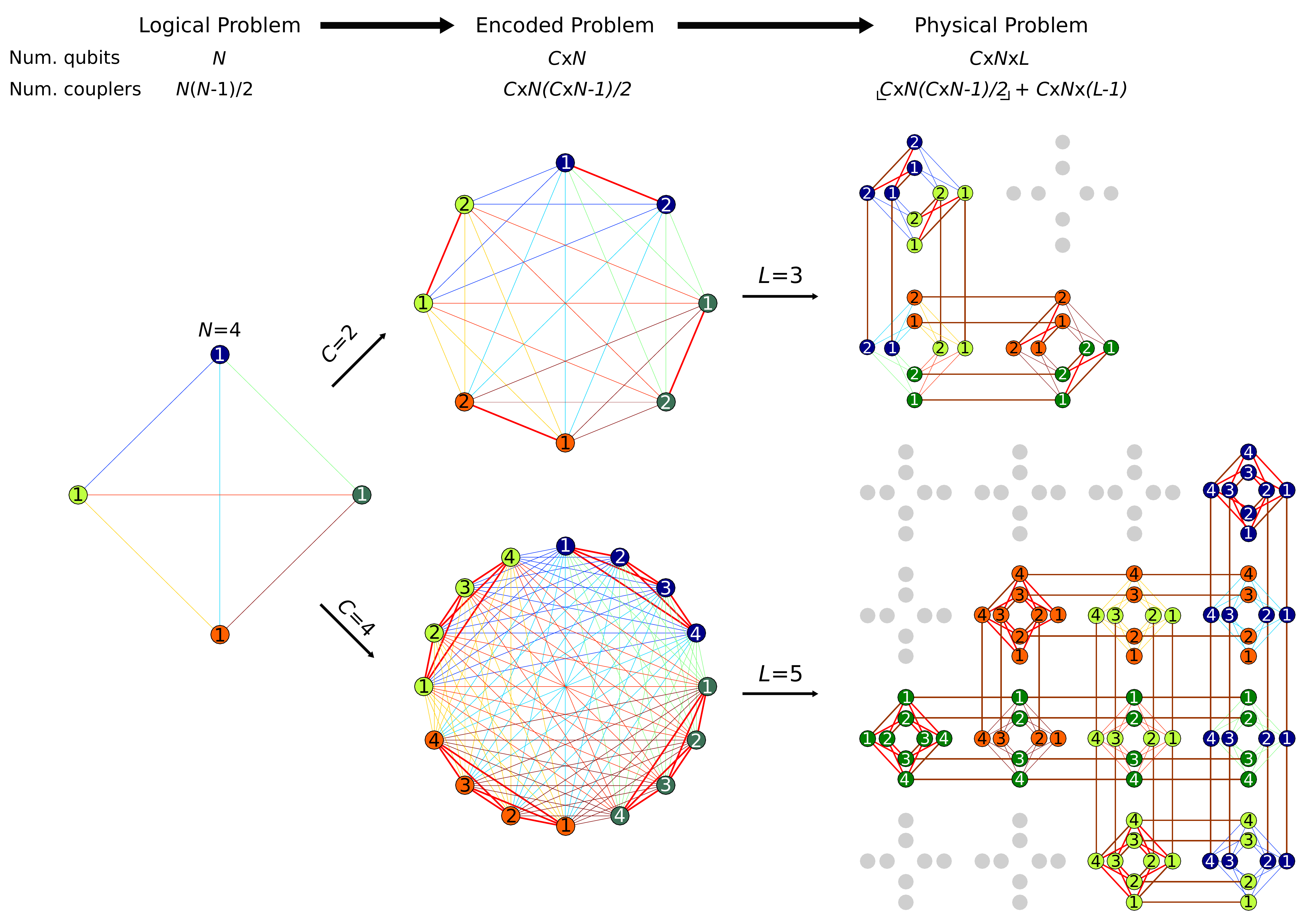}\caption{Illustration of NQAC. The left panel represents the logical problem, the middle panels represent the encoded Hamiltonian with nesting levels $C=2,4$, and the right panel represents the encoded physical problem embedded onto the DW Chimera graph. In going from the left panel to the middle panel, each logical qubit is replaced by $C$ copies that are ferromagnetically coupled; the thick bright red lines in the middle and right panels represent $\gamma_1$, the ferromagnetic coupling between code qubits in the nested Hamiltonian, which we call the `code penalty'. In going from the middle to the right panel, each code qubit (middle panel) is replaced by a chain of physical qubits; brown lines in the right panel represent $\gamma_2$, the ferromagnetic coupling between physical qubits in the encoded problem, which we  call the `minor embedding penalty'. $L$ is the number of qubits per chain (the chain length). The general scaling of the number of qubits and couplers with each step is shown at the top. }
 \label{fig:nqac_overview}
\end{figure*}

NQAC encodes the logical Hamiltonian in two steps: (1) an ``encoding'' step that maps logical qubits to code qubits, (2) an ``embedding'' step that maps code qubits to the physical qubits, e.g., those on the Chimera graph of the DW quantum annealer. These two steps are depicted in Fig.~\ref{fig:nqac_overview} for a fully connected problem of $4$ logical spins. 

For the encoding step, the repetition code has length $C$, which we also refer to as the ``nesting level'' for reasons apparent from Fig.~\ref{fig:nqac_overview}. $C$ determines the amount of hardware resources (qubits, couplers, etc.) used and allows the error correction method to be scaled, and to provide protection against thermal and control errors. Each logical qubit $i$ $(i = 1,\dots,N)$ in $H_P$ is represented by a $C$-tuple of code qubits $(i,c)$, with $c = 1,\dots,C$. The values of the `nested couplers' $\tilde{J}_{(i,c),(j,c')}$ and ``nested local fields" $\tilde{h}_{(i,c)}$ are given as follows in terms of the logical problem parameters and $C$: 

\begin{subequations}
\label{eq:Jh}
\begin{alignat}{3}
 \tilde{J}_{(i,c),(j,c')} &=& J_{ij}, \quad & \forall c,c', i\neq j,\\
 \tilde{h}_{(i,c)} &=& Ch_i, \quad&\forall c,i,\\
 \tilde{J}_{(i,c),(i,c')} &=& {-}\gamma_1, \quad& \forall c\neq c'.
\end{alignat}
\end{subequations}

Code qubits $c$ and $c'$ that belong to different logical qubits $i$ and $j$ are coupled with the strength of the original logical problem couplings $J_{ij}$. Since each logical qubit is encoded in $C$ code qubits, there are a total of $C^2$ nested couplers $\tilde{J}_{(i,c),(j,c')}$ between all the $C$ code qubits that comprise two logical qubits $i$ and $j$. In order to uniformly scale the overall energy scale by $C^2$, each nested local field $\tilde{h}_{(i,c)}$ is set to be $C$ times as large as the logical problem local field. Finally, in order to facilitate alignment of the $C$ code qubits that make up a logical qubit $i$, $C(C-1)/2$ ferromagnetic couplings $\tilde{J}_{(i,c),(i,c')}$ (couplings between code qubits corresponding to the same logical qubit $i$) are introduced, the strength of which is given by $\gamma_1 > 0$. We refer to $\gamma_1$ as the `code penalty'.

The encoded problem must then be implemented on physical QA hardware, which typically has less than full connectivity. Because we collected our results in this work on the DW quantum annealers (described in Appendix~\ref{app:DW}), we used a minor embedding \cite{Choi2,klymko_adiabatic_2012,Cai:2014nx} onto the Chimera hardware graph. The minor embedding step replaces each code qubit by a ferromagnetically coupled chain of physical qubits, with the intra-chain coupling for physical qubits given by another penalty term $\gamma_2 > 0$, which helps keep physical qubits aligned. We refer to $\gamma_2$ as the `minor embedding penalty'. Finally, in order to extract states in terms of the original logical problem, a proper decoding scheme should be used. In this work, we used a majority vote to go from physical qubits to code qubits, and another majority vote to go from the physical problem states to the logical problem states. Other decoding strategies, such as local energy minimization, are also possible~\cite{Vinci:2015jt}. In Appendix~\ref{app:decoding} we study the case of no decoding (i.e., only using  unbroken chains) and find that learning suffers, so a decoding strategy is needed for training to be effective at higher nesting levels. 

It is important to note that all local fields and couplers must satisfy $|h_i| \leq 2,|J_{ij}|\leq 1$ as hard constraints imposed by the physical processor. For the same reason, we must have $\gamma_1,\gamma_2\leq 1$. This implies that in some cases we will find NQAC to be `penalty-limited'~\cite{Vinci:2017ab}, meaning that the optimal values of $\gamma_1$ and $\gamma_2$ exceed $1$ and thus cannot be attained in practice. As we shall see, this has important performance implications.

\subsection{Boltzmann machines}
We combined NQAC with training of a Boltzmann machine. Boltzmann machines are a class of probabilistic graphical models that can be stacked together as deep belief networks or as deep Boltzmann machines \cite{Hinton:2006fv,Salakhutdinov:10a}. Deep learning methods such as these have the potential to learn complex representations of the data, which is useful for applications such as image processing or speech recognition. 
Typically, Boltzmann machines include both visible and hidden units, which give greater modeling power. Here we restrict ourselves to examining fully-visible Boltzmann machines, which are less powerful \cite{Mackay:2003}, but allow us to demonstrate the effect of NQAC more directly. 

A Boltzmann machine is defined on a graph $\mathcal{G} = (\mathcal{V},\mathcal{E})$
with binary variables on the vertices $\mathcal{V}$. As before, let $N=|\mathcal{V}|$ be the number of vertices. In exact analogy to Eq.~\eqref{eq:H_P}, the energy of a particular configuration $\mathbf{x} = (x_1,\dots,x_N)$ is 
\begin{align}
E(\mathbf{x}) = \sum_{i\in \mathcal{V}}b_ix_i + \sum_{(i,j)\in \mathcal{E}}w_{ij}x_ix_j  ,
\label{eq:costf}
\end{align}
where $b_i$ (the biases) and $w_{ij}$ (the weights) are the parameters of the model to be learned from training. Each $x_i$ is a binary variable, and to be consistent with quantum annealing conventions, we define $x_i = \pm 1$ (spins) instead of $0$ or $1$, so that $\mathbf{x}\in \{-1,1\}^N$. The associated Boltzmann or Gibbs distribution is: 
\begin{align}
 P(\mathbf{x}) = \frac{\exp[-\beta E(\mathbf{x})]}{Z}, \quad Z = \sum_{\mathbf{x}} \exp[-\beta E(\mathbf{x})] ,
 \label{eq:boltzprob}
\end{align}
where $\beta={1}/{T}$ is the inverse temperature and $Z$ is the partition function.
To use a Boltzmann machine for machine learning, we assume that a given \textit{training dataset} is generated according to some probability distribution. The goal of the machine learning procedure is to find the set of parameters (i.e., the $b_i$'s and $w_{ij}$'s) that best models this distribution. Typically this is done by maximizing the log-likelihood over the data distribution (or minimizing the negative log-likelihood):
\begin{align}
 \mc{L} = \frac{1}{|\mathcal{D}|}\sum_{\mathbf{x}\in \mathcal{D}}\log P(\mathbf{x}),\label{eq:ll}
\end{align}
where $\mathcal{D}$ represents the target training data distribution. 

We can derive the following update rule for the model parameters by taking gradients with respect to the model parameters:
\bes
\label{eq:b-and-w}
\begin{align}
 b_i(t+1) &= b_i(t) + \eta\frac{\partial \mc{L}}{\partial b_i} , \\
 w_{ij}(t+1) &= w_{ij}(t) + \eta\frac{\partial \mc{L}}{\partial w_{ij}},
\end{align}
\ees
where $\eta$ is the learning rate that controls the size of the update, and the gradients are:
\bes
\label{eq:updates}
\begin{align}
 \frac{\partial \mc{L}}{\partial b_i} &= \langle x_i\rangle_{\mathcal{D}} - \langle x_i\rangle_\beta \label{eq:b_updates}\\
 \frac{\partial \mc{L}}{\partial w_{ij}} &= \langle x_ix_j\rangle_{\mathcal{D}} - \langle x_ix_j\rangle_\beta, \label{eq:w_updates}
\end{align}
\ees
where $\langle \cdot \rangle$ is the expectation value with respect to the appropriate distribution. In practice, rather than updating the model parameters based on averages over the entire dataset, ``mini-batches'' of a fixed number of training samples are used to reduce computational cost and insert some randomness. One pass through all the training samples is referred to as an \textit{epoch}. In general, we implemented the training procedure as follows:\\

\noindent \textbf{Training}
\begin{enumerate}
\item Initialize the problem parameters (biases and weights $b_i$ and $w_{ij}$) to some random values;
\item Compute the averages needed in Eq.~\eqref{eq:updates} by sampling from the Gibbs distribution computed in the previous step;
\item Update the problem parameters using Eq.~\eqref{eq:b-and-w};
\item Repeat steps (2) and (3) until convergence (e.g., of the gradients) to within some tolerance or until a certain number of epochs is reached.
\end{enumerate}
The first term in the gradient step, $\langle\cdot\rangle_{\mathcal{D}}$, is referred to as the \textit{positive} phase, and is an average of the training data distribution. The second term,  $\langle\cdot\rangle_\beta$, is referred to as the \textit{negative} phase and is an average over the current Boltzmann model. The positive phase serves to move the probability density towards the training data distribution, and the negative phase moves probability away from states that do not match the data distribution. Calculating \emph{exact} averages for the negative phase becomes intractable as the number of possible states grows exponentially in $N$. Nevertheless, the averages can still be estimated if there is an efficient way to perform importance sampling. Classically, this is done by Markov chain Monte Carlo methods, or by replacing the likelihood function altogether by a different objective function, such as contrastive divergence (whose derivatives with regard to the parameters can be approximated accurately and efficiently)~\cite{Hinton:2002aa} for restricted Boltzmann machines. 

Alternatively, one may try to use a quantum annealer to perform importance sampling by directly preparing a Gibbs distribution on a physical device~\cite{Amin:2016,Benedetti:2016oz,2012arXiv1204.2821S}. The hope is that this may speed up the bottleneck of computing the negative phase over classical methods for Gibbs sampling (we stress that this uses quantum annealing not for optimization, as originally envisioned~\cite{kadowaki_quantum_1998}, but for sampling). However, using a quantum annealer introduces at least two new complications. The first is that the Gibbs distribution sampled from need not be the one associated with the cost function, due the quasi-static evolution phenomenon alluded to in the Introduction. The reason is that under such evolution the Gibbs distribution prepared by the annealer freezes in a state associated with an intermediate, rather than the final Hamiltonian. The second is the temperature of the distribution. In the classical method, $\beta$ in Eq.~\eqref{eq:boltzprob} is set to $1$, as it amounts to an overall change in the energy scale of the cost function (Eq.~\eqref{eq:costf}), which can be easily rescaled. For quantum annealers, the magnitude of the problem parameters depends on physical quantities, such as a biasing current; hence, the model parameters cannot become arbitrarily large, and both the freezing point and the physical $\beta$ play an important role. Because of these factors, one might expect to find a sweet-spot in effective temperature for quantum annealers. When the effective temperature of a sampler is too high, the samples drawn will essentially look random and will not provide any meaningful training. When the effective temperature is too low, the samples from this distribution will resemble delta-functions on the training example, and thus may fail to generalize to data not within the training set. 

\subsection{NQAC and Boltzmann Machines}
NQAC is a method that may offer more control on the effective temperature, but there are important caveats to keep in mind.  First, the low-lying spectrum associated with the physical NQAC Hamiltonian $\bar{H}_P$ may not be a faithful representation of the spectrum of the associated logical Hamiltonian $H_P$.  We only expect this to be the case for sufficiently strong penalty strengths $\gamma_1, \gamma_2$, but these parameters cannot be made arbitrarily large in practice.  Furthermore, in practice these penalties are picked to maximize some performance metric, which does not guarantee  a faithful representation of the logical spectrum. 
Therefore, after decoding the physical problem states to the encoded problem states and then to the logical problem states, the Gibbs distribution of $\bar{H}_P$ at an inverse-temperature $\beta$ may not be close to the Gibbs distribution of $H_P$ at any inverse-temperature.

Second, there is no expectation that NQAC will be effective in solving the issues associated with using a quantum annealer for preparing a Gibbs state described earlier.  In fact, already the first work on QAC suggested that freezing may occur even earlier~\cite{PAL:13}. Note that this need not be a disadvantage for machine learning purposes, since the (quantum) Gibbs distribution associated with the intermediate state may be preferable for sampling and updates.

 We shall see all these considerations play out when we discuss our results in Section~\ref{sec:results}.

\section{Methods}
\label{sec:methods}

\subsection{Datasets and performance metrics}

In this work we explored both supervised and unsupervised machine learning with two different datasets. 

\subsubsection{Supervised machine learning of MNIST}

\begin{figure}[t]
 \includegraphics[scale=0.29]{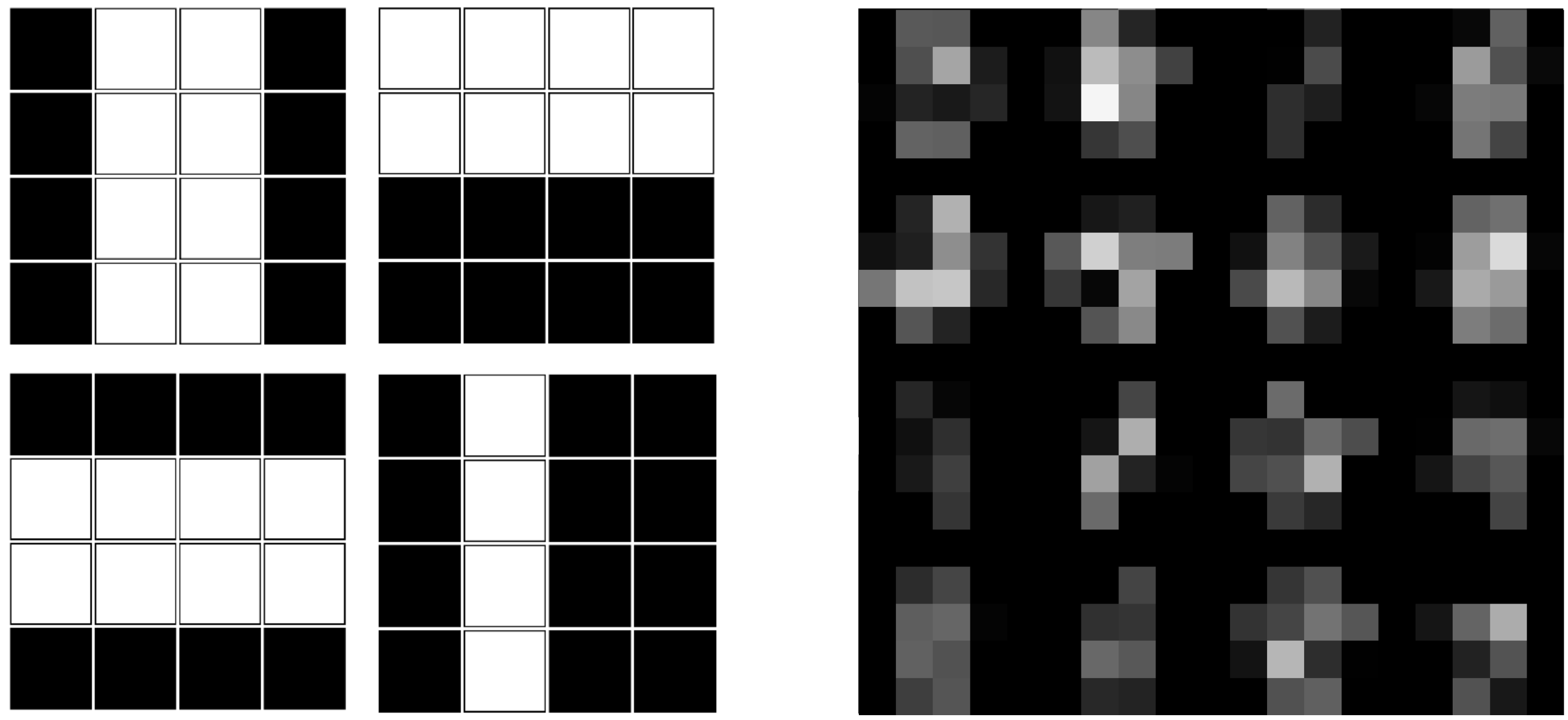}\caption{Some representative images of training dataset. BAS (left) and coarse-grained MNIST (right). Although it is difficult for a human to tell, starting from the upper-left and going across the rows and down the columns, the digits are $\{3, 6, 1, 5\}, \{2, 6, 7, 8\}, \{3, 1, 6, 4\}, \{0, 6, 8, 7\}$.}
 \label{fig:data}
\end{figure}

We trained a Boltzmann machine to do supervised machine learning with a coarse-grained version of the MNIST dataset \cite{MNIST}, a set of hand-written digits. In supervised machine learning, the dataset consists of input states and response variables (labels), i.e., $\mathcal{D}^\textrm{TRAIN} = \{(\mathbf{x}_n,y_n)\}_{n=1}^S$, where $S$ is the size of the training dataset. The goal of the supervised machine learning task is to learn some function $f$ that maps inputs to response variables such that $y_n = f(\mathbf{x}_n).$ For our purposes $\mathbf{x}_n$ corresponds to the $n$th coarse-grained image and $y_n$ is the corresponding label of the image (i.e., an integer digit). Our metric of performance for the MNIST dataset is classification accuracy. Due to limitations on the number of qubits available on the DW processor, we coarse-grained the original MNIST digits, which are $28\times 28$ pixels, to $4\times 4$ images which bear little resemblance to the original digits (see Fig.~\ref{fig:data} for some examples of coarse-grained images). We then removed the corners of the coarse-grained images and only selected images that were labeled $1$-$4$, adding four ``label'' bits to each image (i.e., the last four digits set to $1000$ means the image is labeled with ``$1$'', $0100$ represents ``$2$'', $0010$ represents ``$3$'', and $0001$ represents ``$4$''). After this, each $\mathbf{x}_n$ is a vector of length $16$. Of the $50,000$ training images in the original dataset, we used $5000$ samples for training, and about $2500$ for testing. Minibatches of 50 images were used, so each training epoch consisted of 100 updates to the model parameters. The training examples were randomly permuted at the start of each epoch. Classification accuracies were evaluated on the held-out test dataset, which were unseen during training. Labels were predicted by ``clamping'' the values in the logical problem Hamiltonian to the input \cite{Amin:2016,Benedetti:2016oz} and sampling from the clamped Hamiltonian $H_\textrm{clamp} = H_P(\sigma_i^z=x_i,\sigma_j^z=x_j)$, where $i,j\in\{1,\dots,12\}$ and $x_i,x_j$ are the image pixels (recall that the last four bits were used as label bits). 
Because we obtain $H_\textrm{clamp}$ by setting the values $\sigma_i^z$ and $\sigma_j^z$ of $H_P$ [Eq.~\eqref{eq:H_P}] to the value of the pixels in the image, $H_\textrm{clamp}$ is only defined on the last four label qubits, and the effect of clamping is to add a local field to the remaining unclamped label qubits. In order to determine the predicted label $\hat{y}_n$ of the $n$th image, we then sampled from the four unclamped label qubits, as we describe next. 

We acquired $K=100$ anneals from DW per image, with the $k$th anneal producing a vector $\mathbf{l}_k$, of length four, corresponding to the four possible labels. Note that in principle, depending on the value of $H_\textrm{clamp}$ and due to the probabilistic nature of quantum annealing, $\mathbf{l}_k$ could have more than one non-zero entry.
We assigned the predicted label via $\hat{y}_n = \argmax\frac{1}{K}\sum_{k=1}^K \mathbf{l}_k$, where the $\argmax$ is over the four indices of $\mathbf{l}_k$ (i.e., we selected the label qubit with the largest appearance frequency).  
Once we obtained the prediction $\hat{y}_n$ of a class label, we compared it to the true label $y_n$ and calculated the corresponding classification accuracy $A$:
\begin{align}
A = \frac{1}{S_{\mathrm{test}}}\sum_{n=1}^{S_{\mathrm{test}}} \delta_{y_n,\hat{y}_n},
 \label{eq:cl_acc}
\end{align}
where $\delta$ is the Kronecker delta and $S_{\mathrm{test}}$ is the size of the test dataset. 

\subsubsection{Unsupervised machine learning of Bars and Stripes}


In unsupervised machine learning, the data is provided without a response variable (i.e., $\mathcal{D} = \{\mathbf{x}_n\}_{n=1}^S$), and the goal is to learn the underlying structure within the data. Our main focus in this work was on exploring unsupervised machine learning with a ``bars and stripes'' (BAS) dataset \cite{Mackay:2003}, which has a very simple underlying probability distribution. A BAS dataset is generated by first randomly selecting an orientation with probability $1/2$ (rows or columns). Once an orientation is selected, all pixels in a row or column are set to 1 with probability $1/2$ (see Fig.~\ref{fig:data} for some examples). A BAS dataset of size $D\times D$ has a total of $|\mathcal{D}|=2^{D+1}$ images ($2^D$ images for each orientation times $2$ orientations) of which $2^{D+1}-2$ are unique (there are $2$ duplicate all-white and all-black images). Ideally, each of the $2^{D+1}-4$ non all-white or all-black image would be generated with probability $p_0 = 1/|\mathcal{D}|$ with the all-white and all-black images generated with probability $2p_0$. Therefore for an ideally distributed BAS dataset, all-black and all-white images (or their corresponding bit representations) appear with probability $2 p_0$, and the probability of all other bar and stripe images is $p_0$.  All other non-bar and non-stripe images appear with probability $0$.  Thus, for an ideally distributed BAS dataset, the log-likelihood [Eq.~\eqref{eq:ll}] is $\mathcal{L}=p_0[4\log(2p_0)+(2^{D+1}-4)\log(p_0)]$. We constructed a dataset of $4\times4$ images (thus containing $30$ distinct images), generating $|\mathcal{D}|= S = 5000$ images for the dataset. The frequencies of each type of image appearing in our randomly generated dataset do not exactly match the expected average number due to finite sampling, so the log-likelihood on the test dataset was $-3.37$ (compared to $\mathcal{L}=-3.38$ of an ideally distributed BAS dataset with $D=4$). As with the MNIST dataset, we used minibatches of $50$ images.

Our metric of performance for the BAS is the ``empirical'' log-likelihood; i.e., the log-likelihood of the distribution returned by drawing actual samples from DW. The goal (for DW) was to maximize the empirical log-likelihood by finding the optimal values of the biases and weights. In doing so, it would have ideally discovered the distribution of images described above. More formally, let $Q(\mathbf{x})$ be the empirical probability of an image (or state) $\mathbf{x}$ in the distribution returned from a quantum annealer. We define the empirical log-likelihood as 
\begin{align}
 \tilde{L} = \frac{1}{|\mathcal{D}|} \sum_{\mathbf{x}\in\mathcal{D}} \log Q(\mathbf{x}) ,
 \label{eq:emp-ll}
\end{align}
where the sum is over the images in the dataset $\mathcal{D}$.
Note that in contrast, the ``exact'' log-likelihood in Eq.~\eqref{eq:ll} is defined in terms of the exact Boltzmann distribution on the current model parameters. To calculate the empirical log-likelihood, we sampled from DW $10^4$ times per gradient step update (since we used minibatches of 50, and the training data was 5000 samples, there were a total of 100 such gradient step updates per epoch.), i.e., $Q(\mathbf{x})$ was determined by finding frequencies of the images of the training distribution in the $10^4$ samples returned by DW. 

\subsection{Temperature estimation}
\label{sec:temp-est}

Based on the intuition that higher nesting level should result in a lower effective temperature, we estimated the temperature by comparing the distribution of energies of the decoded logical problem to an exact Gibbs distribution on the logical problem Hamiltonian. To do so, we found the value of $\beta$ such that the total variation distance 
\begin{align} 
d(\beta) = \frac{1}{2}\sum_{E_i}|Q(E_i)- P(E_i;\beta)|\ ,
\label{eq:tvd} 
\end{align}
between the empirical probability distribution $Q$ and the exact Gibbs distribution $P$ [Eq.~\eqref{eq:boltzprob}] is minimized, i.e.,
\begin{align} 
\beta_{\textrm{eff}} = \argmin_{\beta} d(\beta)\ .
\label{eq:beta} 
\end{align}
We calculated the distribution distance in terms of energy levels, not in terms of distinct states, in order to speed up calculation. We also note that this method of estimating the temperature can only be done for sufficiently small problems, because of the challenge of computing the partition function. The data we used was composed of $16$ pixels, and thus calculating the exact Boltzmann probabilities is still feasible. We emphasize that simply because we associate an effective inverse-temperature with the distribution $Q$, the distribution is not necessarily close to the corresponding Gibbs state of the Hamiltonian $H_P$\footnote{This is due to the restriction to a single fitting parameter $\beta$; the existence of a generalized Gibbs distribution, with $\beta \mapsto \vec{\beta}=\{\beta_i\}$, such that $d(\vec{\beta})=0$, is guaranteed by Jaynes' principle~\cite{Jaynes:1957aa}, but in general this would require higher weight classical terms (of which the $\beta_i$ would be prefactors). Similarly, in \cite{Nishimori:2015dp} it was shown that the ground state of quantum annealing maps to classical thermal states but with many-body terms.}.

We also consider the dimensionless inverse-temperature:
\begin{align} 
\hat{\beta}_{\textrm{eff}} = \beta_{\textrm{eff}} \| H_P \| \ ,
\label{eq:betaNorm} 
\end{align}
where $\|\cdot \|$ denotes the operator norm (largest singular value), which allows us to also take into account the magnitude of the Ising parameters in the logical Hamiltonian.

\subsection{Distance from data}
\label{sec:distance_from_data}

For some of our results, we also plot the distance to the target data distribution, instead of the distance from a Gibbs distribution at a given inverse temperature. This quantity is defined as follows:
\begin{align}
  d_\textrm{data} = \frac{1}{2}\sum_{E_i}| Q(E_i) - R(E_i)|,\label{eq:dist_from_data}
\end{align}
where $R(E_i)$ is the frequency of finding a state with energy $E_i$ in the data distribution; i.e., we first calculated the energies of every state that appeared in the data distribution under the current model parameters. However, unlike the distance from Gibbs, the distance from data does not depend on a value of the effective temperature; we calculated the distance this way to be consistent with how the distribution distance from Gibbs is calculated.

The distance to a target data distribution gives more information about how well an algorithm is learning information about the data distribution (and thus has a similar interpretation as the empirical log-likelihood), and the distance from a Gibbs distribution provides more information about how close the distribution of the samples is to the Gibbs distribution of $H_P$. With a perfect noiseless thermal sampler and the assumption that the data can be modeled by a Gibbs distribution of $H_P$, these two quantities should be perfectly correlated. However, as will be seen in our results, this is not always the case when dealing with imperfect samplers; the sampled distributions may be somewhat far away from Gibbs and yet closely resemble the target data distribution. 

\subsection{Training procedure}
\label{sec:TP}
For the sake of comparison, we used a fixed set of initial weights $\{w_{ij}\}$ and biases $\{b_i\}$ (the model parameters). First, to compare the best performance at each nesting level $C$, we need to find the optimal values of the penalty terms $\gamma_1$ and $\gamma_2$. To do so, we did a grid search from $0.2$ to $1.0$ in steps of $0.2$ to find the combination of $\gamma_1$ and $\gamma_2$ 
that gave the best empirical log-likelihood for the BAS dataset and the best classification accuracy for each nesting level. The optimal $\gamma_1$ and $\gamma_2$ may be different at different nesting levels, but recall that $\gamma_1,\gamma_2\leq 1$ due to the upper limit set by the hardware. After finding the optimal $\gamma_1(C)$ and $\gamma_2(C)$, we reinitialized the model parameters and trained for $10$ epochs. All runs with DW were done with an anneal time $t_f$ of  $10\mu$s, except when we studied the dependence on $t_f$, and the learning rate $\eta$ [Eq.~\eqref{eq:b-and-w}] was set to $0.01$.

As we will see later, the training performs reasonably well without the use of NQAC, indicating that the current temperature of the DW device does not hamper performance for the BAS data set at the sizes we consider. Therefore, to mimic an effectively higher temperature device for which NQAC might be useful (as the device gets larger and data sets get more complex, we expect the same temperature to be too high), we introduced a scaling parameter $\alpha$ for the logical problem Hamiltonian, i.e., 
\begin{equation}
H_P \mapsto \alpha H_P\ , \quad 0 < \alpha \le 1\ .
\label{eq:alpha}
\end{equation}


Smaller $\alpha$ emulates a higher temperature for the Ising Gibbs state, but it also increases the detrimental effect of control errors on the programmed Hamiltonian. If the rise in temperature could be alleviated by increasing the magnitude of the Ising parameters, then training would not be hindered. However, on analog physical annealers, there is an upper bound on the strength of the Ising parameters. As the effective temperature increases, learning becomes difficult if not impossible, because the probability distribution from which samples are taken becomes too uniform. Similarly, as control errors increase in magnitude, sampling at low temperatures begins to resemble sampling at higher temperatures~\cite{albash_2019_analogerrors}. By reducing $\alpha$, we (artificially) probe a scenario where the temperature and control errors increase, in which case NQAC may be one viable approaches to effectively reduce these error sources. 

\subsection{Classical repetition vs NQAC}
\label{sec:methods_classical_repetition}

Because higher nesting levels use more qubits than $C=1$ it is fairer to compare the performance at a nesting level $ C > 1$ with NQAC to the $C=1$ problem replicated to use approximately the same amount of physical resources [recall from Fig.~\ref{fig:nqac_overview} that the physical problem uses at least $C N(C N -1)/2 + C N (L-1)$ qubits, where $C$ is the nesting level, $N$ is the size of the logical problem, and $L$ is the length of the chain needed for the embedding]. To implement this, we created $M_C$ replicas, where $M_C$ is the closest integer multiple of the number of physical qubits needed for nesting level $C$ compared to $C=1$;
we find that $M_2 = 4$ and $M_3 = 8$. Then, at each training iteration, we generated $M_C$ replicas of the $C=1$ encoded physical problem and performed the update of the parameters according to Eqs.~\eqref{eq:b_updates} and \eqref{eq:w_updates}. We then sampled from the updated parameters and selected the replica whose parameters gave the best empirical log-likelihood and set the parameters of the remaining replicas to the best-performing replica. We then repeated the process until convergence was reached. 

\subsection{Quenching}
\label{sec:quench}
In addition to evaluating the effect of NQAC on machine learning performance, we also explored various control parameters affecting the anneal. Specifically, we used the ``annealing schedule variation'' feature of the DW2000Q devices~\cite{DW2KQ}, which allows for different annealing rates along different segments of the annealing schedule. We can therefore approximate a quench by choosing a faster annealing rate for $s > s_{\mathrm{int}}$ than for $s \leq s_{\mathrm{int}}$, where $0 < s_\textrm{int} < 1$. To probe the distributions at intermediate points in the anneal, we quenched between $s_\textrm{int}=0.1$ and $0.9$ in steps of $0.1$. Because the annealing parameters cannot be changed instantaneously, we set the quench time to 1 $\mu$s. We emphasize that we do not expect this quench to be sufficiently fast for the measurement outcomes to be an accurate reflection of the state at the beginning of the quench.  For each quench point, we performed $20$ programming cycles. 

\subsection{SQA and SVMC Simulations}
\label{sec:SQAandSVMC}
We also compared performance to simulated quantum annealing (SQA) \cite{Santoro} and spin vector Monte Carlo (SVMC) \cite{SSSV} at intermediate points in the anneal. SQA is a version of the quantum Monte Carlo method that can capture thermal averages at intermediate points in the anneal but is unable to simulate unitary quantum dynamics. SVMC is also a Monte Carlo algorithm but replaces the qubits with classical angles; it can be considered the semi-classical limit of quantum annealing, where each qubit is a pure state on the $x$-$z$ plane of the Bloch sphere, but correlations between qubits are purely classical~\cite{Albash:2014if,Crowley:2016aa}. For both SQA and SVMC, we used a temperature of $12$mK. To simulate the effect of quenching described above, we used $500$ sweeps to change the annealing parameters from their intermediate value to the final value. For each Hamiltonian at each quench point, we ran both SQA and SVMC with between $2500$ and about $10^7$ sweeps (incremented on a logarithmic scale) on the same encoded physical problem Hamiltonian sent to DW. We then selected the number of sweeps that gave an effective inverse temperature on the logical problem distribution that was closest to DW's for each Hamiltonian at each quench point. Because each (noisy) Hamiltonian realization generated by the DW programming cycle produces a slightly different distribution resulting in a slightly different effective temperature and distance from Gibbs, instead of adding noise to the SQA and SVMC simulations we ran SQA and SVMC once at each quench point for each final Hamiltonian and selected the number of sweeps that gave the closest effective temperature for each of the outputs of the $20$ noisy realizations of the DW runs. 

\begin{figure}[t]
\includegraphics[width=\columnwidth]{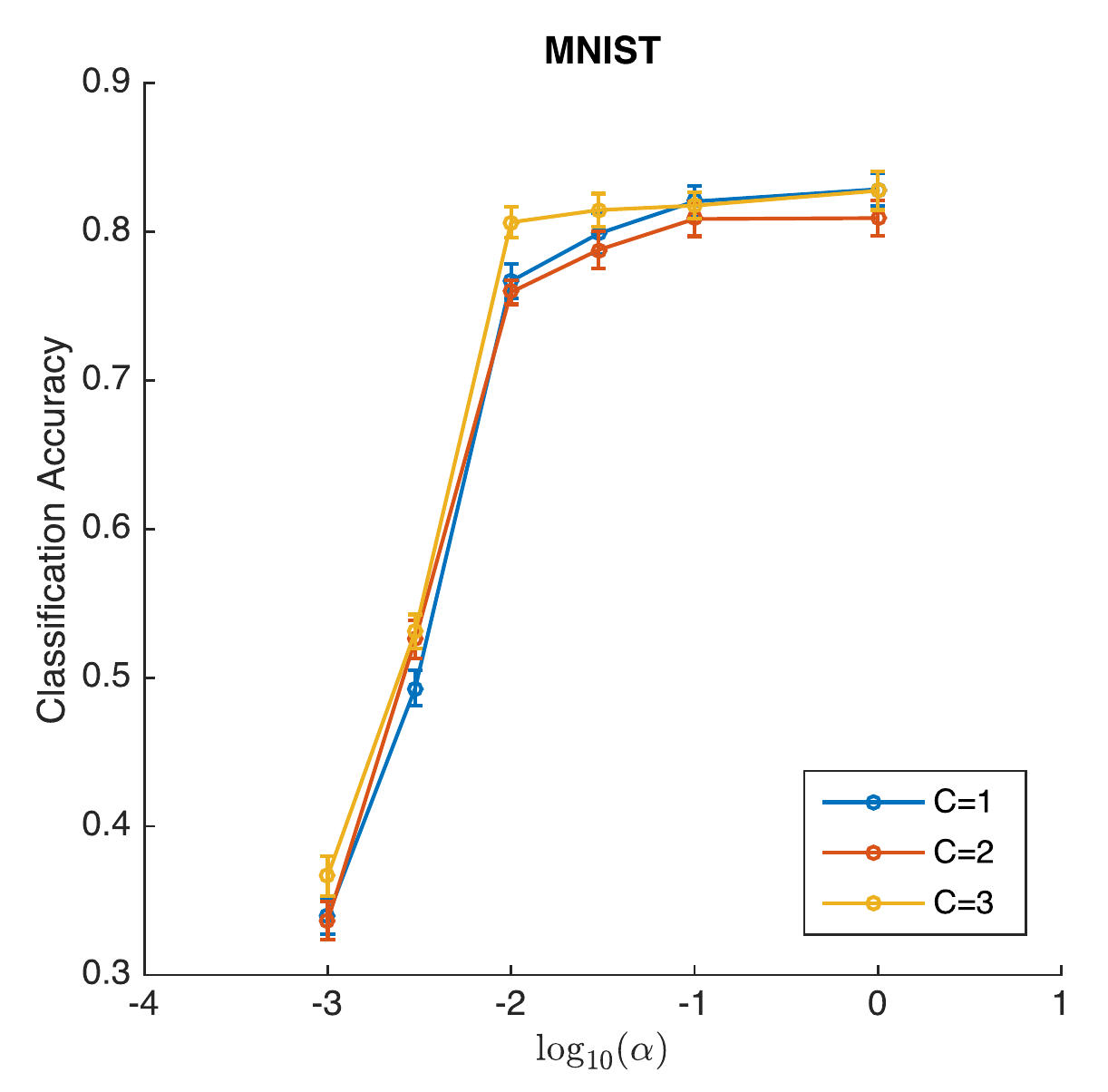}
\caption{Training performance using NQAC on the coarse-grained MNIST dataset as a function of the scaling parameter $\alpha$, for different nesting levels $C$ and $t_f = 10 \mu s$. Data points correspond to the mean performance over $50$ tests, where each test used $100$ random images. Performance across $C$ values remains consistent across the entire range of $\alpha$, although $C=3$ gains a statistically significant advantage at $\alpha = 10^{-2}$.
Here and in all subsequent figures the error bars represent two standard deviations ($95\%$ confidence intervals), unless explicitly stated otherwise.}
\label{fig:MNISTResultsvsAlpha}
\end{figure}



\begin{figure*}[t]
\includegraphics[width=\textwidth]{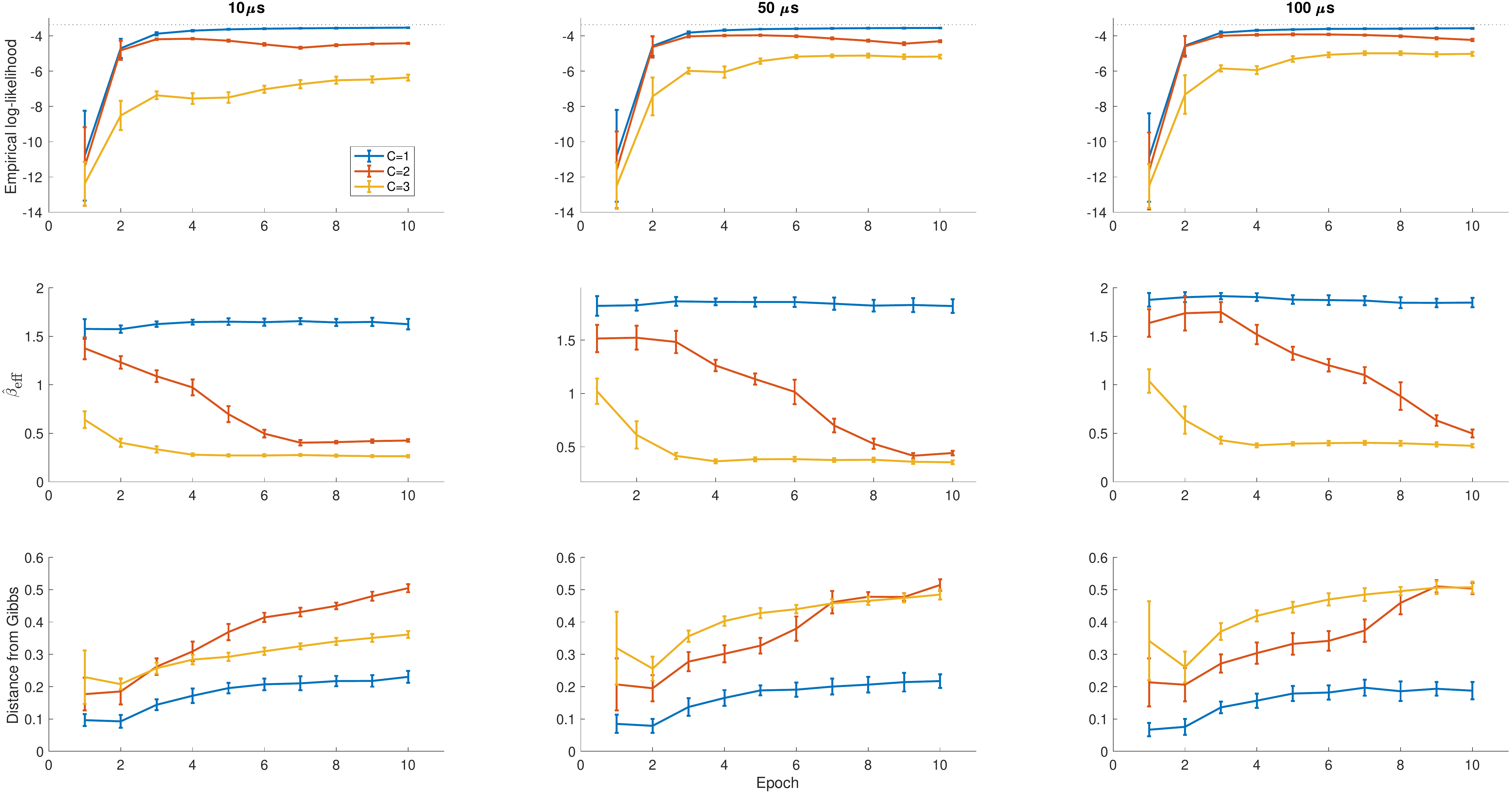}
\caption{NQAC training results for the BAS dataset at $\alpha=1$, with different anneal times  $t_f$.   The empirical log-likelihood is computed using Eq.~\eqref{eq:emp-ll}, whose maximum possible value in this case is $-3.41$, shown with a dotted line. Training consistently improves the empirical log-likelihood for $C=1$, but $C=2,3$ exhibit fluctuations. The dimensionless temperature $1/{\beta}_{\textrm{eff}}$ is constant for $C=1$ but increases for $C=2,3$. Likewise, the distance from the Gibbs distribution increases. For $C=1$, the empirical log-likelihood exhibits minimal variation with $t_f$, while the maximum ${\beta}_{\textrm{eff}}$ exhibits some growth with $t_f$. For $C=2, 3$, the performance is hampered by sub-optimality of the penalty strength, which reaches the maximum attainable value imposed by the hardware. This is reflected in the dimensionless inverse-temperature, in that the training for $C=2,3$ is unable to reach the same values as the $C=1$ case. 
}
\label{fig:ResultsvsAT_alpha=1}
\end{figure*}

\begin{figure*}[t]
\begin{center}
 \includegraphics[width=\linewidth]{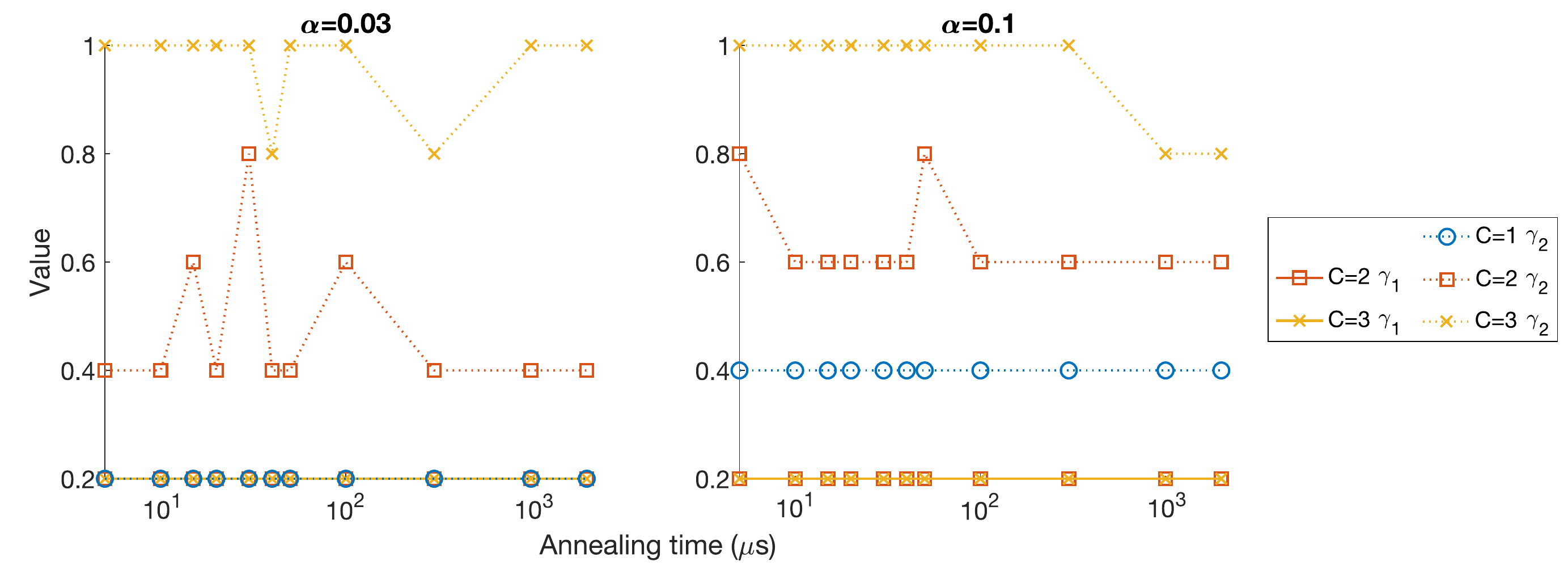}
\caption{Optimal penalty values found for the training results on the BAS dataset at $\alpha = 0.03$ and $0.1$ for different annealing times.}
\label{fig:optimalgamma}
\end{center}
\end{figure*}

\begin{figure*}[t]
\subfigure[\ $\alpha=0.03$]{\includegraphics[width=\textwidth]{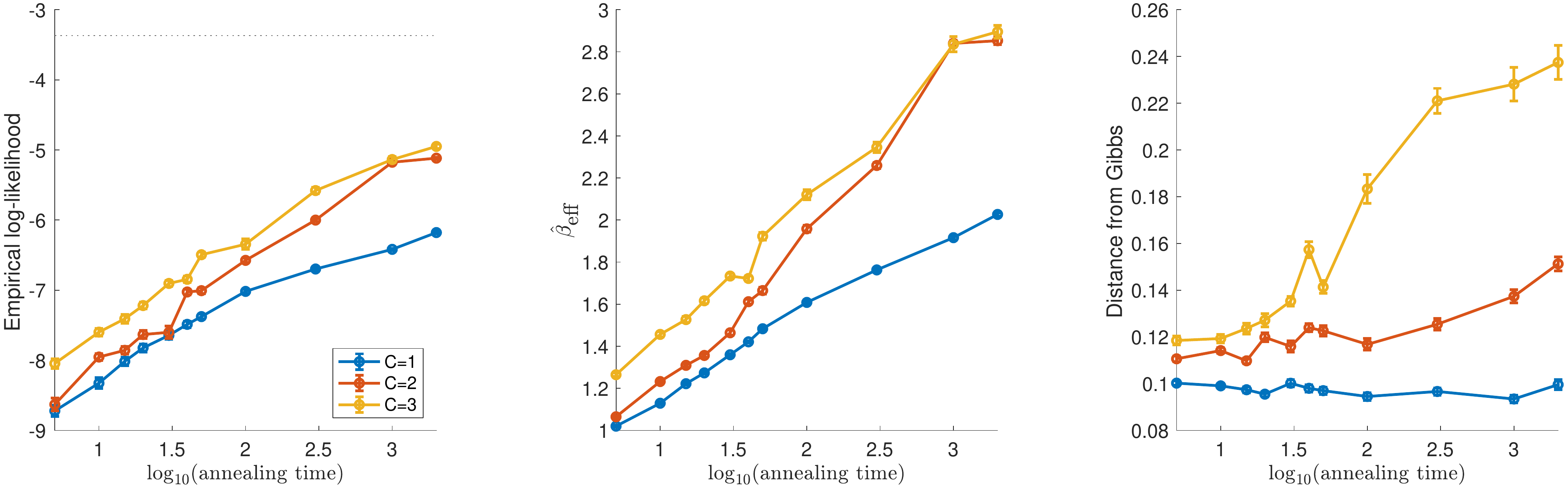}\label{fig:ResultsvsAT_alpha=0.03}}
\subfigure[\ $\alpha=0.1$]{\includegraphics[width=\textwidth]{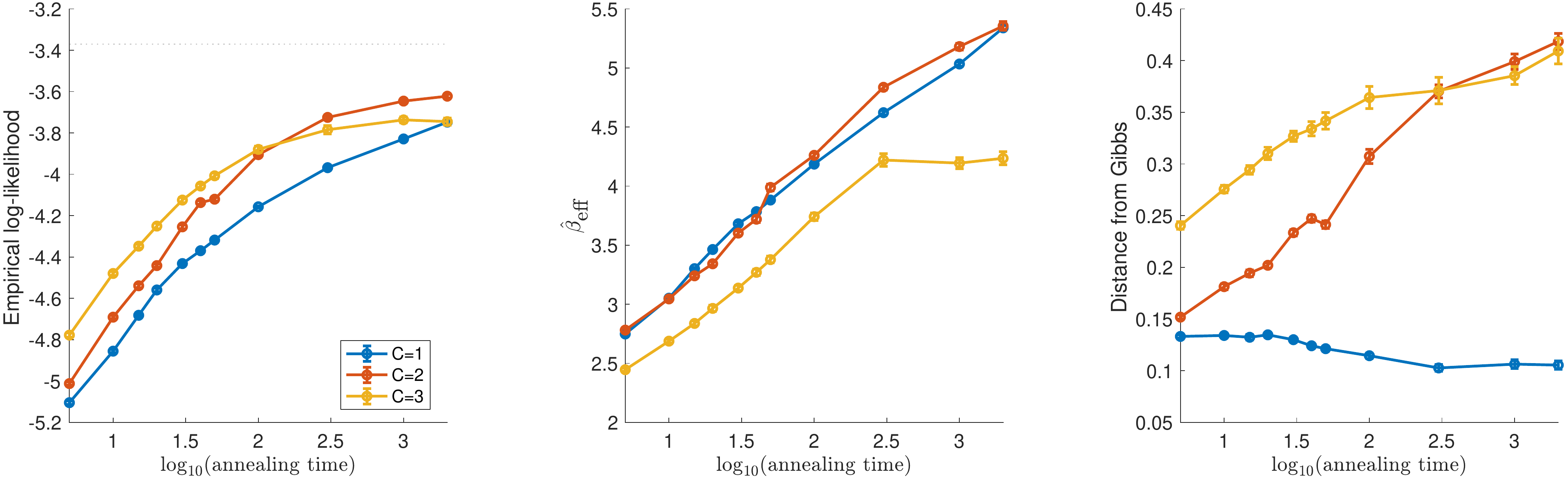}\label{fig:ResultsvsAT_alpha=0.1}}
\caption{Effect of using NQAC with different anneal times for the BAS dataset, measured in microseconds, for $\alpha=0.03$ (top row) and $\alpha=0.1$ (bottom row). The maximum possible value of the empirical log-likelihood in this case is $-3.41$, shown with a dotted line. Both the empirical log-likelihood and inverse temperature exhibit statistically significant growth with anneal time $t_f$ and nesting level $C$. The distance from Gibbs grows with $C$ and also with $t_f$ for $C=2,3$.}
\label{fig:ResultsvsAT}
\end{figure*}

\begin{figure*}[t]
\includegraphics[width=\textwidth]{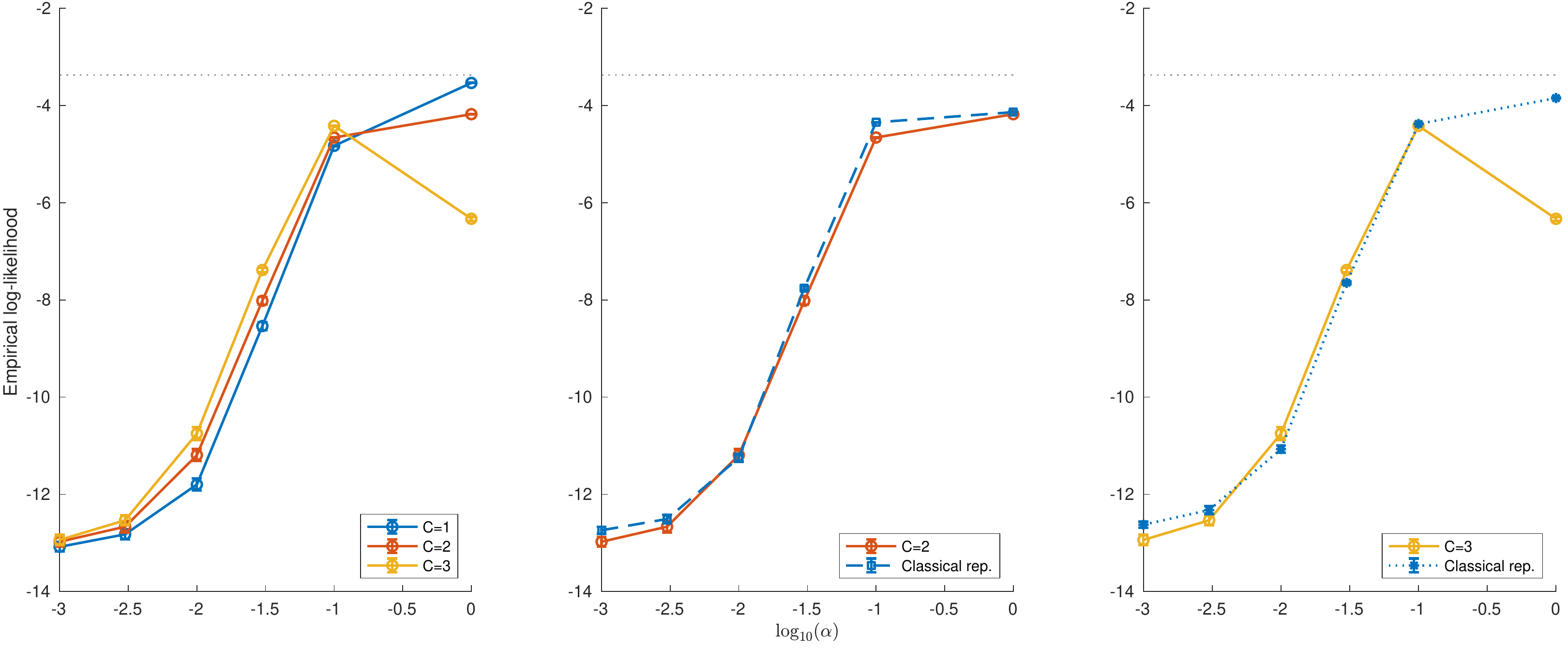}
 \caption{Training performance using NQAC on the BAS dataset after $10$ training epochs. A small advantage with increasing nesting level $C$ is noticeable for the BAS dataset for intermediate values of the scaling parameter $\alpha$. For the result shown here, all runs were done with an anneal time $t_f$ of $10 \mu$s. 
The maximum possible value of the empirical log-likelihood for the BAS dataset used is $-3.41$, indicated by the dotted lines. The left panel compares different nesting levels. The middle and left panels compare the performance of a classical repetition versus NQAC (see Sec.~\ref{sec:methods_classical_repetition} for a description). 
}
\label{fig:BASResultsvsAlpha}
\end{figure*}

\begin{figure*}[t]
\subfigure[\ $C=2$]{\includegraphics[width=\linewidth]{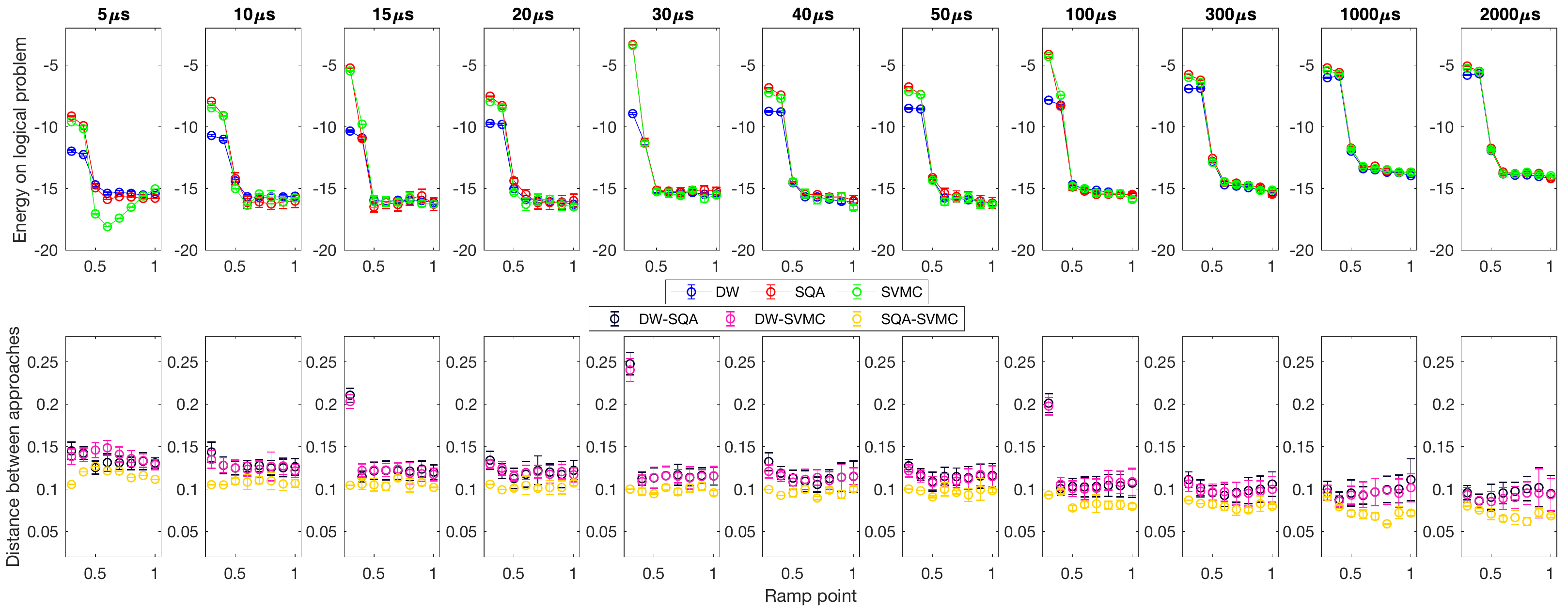}\label{fig:dw_sqa_svmc_closest_to_boltzmann_C=2}}
\subfigure[\ $C=3$]{\includegraphics[width=\linewidth]{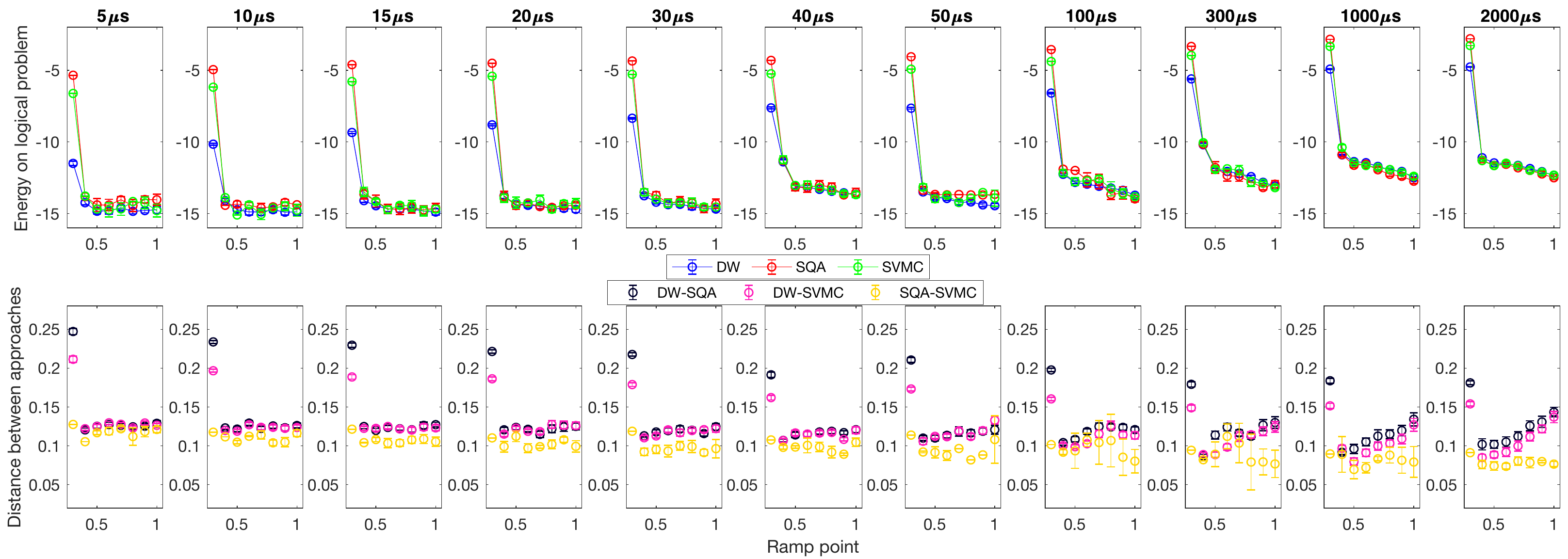}\label{fig:dw_sqa_svmc_closest_to_boltzmann_C=3}}
\caption{Direct comparison between DW, SQA, and SVMC for (a) $C=2$ and (b) $C=3$ at $\alpha=0.03$. In the first row, blue, red and green represent the average energy of DW, SQA and SVMC, respectively. In the second row, black, pink and yellow represent the distribution distances between DW and SQA, DW and SVMC, and SQA and SVMC, respectively. We chose the number of sweeps of SQA and SVMC to match DW's $\beta_{\mathrm{eff}}$ at each anneal time and quench point (see Appendix~\ref{app:sweeps}).}
\label{fig:dw_sqa_svmc_closest_to_boltzmann}
\end{figure*}

\section{Results}
\label{sec:results}

In this section we present our results testing NQAC's ability to improve machine learning performance on the BAS dataset and the coarse-grained MNIST dataset. We consider performance as a function of the scaling parameter $\alpha$ [Eq.~\eqref{eq:alpha}] and the anneal time $t_f$, for different NQAC nesting levels. We also study the dependence on the quench point, and compare SQA and SVMC simulations.


\subsection{Training performance as a function of scaling parameter $\alpha$}
\label{sec:perf-alpha}

\subsubsection{MNIST dataset}

For the MNIST data, we show in Fig.~\ref{fig:MNISTResultsvsAlpha} the classification accuracy over $50$ different groups of $100$ random images. We observe two distinct regimes in the classification accuracy as a function of the scaling parameter $\alpha$.
For small $\alpha$, where we expect the noise to be high, the classification accuracy grows almost linearly with $\log \alpha$ before entering a regime of slower growth.  In both regimes, we observe that there is no significant performance difference between $C=1$ and higher $C$ values, although for $\alpha \leq 10^{-2}$ (when the effect of noise should be very high) we begin to observe a small improvement for $C=3$, which is most pronounced at $\alpha=10^{-2}$. As $\alpha$ increases, NQAC becomes less and less effective and eventually $C=1$ overtakes $C=2$ before it catches up with $C=3$. The performance results shown always use optimal penalty values for both $\gamma_1$ and $\gamma_2$, except for $C=1$ where only the minor-embedding penalty strength $\gamma_2$ needs to be optimized. The hardware-imposed limit $\gamma_1,\gamma_2\leq 1$ is not reached, except for $C=2$ at the highest $\alpha$ value, which may explain why $C=1$ overtakes $C=2$; see Appendix~\ref{app:opt-gamma} for further details. We stress, however, that because the results for each $\alpha$ correspond to a single Boltzmann machine instance, we cannot conclude whether this is the typical classification accuracy over different training runs.

Note that simple out-of-the-box classical methods like a logistic regression or a nearest neighbor classifier get accuracies of more than $90\%$ (e.g., using the scikit-learn package in Python \cite{scikit-learn}). The results shown in Fig.~\ref{fig:MNISTResultsvsAlpha} are not aimed at comparing to classical methods, but to test whether the use of NQAC improves machine learning performance with a quantum annealer. 
Despite the small but statistically significant performance improvement seen in Fig.~\ref{fig:MNISTResultsvsAlpha} for $C=3$ for $\alpha \leq 10^{-2}$, it is certainly not the case that NQAC affords a compelling advantage. This is likely due to the fact that 
the minor-embedding chains grow longer as a function of $C$, making the system more susceptible to `chain-breaking' errors. Indeed, we find that the associated penalty parameter $\gamma_2$ is generally larger as $C$ grows (see Appendix~\ref{app:opt-gamma}), as would be required to suppress such errors. Hence, our results suggest that the performance on the MNIST database is dominated by the cost of minor-embedding, and there may only be a small window where NQAC is effective at overcoming this cost. We likely need to await the next generation of quantum annealers with better qubit connectivity~\cite{DWave-Pegasus-techreport} to test whether an increase in nesting level can provide a significant advantage for the MNIST dataset.

\subsubsection{BAS dataset} 
\label{sec:BAS-ta}

We now consider  the BAS dataset and training at  several anneal times.  At each $t_f$, we repeated the same procedure: we did a grid search in steps of $0.2$ for the values of $\gamma_1$ and $\gamma_2$ and selected the values that gave the best performance at the end of $5$ epochs. We then took those optimal values of the penalties and trained on the entire dataset for $10$ epochs.\footnote{We note that there could be a drift in the optimal penalty values after $5$ epochs. However, in most machine learning contexts, hyperparameters are trained for a few iterations and then assumed to remain constant. We adopt the same strategy here to avoid what could otherwise become an optimization over the entire space of parameters.} 

Figure~\ref{fig:ResultsvsAT_alpha=1} shows our results on the BAS dataset for $\alpha = 1$ and several anneal times. The case $\alpha=1$ is the one of most practical interest.

We first discuss the case of $C = 1$, i.e., without nesting.
For $C=1$, we observe that the dimensionless inverse-temperature $\hat{\beta}_{\textrm{eff}}$ [Eq.~\eqref{eq:betaNorm}] remains effectively constant, even as far back as the first few epochs of training.  Since the early training Hamiltonians have small norm, the effective temperature $1/{\beta}_{\textrm{eff}}$ of the distribution is initially small and grows with increasing training epochs.  This suggests two things: first, lower temperatures are likely to be useful earlier in the training, and second, the current DW device provides an adequate temperature range for training on the BAS dataset for the sizes considered here. 

As the anneal time increases, the $C=1$ distributions attain a lower dimensionless effective temperature as well as a slight reduction in distance from the associated Gibbs distribution.  The latter is consistent with the system having more time to thermalize.  However, the Gibbs distance consistently increases with training epochs, indicating that as the trained Hamiltonian becomes more complex, thermalization to the logical Gibbs state becomes more difficult.

We now consider the effect of nesting ($C>1$).  We observe that the training with NQAC underperforms, with higher $C$ values performing worse than lower $C$ values consistently as a function of both training time and anneal times used.  In particular we note that $C=1$ consistently has a lower dimensionless temperature, suggesting (perhaps counterintuitively) that the distributions generated for $C>1$ are effectively too warm.
This is very likely due to entering the `penalty-limited regime,' as was also the case in previous NQAC work~\cite{Vinci:2017ab}. In this regime, the optimal strength of $\gamma_1$ and $\gamma_2$ is greater than $1$, 
but as noted earlier, hardware limits prevent the penalties from reaching their optimal value. In this regime, we might expect that the decoding of the NQAC states does not preserve the ordering of the low-lying states of $H_P$, resulting in a distribution that appears warm. As can be seen in Appendix~\ref{app:opt-gamma}, the minor embedding penalty indeed becomes hardware limited for $\alpha=1$.

Next, we consider the same performance metrics for smaller $\alpha$ values, a proxy for increased physical temperature. We also expect datasets with larger problem sizes to require lower effective temperatures~\cite{Albash:2017ab}, and in this case one would expect NQAC to provide an advantage by lowering the effective temperature~\cite{Matsuura:2018,Vinci:2017ab,vinci2015nested}.

We first show in Fig.~\ref{fig:optimalgamma} the optimal penalty values at two smaller $\alpha$ values, $\alpha =0.03,0.1$.  We find that while reducing the overall energy scale of the problem Hamiltonian allows the $C=2$ case to have optimal penalties below 1 for all anneal times, the case of $C=3$ continues to require penalty values greater than $1$. We therefore expect that our results for $C=3$ to also be penalty limited.  

We note further that for both $\alpha$ values shown, for $C=2,3$ the code penalty $\gamma_1$ attains the minimum value of $0.2$ in the set of values we tried. This appears counterintuitive since this penalty is expected to energetically suppress bit flip errors.  
However, the fact that the minor embedding penalty $\gamma_2$ is maximized at the same time, suggests a tradeoff whereby the low $\gamma_1$ value allows for clusters of code qubits to be flipped more easily, while the large $\gamma_2$ value reduces the number of broken chains to ensure successful chain decoding. It is also possible that the values found by our optimization procedure reflect a local optimum, and that a more balanced combination of $\gamma_1$ and $\gamma_2$ could be found via a more exhaustive search.

Figure~\ref{fig:ResultsvsAT} shows the empirical log-likelihood for the BAS dataset versus $t_f$, for different nesting levels $C$ and two smaller $\alpha$ values. As shown in the left panels, the empirical log-likelihood for the BAS dataset exhibits a small but statistically significant (at the $2\sigma$ level) improvement over a wide range of anneal times.  We note that for $\alpha = 0.03$ [Fig.~\ref{fig:ResultsvsAT_alpha=0.03}], the advantage grows with increasing anneal time, although there is only a small difference between $C=2$ and $C=3$.  For $\alpha = 0.1$ [Fig.~\ref{fig:ResultsvsAT_alpha=0.1}], the advantage for $C=2$ and $C=3$ is present at almost all anneal times, although the advantage starts to decrease for the largest anneal times.  For $C=3$, the advantage vanishes at the longest anneal time.

In order to better understand this performance behavior, we can consider the middle panels of Fig.~\ref{fig:ResultsvsAT}. We first note that the dimensionless effective inverse temperature increases steadily with increasing anneal time.  This indicates that training with longer anneal times results in a distribution that is effectively colder, and based on the results from the left panels, in better performance.  For $\alpha = 0.03$ [Fig.~\ref{fig:ResultsvsAT_alpha=0.03}], the dimensionless effective temperature for $C=3$ is consistently lower than for $C=2$, which in turn is consistently lower than for $C=1$.  This suggests a correlation between improved performance and the ability to achieve effectively colder distributions.  However, the right panel provides an important caveat.  The sampled states for $C=3$ are significantly further away from the corresponding Gibbs distribution relative to $C=2$ for long anneal times.  Together, the three panels suggest that while a colder effective distribution is generically useful in the high noise regime and can be achieved with a higher $C$, there is a price to be paid with increasing distance from Gibbs, which can then begin to hurt performance. 

Similar observations can be made for $\alpha = 0.1$ [Fig.~\ref{fig:ResultsvsAT_alpha=0.1}].  For $C=2$, the effective temperature is consistently lower than that of $C=1$, and the performance is consistently better than $C=1$ for all anneal times.  We observe a decreasing advantage for large $t_f$, which may be attributed again to the growing distance from Gibbs. For $C=3$, the interpretation of the results is complicated by the fact that we are penalty limited for most annealing times (see Fig.~\ref{fig:optimalgamma}).  We believe this is the reason that we observe a higher dimensionless effective temperature even if the performance beats $C=2$ at low anneal times.


For $C=2$, the growing distance from Gibbs cannot be attributed to being `penalty-limited', since the optimal penalties are always below the device limit of $1$, as seen in Fig.~\ref{fig:optimalgamma} (except at $\alpha=1$; see Appendix~\ref{app:opt-gamma}).  The growing deviation from Gibbs may appear unexpected in light of the adiabatic theorem for open systems~\cite{oreshkov_adiabatic_2010,Avron:2012tv,Venuti:2015kq}, but we recall that the optimal penalty is chosen to maximize the log-likelihood and not to minimize the distance from Gibbs. It is likely that in the high noise (low $\alpha$) case, the optimal penalty values are such that the decoding procedure does not map the low lying spectrum of the implemented Hamiltonian to the logical Hamiltonian. This would explain why the distance from Gibbs grows, since the latter is measured relative to the Gibbs distribution of the logical problem. The mechanism could be that the optimal penalty values favor weakly coupled clusters of spins to help them flip. This is likely an outcome of decoherence that prevents coherent multi-qubit spin flips.

Taken together, these results indicate a complicated relationship between performance and anneal time for NQAC. For $C=1$, the trend is clear: longer anneal times result in colder and more Gibbs-like distributions.  For NQAC, the decoding procedure likely distorts this picture: better performance can be achieved by lowering the effective temperature (maybe by facilitating large-cluster spin flips via low penalties) even if the resulting distribution departs from being Gibbs (because the ordering of states is not preserved). Nevertheless, \emph{despite the increase in distance from the Gibbs distribution with $C$, the measures of machine learning performance improve with longer anneal times}, indicating that some meaningful learning is still taking place. However, the advantage this can give does not appear to be indefinite and appears to be related to the noise level, as possibly indicated by the leveling of the performance results for $\alpha = 0.1$.


So far, our training performance comparison has ignored the extra qubit resources required by NQAC. Recall, as explained in Sec.~\ref{sec:methods_classical_repetition}, that a fair comparison should account for these resources, and compare NQAC at $C>1$ to a classical repetition of $C=1$ with the same number of physical qubits. The result is shown in Fig.~\ref{fig:BASResultsvsAlpha}, for $t_f=10\mu$s. First, in the left panel we compare different nesting levels and observe that increasing $C$ helps for sufficiently low $\alpha$, where the results are not penalty limited. The middle and right panels then address the equalization of resources question, by comparing $C=2$ and $C=3$ to repetitions of $C=1$. In the repetition case we simply used $C=1$ multiple times and took the best of these repetitions, as in previous work~\cite{PAL:13,PAL:14,vinci2015nested}. At very low values of $\alpha$, classical repetition performs better than NQAC. There is no improvement for $C=2$ (middle panel of Fig.~\ref{fig:BASResultsvsAlpha}). However, for $C=3$ there is a small, but statistically significant advantage at the 95\% confidence level in using NQAC at intermediate values of $\alpha$. Given the sub-optimal penalty values for $C=3$ (Fig.~\ref{fig:optimalgamma}), it is unclear whether the enhancement can actually be more substantial. What is clear is that at the highest $\alpha$ values, the performance of $C=3$ is seriously hindered.

\subsection{Comparison to SQA and SVMC} 
\label{sec:comp_SQA-SVMC}

Our discussion so far has only relied on the output distribution of the DW processor at the end of the anneal.  We can attempt to better understand the origin of this distribution by using the `annealing schedule variation' feature of the DW2000Q devices (see Sec.~\ref{sec:quench} for more details) and comparing it to the distributions generated by SQA and SVMC. Recall that SVMC is a purely classical model of interacting $O(2)$ rotors subject to the semiclassical DW Hamiltonian, while SQA is a quantum Monte Carlo algorithm that is designed to converge to the instantaneous quantum Gibbs distribution (of the actual DW Hamiltonian) for a sufficiently large number of sweeps (see Sec.~\ref{sec:SQAandSVMC} for more details).  Figure~\ref{fig:dw_sqa_svmc_closest_to_boltzmann} shows the average energy and the distribution distance between the DW, SQA, and SVMC for $C=2$ and $C=3$ on the $11$ trained models (each corresponding to using a different $t_f$). We observe almost identical behavior among the three methods for different quench points, in particular a large change in the average energy as we increase the position of the quench.  This change is likely associated with crossing the minimum gap of the system, and the associated freezing transition.  This point moves to earlier in the anneal for larger $C$, which is consistent with the encoding raising the Ising energy scale and hence moving the global minimum gap to an earlier point in the anneal~\cite{Choi:19}.  We also notice that the larger $C$ value seems to make the transition sharper.

There is a consistent discrepancy between the approaches at the smallest quench points, where DW has a lower average energy than SVMC and SQA.  Since these small quench points have the system crossing the minimum gap, the discrepancy is likely due to using too few sweeps in SVMC and SQA to accurately represent the quench. The number of sweeps at each quench point is presented in Appendix~\ref{app:sweeps}.

We note that after crossing the minimum gap, the average energy of the decoded state does not change significantly anymore, indicative of the freeze-out point. Recall that this point is associated with the quasi-static regime effect~\cite{Amin:2015qf,Venuti:2015kq,Marshall:2017aa,Chancellor:2016pa,Albash:2017ab}, whereby the dynamics of a quantum annealer are dramatically slowed down after an intermediate point in the anneal.  This has the important consequence that if the system reaches a Gibbs state at an intermediate point in the anneal and freezes there, then we should not expect  agreement with the Gibbs state at the end of the anneal (which would correspond to the Gibbs distribution over the logical problem if the ordering of states is preserved).

The strong agreement between DW and SVMC suggests that quantum effects are not playing a significant role in determining the final distribution. Inasmuch as we trust SQA and SVMC as realistic simulations of the underlying physics, the fact that SQA and SVMC yield the smallest distance consistently strongly suggests that the temperature is simply too high to observe meaningful quantum effects like tunneling. This does not necessarily mean that tunneling cannot play a role at lower device temperatures. In Ref.~\cite{Albash:2017aa}, simulations demonstrated that at high temperature SQA and SVMC performed similarly, but at lower temperatures the two performed differently, most likely because SQA is able to simulate tunneling events, whereas SVMC can only make use of thermal hopping. 

Additional results comparing SQA and SVMC to DW are provided in Appendix~\ref{app:D}. In particular, we study the freezing transition in more detail using other metrics. 


\section{Discussion}
\label{sec:conclusions}

Previous work has demonstrated that NQAC can offer an effective temperature reduction by trading more physical qubits to create `colder' code qubits in optimization problems and estimation of gradients in the training of Boltzmann machines~\cite{Vinci:2017ab}. This previous result used a fixed Hamiltonian for a given problem size. Here, we have incorporated NQAC into training a machine learning model, where the parameters of the Hamiltonian are iteratively adjusted to match some target distribution. In order to be able to study the performance of NQAC on both the BAS and MNIST datasets, we had to rescale the Ising Hamiltonian by the scaling parameter $\alpha$, in order to reach accessible optimal penalty values.  Even so, we found that only for $C=2$ the optimal penalty value was generally within the accessible range of the DW2000Q device. Therefore any conclusions we draw about the performance with increasing $C$ are subject to this caveat.

We have shown, using both the BAS and MNIST datasets, that for intermediate values of $\alpha$ NQAC offers an increase in training performance with higher nesting level $C$ (in agreement with previous results), but when $\alpha$ is close to $1$, encoding is limited by the strength of the penalties and does not provide an advantage. For these intermediate values of $\alpha$, we tested whether NQAC outperforms classical repetition for the BAS dataset, and found this to be the case at the largest accessible nesting level ($C=3$), though the effect is small. However, we have also shown that the advantage in using NQAC increases with longer anneal times for $\alpha=$0.03 [see Fig.~\ref{fig:ResultsvsAT_alpha=0.03}].
Additional results at different values of $\alpha$ are presented in Appendix~\ref{app:alpha_at}, where trends are not as consistent. 

In the unsupervised case (BAS dataset) we also studied how the improvement in performance correlates with the effective temperature of the associated Gibbs distribution; our results show that the improved learning is associated with a decrease in the effective temperature. In addition, we found that longer anneal times increase machine learning training performance across all nesting levels, but --- counterintuitively --- the distribution distance from a Gibbs distribution with respect to the final Hamiltonian increased with anneal time. This shows that improved unsupervised learning can happen without improved equilibration with respect to the target Hamiltonian. 

Our results indicate a complicated relationship between performance, penalty strengths, and anneal time for NQAC. It is clear that improved performance can be had at the expense of closeness to the Gibbs distribution for short and mid-range annealing times, but this gain diminishes at large annealing times. 

To understand the details of the changes in the effective temperature and distribution distance, we used the ``annealing schedule variation'' feature of the DW2000Q device to probe the state of the system at intermediate points in the anneal, and made comparisons to SQA and SVMC to gain some physical insight into the sampling process. Overall, our results agree quite well with previous understanding of the dynamics of a quantum annealer in the quasi-static regime: at some intermediate point in the anneal, dynamics are frozen and the distribution (and various expectation values of the distribution) do not change significantly after this freezing point. Strong agreement with SQA and SVMC suggests that DW is sampling from the same semiclassical distribution at intermediate points in the anneal, though this agreement weakens at longer anneal times.

Caveats about our conclusions include generalizability to problems of larger size and other datasets, and whether increasing the nesting level will continue to provide an advantage. For the size of the problems we have examined here, misspecification of the Ising Hamiltonian parameters, which are typically estimated to be around $0.03$ in the units of the coupling parameters $J_{ij}$ of the DW device, do not impact results to the point where learning cannot be achieved. The effect of misspecification errors is to yield distributions that can resemble warmer Boltzmann distributions, at least in terms of the distribution of energy eigenstates~\cite{albash_2019_analogerrors}. For sufficiently large misspecification errors and temperature of the distribution, we should not expect any meaningful learning to take place. Significant mitigation against misspecification errors can be achieved even with simple QAC at $C=1$~\cite{Pearson:2019aa}, but we do not expect the behavior shown in here to be scalable unless we can increase the code distance without entering the penalty-limited regime and simultaneously avoid affecting the thermalization. 

One additional complication that we did not discuss here in detail is the effect of the decoding strategy for broken chains. Majority voting can distort the distribution and has been shown to be a suboptimal decoding strategy for thermal sampling problems~\cite{1909.12184}; replacing it in the context of our work with a better strategy is an interesting topic for future investigations. Nonetheless, without any decoding, learning suffers significantly for higher $C$. In Appendix~\ref{app:decoding} we present additional results where no decoding is used (i.e., broken chains are discarded) and find worse performance. These caveats notwithstanding, we remain hopeful that building upon the results demonstrated here, NQAC, or improved error suppression and correction methods, may help in finding a machine learning advantage for noisy intermediate-scale quantum annealers. 

\section{Acknowledgements}
Computation for the work described in this paper was supported by the University of Southern California’s Center for High-Performance Computing (hpcc.usc.edu). This research is based upon work (partially) supported by the Office of the Director of National Intelligence (ODNI), Intelligence Advanced Research Projects Activity (IARPA), via the U.S. Army Research Office contract W911NF-17-C-0050. The views and conclusions contained herein are those of the authors and should not be interpreted as necessarily representing the official policies or endorsements, either expressed or implied, of
the ODNI, IARPA, DARPA, or the U.S. Government. The U.S. Government is authorized to reproduce and distribute reprints for Governmental purposes notwithstanding any copyright annotation thereon. This work is partially supported by DOE/HEP QuantISED program grant, Quantum Machine Learning and Quantum Computation Frameworks (QMLQCF) for HEP, award number DE-SC0019227.

\bibliography{refs}

\begin{thebibliography}{65}%
\makeatletter
\providecommand \@ifxundefined [1]{%
 \@ifx{#1\undefined}
}%
\providecommand \@ifnum [1]{%
 \ifnum #1\expandafter \@firstoftwo
 \else \expandafter \@secondoftwo
 \fi
}%
\providecommand \@ifx [1]{%
 \ifx #1\expandafter \@firstoftwo
 \else \expandafter \@secondoftwo
 \fi
}%
\providecommand \natexlab [1]{#1}%
\providecommand \enquote  [1]{``#1''}%
\providecommand \bibnamefont  [1]{#1}%
\providecommand \bibfnamefont [1]{#1}%
\providecommand \citenamefont [1]{#1}%
\providecommand \href@noop [0]{\@secondoftwo}%
\providecommand \href [0]{\begingroup \@sanitize@url \@href}%
\providecommand \@href[1]{\@@startlink{#1}\@@href}%
\providecommand \@@href[1]{\endgroup#1\@@endlink}%
\providecommand \@sanitize@url [0]{\catcode `\\12\catcode `\$12\catcode
  `\&12\catcode `\#12\catcode `\^12\catcode `\_12\catcode `\%12\relax}%
\providecommand \@@startlink[1]{}%
\providecommand \@@endlink[0]{}%
\providecommand \url  [0]{\begingroup\@sanitize@url \@url }%
\providecommand \@url [1]{\endgroup\@href {#1}{\urlprefix }}%
\providecommand \urlprefix  [0]{URL }%
\providecommand \Eprint [0]{\href }%
\providecommand \doibase [0]{http://dx.doi.org/}%
\providecommand \selectlanguage [0]{\@gobble}%
\providecommand \bibinfo  [0]{\@secondoftwo}%
\providecommand \bibfield  [0]{\@secondoftwo}%
\providecommand \translation [1]{[#1]}%
\providecommand \BibitemOpen [0]{}%
\providecommand \bibitemStop [0]{}%
\providecommand \bibitemNoStop [0]{.\EOS\space}%
\providecommand \EOS [0]{\spacefactor3000\relax}%
\providecommand \BibitemShut  [1]{\csname bibitem#1\endcsname}%
\let\auto@bib@innerbib\@empty
\bibitem [{\citenamefont {Harris}\ \emph {et~al.}(2010)\citenamefont {Harris},
  \citenamefont {Johnson}, \citenamefont {Lanting}, \citenamefont {Berkley},
  \citenamefont {Johansson}, \citenamefont {Bunyk}, \citenamefont {Tolkacheva},
  \citenamefont {Ladizinsky}, \citenamefont {Ladizinsky}, \citenamefont {Oh},
  \citenamefont {Cioata}, \citenamefont {Perminov}, \citenamefont {Spear},
  \citenamefont {Enderud}, \citenamefont {Rich}, \citenamefont {Uchaikin},
  \citenamefont {Thom}, \citenamefont {Chapple}, \citenamefont {Wang},
  \citenamefont {Wilson}, \citenamefont {Amin}, \citenamefont {Dickson},
  \citenamefont {Karimi}, \citenamefont {Macready}, \citenamefont {Truncik},\
  and\ \citenamefont {Rose}}]{Harris:2010kx}%
  \BibitemOpen
  \bibfield  {author} {\bibinfo {author} {\bibfnamefont {R.}~\bibnamefont
  {Harris}}, \bibinfo {author} {\bibfnamefont {M.~W.}\ \bibnamefont {Johnson}},
  \bibinfo {author} {\bibfnamefont {T.}~\bibnamefont {Lanting}}, \bibinfo
  {author} {\bibfnamefont {A.~J.}\ \bibnamefont {Berkley}}, \bibinfo {author}
  {\bibfnamefont {J.}~\bibnamefont {Johansson}}, \bibinfo {author}
  {\bibfnamefont {P.}~\bibnamefont {Bunyk}}, \bibinfo {author} {\bibfnamefont
  {E.}~\bibnamefont {Tolkacheva}}, \bibinfo {author} {\bibfnamefont
  {E.}~\bibnamefont {Ladizinsky}}, \bibinfo {author} {\bibfnamefont
  {N.}~\bibnamefont {Ladizinsky}}, \bibinfo {author} {\bibfnamefont
  {T.}~\bibnamefont {Oh}}, \bibinfo {author} {\bibfnamefont {F.}~\bibnamefont
  {Cioata}}, \bibinfo {author} {\bibfnamefont {I.}~\bibnamefont {Perminov}},
  \bibinfo {author} {\bibfnamefont {P.}~\bibnamefont {Spear}}, \bibinfo
  {author} {\bibfnamefont {C.}~\bibnamefont {Enderud}}, \bibinfo {author}
  {\bibfnamefont {C.}~\bibnamefont {Rich}}, \bibinfo {author} {\bibfnamefont
  {S.}~\bibnamefont {Uchaikin}}, \bibinfo {author} {\bibfnamefont {M.~C.}\
  \bibnamefont {Thom}}, \bibinfo {author} {\bibfnamefont {E.~M.}\ \bibnamefont
  {Chapple}}, \bibinfo {author} {\bibfnamefont {J.}~\bibnamefont {Wang}},
  \bibinfo {author} {\bibfnamefont {B.}~\bibnamefont {Wilson}}, \bibinfo
  {author} {\bibfnamefont {M.~H.~S.}\ \bibnamefont {Amin}}, \bibinfo {author}
  {\bibfnamefont {N.}~\bibnamefont {Dickson}}, \bibinfo {author} {\bibfnamefont
  {K.}~\bibnamefont {Karimi}}, \bibinfo {author} {\bibfnamefont
  {B.}~\bibnamefont {Macready}}, \bibinfo {author} {\bibfnamefont {C.~J.~S.}\
  \bibnamefont {Truncik}}, \ and\ \bibinfo {author} {\bibfnamefont
  {G.}~\bibnamefont {Rose}},\ }\bibfield  {title} {\enquote {\bibinfo {title}
  {Experimental investigation of an eight-qubit unit cell in a superconducting
  optimization processor},}\ }\href {\doibase 10.1103/PhysRevB.82.024511}
  {\bibfield  {journal} {\bibinfo  {journal} {Phys. Rev. B}\ }\textbf {\bibinfo
  {volume} {82}},\ \bibinfo {pages} {024511} (\bibinfo {year}
  {2010})}\BibitemShut {NoStop}%
\bibitem [{\citenamefont {Johnson}\ \emph {et~al.}(2011)\citenamefont
  {Johnson}, \citenamefont {Amin}, \citenamefont {Gildert}, \citenamefont
  {Lanting}, \citenamefont {Hamze}, \citenamefont {Dickson}, \citenamefont
  {Harris}, \citenamefont {Berkley}, \citenamefont {Johansson}, \citenamefont
  {Bunyk}, \citenamefont {Chapple}, \citenamefont {Enderud}, \citenamefont
  {Hilton}, \citenamefont {Karimi}, \citenamefont {Ladizinsky}, \citenamefont
  {Ladizinsky}, \citenamefont {Oh}, \citenamefont {Perminov}, \citenamefont
  {Rich}, \citenamefont {Thom}, \citenamefont {Tolkacheva}, \citenamefont
  {Truncik}, \citenamefont {Uchaikin}, \citenamefont {Wang}, \citenamefont
  {Wilson},\ and\ \citenamefont {Rose}}]{Dwave}%
  \BibitemOpen
  \bibfield  {author} {\bibinfo {author} {\bibfnamefont {M.~W.}\ \bibnamefont
  {Johnson}}, \bibinfo {author} {\bibfnamefont {M.~H.~S.}\ \bibnamefont
  {Amin}}, \bibinfo {author} {\bibfnamefont {S.}~\bibnamefont {Gildert}},
  \bibinfo {author} {\bibfnamefont {T.}~\bibnamefont {Lanting}}, \bibinfo
  {author} {\bibfnamefont {F.}~\bibnamefont {Hamze}}, \bibinfo {author}
  {\bibfnamefont {N.}~\bibnamefont {Dickson}}, \bibinfo {author} {\bibfnamefont
  {R.}~\bibnamefont {Harris}}, \bibinfo {author} {\bibfnamefont {A.~J.}\
  \bibnamefont {Berkley}}, \bibinfo {author} {\bibfnamefont {J.}~\bibnamefont
  {Johansson}}, \bibinfo {author} {\bibfnamefont {P.}~\bibnamefont {Bunyk}},
  \bibinfo {author} {\bibfnamefont {E.~M.}\ \bibnamefont {Chapple}}, \bibinfo
  {author} {\bibfnamefont {C.}~\bibnamefont {Enderud}}, \bibinfo {author}
  {\bibfnamefont {J.~P.}\ \bibnamefont {Hilton}}, \bibinfo {author}
  {\bibfnamefont {K.}~\bibnamefont {Karimi}}, \bibinfo {author} {\bibfnamefont
  {E.}~\bibnamefont {Ladizinsky}}, \bibinfo {author} {\bibfnamefont
  {N.}~\bibnamefont {Ladizinsky}}, \bibinfo {author} {\bibfnamefont
  {T.}~\bibnamefont {Oh}}, \bibinfo {author} {\bibfnamefont {I.}~\bibnamefont
  {Perminov}}, \bibinfo {author} {\bibfnamefont {C.}~\bibnamefont {Rich}},
  \bibinfo {author} {\bibfnamefont {M.~C.}\ \bibnamefont {Thom}}, \bibinfo
  {author} {\bibfnamefont {E.}~\bibnamefont {Tolkacheva}}, \bibinfo {author}
  {\bibfnamefont {C.~J.~S.}\ \bibnamefont {Truncik}}, \bibinfo {author}
  {\bibfnamefont {S.}~\bibnamefont {Uchaikin}}, \bibinfo {author}
  {\bibfnamefont {J.}~\bibnamefont {Wang}}, \bibinfo {author} {\bibfnamefont
  {B.}~\bibnamefont {Wilson}}, \ and\ \bibinfo {author} {\bibfnamefont
  {G.}~\bibnamefont {Rose}},\ }\bibfield  {title} {\enquote {\bibinfo {title}
  {Quantum annealing with manufactured spins},}\ }\href
  {https://www.nature.com/nature/journal/v473/n7346/full/nature10012.html}
  {\bibfield  {journal} {\bibinfo  {journal} {Nature}\ }\textbf {\bibinfo
  {volume} {473}},\ \bibinfo {pages} {194--198} (\bibinfo {year}
  {2011})}\BibitemShut {NoStop}%
\bibitem [{\citenamefont {Bunyk}\ \emph {et~al.}(Aug. 2014)\citenamefont
  {Bunyk}, \citenamefont {Hoskinson}, \citenamefont {Johnson}, \citenamefont
  {Tolkacheva}, \citenamefont {Altomare}, \citenamefont {Berkley},
  \citenamefont {Harris}, \citenamefont {Hilton}, \citenamefont {Lanting},
  \citenamefont {Przybysz},\ and\ \citenamefont {Whittaker}}]{Bunyk:2014hb}%
  \BibitemOpen
  \bibfield  {author} {\bibinfo {author} {\bibfnamefont {P.~I}\ \bibnamefont
  {Bunyk}}, \bibinfo {author} {\bibfnamefont {E.~M.}\ \bibnamefont
  {Hoskinson}}, \bibinfo {author} {\bibfnamefont {M.~W.}\ \bibnamefont
  {Johnson}}, \bibinfo {author} {\bibfnamefont {E.}~\bibnamefont {Tolkacheva}},
  \bibinfo {author} {\bibfnamefont {F.}~\bibnamefont {Altomare}}, \bibinfo
  {author} {\bibfnamefont {AJ.}\ \bibnamefont {Berkley}}, \bibinfo {author}
  {\bibfnamefont {R.}~\bibnamefont {Harris}}, \bibinfo {author} {\bibfnamefont
  {J.~P.}\ \bibnamefont {Hilton}}, \bibinfo {author} {\bibfnamefont
  {T.}~\bibnamefont {Lanting}}, \bibinfo {author} {\bibfnamefont {AJ.}\
  \bibnamefont {Przybysz}}, \ and\ \bibinfo {author} {\bibfnamefont
  {J.}~\bibnamefont {Whittaker}},\ }\bibfield  {title} {\enquote {\bibinfo
  {title} {Architectural considerations in the design of a superconducting
  quantum annealing processor},}\ }\href {\doibase 10.1109/TASC.2014.2318294}
  {\bibfield  {journal} {\bibinfo  {journal} {IEEE Transactions on Applied
  Superconductivity}\ }\textbf {\bibinfo {volume} {24}},\ \bibinfo {pages}
  {1--10} (\bibinfo {year} {Aug. 2014})}\BibitemShut {NoStop}%
\bibitem [{\citenamefont {Lanting}\ \emph {et~al.}(2014)\citenamefont
  {Lanting}, \citenamefont {Przybysz}, \citenamefont {Smirnov}, \citenamefont
  {Spedalieri}, \citenamefont {Amin}, \citenamefont {Berkley}, \citenamefont
  {Harris}, \citenamefont {Altomare}, \citenamefont {Boixo}, \citenamefont
  {Bunyk}, \citenamefont {Dickson}, \citenamefont {Enderud}, \citenamefont
  {Hilton}, \citenamefont {Hoskinson}, \citenamefont {Johnson}, \citenamefont
  {Ladizinsky}, \citenamefont {Ladizinsky}, \citenamefont {Neufeld},
  \citenamefont {Oh}, \citenamefont {Perminov}, \citenamefont {Rich},
  \citenamefont {Thom}, \citenamefont {Tolkacheva}, \citenamefont {Uchaikin},
  \citenamefont {Wilson},\ and\ \citenamefont {Rose}}]{DWave-entanglement}%
  \BibitemOpen
  \bibfield  {author} {\bibinfo {author} {\bibfnamefont {T.}~\bibnamefont
  {Lanting}}, \bibinfo {author} {\bibfnamefont {A.~J.}\ \bibnamefont
  {Przybysz}}, \bibinfo {author} {\bibfnamefont {A.~Yu.}\ \bibnamefont
  {Smirnov}}, \bibinfo {author} {\bibfnamefont {F.~M.}\ \bibnamefont
  {Spedalieri}}, \bibinfo {author} {\bibfnamefont {M.~H.}\ \bibnamefont
  {Amin}}, \bibinfo {author} {\bibfnamefont {A.~J.}\ \bibnamefont {Berkley}},
  \bibinfo {author} {\bibfnamefont {R.}~\bibnamefont {Harris}}, \bibinfo
  {author} {\bibfnamefont {F.}~\bibnamefont {Altomare}}, \bibinfo {author}
  {\bibfnamefont {S.}~\bibnamefont {Boixo}}, \bibinfo {author} {\bibfnamefont
  {P.}~\bibnamefont {Bunyk}}, \bibinfo {author} {\bibfnamefont
  {N.}~\bibnamefont {Dickson}}, \bibinfo {author} {\bibfnamefont
  {C.}~\bibnamefont {Enderud}}, \bibinfo {author} {\bibfnamefont {J.~P.}\
  \bibnamefont {Hilton}}, \bibinfo {author} {\bibfnamefont {E.}~\bibnamefont
  {Hoskinson}}, \bibinfo {author} {\bibfnamefont {M.~W.}\ \bibnamefont
  {Johnson}}, \bibinfo {author} {\bibfnamefont {E.}~\bibnamefont {Ladizinsky}},
  \bibinfo {author} {\bibfnamefont {N.}~\bibnamefont {Ladizinsky}}, \bibinfo
  {author} {\bibfnamefont {R.}~\bibnamefont {Neufeld}}, \bibinfo {author}
  {\bibfnamefont {T.}~\bibnamefont {Oh}}, \bibinfo {author} {\bibfnamefont
  {I.}~\bibnamefont {Perminov}}, \bibinfo {author} {\bibfnamefont
  {C.}~\bibnamefont {Rich}}, \bibinfo {author} {\bibfnamefont {M.~C.}\
  \bibnamefont {Thom}}, \bibinfo {author} {\bibfnamefont {E.}~\bibnamefont
  {Tolkacheva}}, \bibinfo {author} {\bibfnamefont {S.}~\bibnamefont
  {Uchaikin}}, \bibinfo {author} {\bibfnamefont {A.~B.}\ \bibnamefont
  {Wilson}}, \ and\ \bibinfo {author} {\bibfnamefont {G.}~\bibnamefont
  {Rose}},\ }\bibfield  {title} {\enquote {\bibinfo {title} {Entanglement in a
  quantum annealing processor},}\ }\href {\doibase 10.1103/PhysRevX.4.021041}
  {\bibfield  {journal} {\bibinfo  {journal} {Phys. Rev. X}\ }\textbf {\bibinfo
  {volume} {4}},\ \bibinfo {pages} {021041--} (\bibinfo {year}
  {2014})}\BibitemShut {NoStop}%
\bibitem [{\citenamefont {Boixo}\ \emph {et~al.}(2016)\citenamefont {Boixo},
  \citenamefont {Smelyanskiy}, \citenamefont {Shabani}, \citenamefont {Isakov},
  \citenamefont {Dykman}, \citenamefont {Denchev}, \citenamefont {Amin},
  \citenamefont {Smirnov}, \citenamefont {Mohseni},\ and\ \citenamefont
  {Neven}}]{Boixo:2014yu}%
  \BibitemOpen
  \bibfield  {author} {\bibinfo {author} {\bibfnamefont {Sergio}\ \bibnamefont
  {Boixo}}, \bibinfo {author} {\bibfnamefont {Vadim~N.}\ \bibnamefont
  {Smelyanskiy}}, \bibinfo {author} {\bibfnamefont {Alireza}\ \bibnamefont
  {Shabani}}, \bibinfo {author} {\bibfnamefont {Sergei~V.}\ \bibnamefont
  {Isakov}}, \bibinfo {author} {\bibfnamefont {Mark}\ \bibnamefont {Dykman}},
  \bibinfo {author} {\bibfnamefont {Vasil~S.}\ \bibnamefont {Denchev}},
  \bibinfo {author} {\bibfnamefont {Mohammad~H.}\ \bibnamefont {Amin}},
  \bibinfo {author} {\bibfnamefont {Anatoly~Yu}\ \bibnamefont {Smirnov}},
  \bibinfo {author} {\bibfnamefont {Masoud}\ \bibnamefont {Mohseni}}, \ and\
  \bibinfo {author} {\bibfnamefont {Hartmut}\ \bibnamefont {Neven}},\
  }\bibfield  {title} {\enquote {\bibinfo {title} {Computational multiqubit
  tunnelling in programmable quantum annealers},}\ }\href
  {http://dx.doi.org/10.1038/ncomms10327} {\bibfield  {journal} {\bibinfo
  {journal} {Nat Commun}\ }\textbf {\bibinfo {volume} {7}} (\bibinfo {year}
  {2016})}\BibitemShut {NoStop}%
\bibitem [{\citenamefont {Albash}\ and\ \citenamefont
  {Lidar}(2018{\natexlab{a}})}]{Albash-Lidar:RMP}%
  \BibitemOpen
  \bibfield  {author} {\bibinfo {author} {\bibfnamefont {Tameem}\ \bibnamefont
  {Albash}}\ and\ \bibinfo {author} {\bibfnamefont {Daniel~A.}\ \bibnamefont
  {Lidar}},\ }\bibfield  {title} {\enquote {\bibinfo {title} {Adiabatic quantum
  computation},}\ }\href
  {https://link.aps.org/doi/10.1103/RevModPhys.90.015002} {\bibfield  {journal}
  {\bibinfo  {journal} {Reviews of Modern Physics}\ }\textbf {\bibinfo {volume}
  {90}},\ \bibinfo {pages} {015002--} (\bibinfo {year}
  {2018}{\natexlab{a}})}\BibitemShut {NoStop}%
\bibitem [{\citenamefont {Jansen}\ \emph {et~al.}(2007)\citenamefont {Jansen},
  \citenamefont {Ruskai},\ and\ \citenamefont {Seiler}}]{Jansen:07}%
  \BibitemOpen
  \bibfield  {author} {\bibinfo {author} {\bibfnamefont {Sabine}\ \bibnamefont
  {Jansen}}, \bibinfo {author} {\bibfnamefont {Mary-Beth}\ \bibnamefont
  {Ruskai}}, \ and\ \bibinfo {author} {\bibfnamefont {Ruedi}\ \bibnamefont
  {Seiler}},\ }\bibfield  {title} {\enquote {\bibinfo {title} {Bounds for the
  adiabatic approximation with applications to quantum computation},}\ }\href
  {http://scitation.aip.org/content/aip/journal/jmp/48/10/10.1063/1.2798382}
  {\bibfield  {journal} {\bibinfo  {journal} {J. Math. Phys.}\ }\textbf
  {\bibinfo {volume} {48}},\ \bibinfo {pages} {102111} (\bibinfo {year}
  {2007})}\BibitemShut {NoStop}%
\bibitem [{\citenamefont {Lidar}\ \emph {et~al.}(2009)\citenamefont {Lidar},
  \citenamefont {Rezakhani},\ and\ \citenamefont {Hamma}}]{lidar:102106}%
  \BibitemOpen
  \bibfield  {author} {\bibinfo {author} {\bibfnamefont {Daniel~A.}\
  \bibnamefont {Lidar}}, \bibinfo {author} {\bibfnamefont {Ali~T.}\
  \bibnamefont {Rezakhani}}, \ and\ \bibinfo {author} {\bibfnamefont
  {Alioscia}\ \bibnamefont {Hamma}},\ }\bibfield  {title} {\enquote {\bibinfo
  {title} {Adiabatic approximation with exponential accuracy for many-body
  systems and quantum computation},}\ }\href
  {http://scitation.aip.org/content/aip/journal/jmp/50/10/10.1063/1.3236685}
  {\bibfield  {journal} {\bibinfo  {journal} {J. Math. Phys.}\ }\textbf
  {\bibinfo {volume} {50}},\ \bibinfo {pages} {102106} (\bibinfo {year}
  {2009})}\BibitemShut {NoStop}%
\bibitem [{\citenamefont {R{\o}nnow}\ \emph {et~al.}(2014)\citenamefont
  {R{\o}nnow}, \citenamefont {Wang}, \citenamefont {Job}, \citenamefont
  {Boixo}, \citenamefont {Isakov}, \citenamefont {Wecker}, \citenamefont
  {Martinis}, \citenamefont {Lidar},\ and\ \citenamefont {Troyer}}]{speedup}%
  \BibitemOpen
  \bibfield  {author} {\bibinfo {author} {\bibfnamefont {Troels~F.}\
  \bibnamefont {R{\o}nnow}}, \bibinfo {author} {\bibfnamefont {Zhihui}\
  \bibnamefont {Wang}}, \bibinfo {author} {\bibfnamefont {Joshua}\ \bibnamefont
  {Job}}, \bibinfo {author} {\bibfnamefont {Sergio}\ \bibnamefont {Boixo}},
  \bibinfo {author} {\bibfnamefont {Sergei~V.}\ \bibnamefont {Isakov}},
  \bibinfo {author} {\bibfnamefont {David}\ \bibnamefont {Wecker}}, \bibinfo
  {author} {\bibfnamefont {John~M.}\ \bibnamefont {Martinis}}, \bibinfo
  {author} {\bibfnamefont {Daniel~A.}\ \bibnamefont {Lidar}}, \ and\ \bibinfo
  {author} {\bibfnamefont {Matthias}\ \bibnamefont {Troyer}},\ }\bibfield
  {title} {\enquote {\bibinfo {title} {{Defining and detecting quantum
  speedup}},}\ }\href {http://science.sciencemag.org/content/345/6195/420}
  {\bibfield  {journal} {\bibinfo  {journal} {Science}\ }\textbf {\bibinfo
  {volume} {345}},\ \bibinfo {pages} {420--424} (\bibinfo {year}
  {2014})}\BibitemShut {NoStop}%
\bibitem [{\citenamefont {Childs}\ \emph {et~al.}(2001)\citenamefont {Childs},
  \citenamefont {Farhi},\ and\ \citenamefont
  {Preskill}}]{childs_robustness_2001}%
  \BibitemOpen
  \bibfield  {author} {\bibinfo {author} {\bibfnamefont {Andrew~M.}\
  \bibnamefont {Childs}}, \bibinfo {author} {\bibfnamefont {Edward}\
  \bibnamefont {Farhi}}, \ and\ \bibinfo {author} {\bibfnamefont {John}\
  \bibnamefont {Preskill}},\ }\bibfield  {title} {\enquote {\bibinfo {title}
  {Robustness of adiabatic quantum computation},}\ }\href
  {http://journals.aps.org/pra/abstract/10.1103/PhysRevA.65.012322} {\bibfield
  {journal} {\bibinfo  {journal} {Phys. Rev. A}\ }\textbf {\bibinfo {volume}
  {65}},\ \bibinfo {pages} {012322} (\bibinfo {year} {2001})}\BibitemShut
  {NoStop}%
\bibitem [{\citenamefont {Ashhab}\ \emph {et~al.}(2006)\citenamefont {Ashhab},
  \citenamefont {Johansson},\ and\ \citenamefont {Nori}}]{PhysRevA.74.052330}%
  \BibitemOpen
  \bibfield  {author} {\bibinfo {author} {\bibfnamefont {S.}~\bibnamefont
  {Ashhab}}, \bibinfo {author} {\bibfnamefont {J.~R.}\ \bibnamefont
  {Johansson}}, \ and\ \bibinfo {author} {\bibfnamefont {Franco}\ \bibnamefont
  {Nori}},\ }\bibfield  {title} {\enquote {\bibinfo {title} {Decoherence in a
  scalable adiabatic quantum computer},}\ }\href
  {http://link.aps.org/doi/10.1103/PhysRevA.74.052330} {\bibfield  {journal}
  {\bibinfo  {journal} {Phys. Rev. A}\ }\textbf {\bibinfo {volume} {74}},\
  \bibinfo {pages} {052330} (\bibinfo {year} {2006})}\BibitemShut {NoStop}%
\bibitem [{\citenamefont {Albash}\ and\ \citenamefont
  {Lidar}(2015)}]{Albash:2015nx}%
  \BibitemOpen
  \bibfield  {author} {\bibinfo {author} {\bibfnamefont {Tameem}\ \bibnamefont
  {Albash}}\ and\ \bibinfo {author} {\bibfnamefont {Daniel~A.}\ \bibnamefont
  {Lidar}},\ }\bibfield  {title} {\enquote {\bibinfo {title} {Decoherence in
  adiabatic quantum computation},}\ }\href
  {http://link.aps.org/doi/10.1103/PhysRevA.91.062320} {\bibfield  {journal}
  {\bibinfo  {journal} {Phys. Rev. A}\ }\textbf {\bibinfo {volume} {91}},\
  \bibinfo {pages} {062320--} (\bibinfo {year} {2015})}\BibitemShut {NoStop}%
\bibitem [{\citenamefont {Amin}(2015)}]{Amin:2015qf}%
  \BibitemOpen
  \bibfield  {author} {\bibinfo {author} {\bibfnamefont {Mohammad~H.}\
  \bibnamefont {Amin}},\ }\bibfield  {title} {\enquote {\bibinfo {title}
  {Searching for quantum speedup in quasistatic quantum annealers},}\ }\href
  {https://link.aps.org/doi/10.1103/PhysRevA.92.052323} {\bibfield  {journal}
  {\bibinfo  {journal} {Physical Review A}\ }\textbf {\bibinfo {volume} {92}},\
  \bibinfo {pages} {052323--} (\bibinfo {year} {2015})}\BibitemShut {NoStop}%
\bibitem [{\citenamefont {Venuti}\ \emph {et~al.}(2016)\citenamefont {Venuti},
  \citenamefont {Albash}, \citenamefont {Lidar},\ and\ \citenamefont
  {Zanardi}}]{Venuti:2015kq}%
  \BibitemOpen
  \bibfield  {author} {\bibinfo {author} {\bibfnamefont {Lorenzo~Campos}\
  \bibnamefont {Venuti}}, \bibinfo {author} {\bibfnamefont {Tameem}\
  \bibnamefont {Albash}}, \bibinfo {author} {\bibfnamefont {Daniel~A.}\
  \bibnamefont {Lidar}}, \ and\ \bibinfo {author} {\bibfnamefont {Paolo}\
  \bibnamefont {Zanardi}},\ }\bibfield  {title} {\enquote {\bibinfo {title}
  {Adiabaticity in open quantum systems},}\ }\href
  {http://link.aps.org/doi/10.1103/PhysRevA.93.032118} {\bibfield  {journal}
  {\bibinfo  {journal} {{Phys. Rev. A}}\ }\textbf {\bibinfo {volume} {93}},\
  \bibinfo {pages} {032118--} (\bibinfo {year} {2016})}\BibitemShut {NoStop}%
\bibitem [{\citenamefont {Marshall}\ \emph {et~al.}(2017)\citenamefont
  {Marshall}, \citenamefont {Rieffel},\ and\ \citenamefont
  {Hen}}]{Marshall:2017aa}%
  \BibitemOpen
  \bibfield  {author} {\bibinfo {author} {\bibfnamefont {Jeffrey}\ \bibnamefont
  {Marshall}}, \bibinfo {author} {\bibfnamefont {Eleanor~G.}\ \bibnamefont
  {Rieffel}}, \ and\ \bibinfo {author} {\bibfnamefont {Itay}\ \bibnamefont
  {Hen}},\ }\bibfield  {title} {\enquote {\bibinfo {title} {Thermalization,
  freeze-out, and noise: Deciphering experimental quantum annealers},}\ }\href
  {\doibase 10.1103/PhysRevApplied.8.064025} {\bibfield  {journal} {\bibinfo
  {journal} {Phys. Rev. Applied}\ }\textbf {\bibinfo {volume} {8}},\ \bibinfo
  {pages} {064025} (\bibinfo {year} {2017})}\BibitemShut {NoStop}%
\bibitem [{\citenamefont {{Chancellor Nicholas}}\ \emph
  {et~al.}(2016)\citenamefont {{Chancellor Nicholas}}, \citenamefont {{Szoke
  Szilard}}, \citenamefont {{Vinci Walter}}, \citenamefont {{Aeppli Gabriel}},\
  and\ \citenamefont {{Warburton Paul A.}}}]{Chancellor:2016pa}%
  \BibitemOpen
  \bibfield  {author} {\bibinfo {author} {\bibnamefont {{Chancellor
  Nicholas}}}, \bibinfo {author} {\bibnamefont {{Szoke Szilard}}}, \bibinfo
  {author} {\bibnamefont {{Vinci Walter}}}, \bibinfo {author} {\bibnamefont
  {{Aeppli Gabriel}}}, \ and\ \bibinfo {author} {\bibnamefont {{Warburton Paul
  A.}}},\ }\bibfield  {title} {\enquote {\bibinfo {title} {{Maximum-Entropy
  Inference with a Programmable Annealer}},}\ }\href {\doibase
  http://dx.doi.org/10.1038/srep22318 10.1038/srep22318} {\bibfield  {journal}
  {\bibinfo  {journal} {Scientific Reports}\ }\textbf {\bibinfo {volume} {6}},\
  \bibinfo {pages} {22318} (\bibinfo {year} {2016})}\BibitemShut {NoStop}%
\bibitem [{\citenamefont {Albash}\ \emph {et~al.}(2017)\citenamefont {Albash},
  \citenamefont {Martin-Mayor},\ and\ \citenamefont {Hen}}]{Albash:2017ab}%
  \BibitemOpen
  \bibfield  {author} {\bibinfo {author} {\bibfnamefont {Tameem}\ \bibnamefont
  {Albash}}, \bibinfo {author} {\bibfnamefont {Victor}\ \bibnamefont
  {Martin-Mayor}}, \ and\ \bibinfo {author} {\bibfnamefont {Itay}\ \bibnamefont
  {Hen}},\ }\bibfield  {title} {\enquote {\bibinfo {title} {Temperature scaling
  law for quantum annealing optimizers},}\ }\href
  {https://link.aps.org/doi/10.1103/PhysRevLett.119.110502} {\bibfield
  {journal} {\bibinfo  {journal} {Physical Review Letters}\ }\textbf {\bibinfo
  {volume} {119}},\ \bibinfo {pages} {110502--} (\bibinfo {year}
  {2017})}\BibitemShut {NoStop}%
\bibitem [{\citenamefont {Perdomo-Ortiz}\ \emph {et~al.}(2018)\citenamefont
  {Perdomo-Ortiz}, \citenamefont {Benedetti}, \citenamefont
  {Realpe-G{\'o}mez},\ and\ \citenamefont {Biswas}}]{Perdomo:17qml}%
  \BibitemOpen
  \bibfield  {author} {\bibinfo {author} {\bibfnamefont {Alejandro}\
  \bibnamefont {Perdomo-Ortiz}}, \bibinfo {author} {\bibfnamefont {Marcello}\
  \bibnamefont {Benedetti}}, \bibinfo {author} {\bibfnamefont {John}\
  \bibnamefont {Realpe-G{\'o}mez}}, \ and\ \bibinfo {author} {\bibfnamefont
  {Rupak}\ \bibnamefont {Biswas}},\ }\bibfield  {title} {\enquote {\bibinfo
  {title} {Opportunities and challenges for quantum-assisted machine learning
  in near-term quantum computers},}\ }\href
  {https://iopscience.iop.org/article/10.1088/2058-9565/aab859} {\bibfield
  {journal} {\bibinfo  {journal} {Quantum Sci. Technol.}\ }\textbf {\bibinfo
  {volume} {3}},\ \bibinfo {pages} {030502} (\bibinfo {year} {2018})},\ \Eprint
  {http://arxiv.org/abs/arXiv:1708.09757} {arXiv:1708.09757} \BibitemShut
  {NoStop}%
\bibitem [{\citenamefont {Venuti}\ \emph {et~al.}(2017)\citenamefont {Venuti},
  \citenamefont {Albash}, \citenamefont {Marvian}, \citenamefont {Lidar},\ and\
  \citenamefont {Zanardi}}]{Venuti:2017aa}%
  \BibitemOpen
  \bibfield  {author} {\bibinfo {author} {\bibfnamefont {Lorenzo~Campos}\
  \bibnamefont {Venuti}}, \bibinfo {author} {\bibfnamefont {Tameem}\
  \bibnamefont {Albash}}, \bibinfo {author} {\bibfnamefont {Milad}\
  \bibnamefont {Marvian}}, \bibinfo {author} {\bibfnamefont {Daniel}\
  \bibnamefont {Lidar}}, \ and\ \bibinfo {author} {\bibfnamefont {Paolo}\
  \bibnamefont {Zanardi}},\ }\bibfield  {title} {\enquote {\bibinfo {title}
  {Relaxation versus adiabatic quantum steady-state preparation},}\ }\href
  {https://link.aps.org/doi/10.1103/PhysRevA.95.042302} {\bibfield  {journal}
  {\bibinfo  {journal} {{Phys. Rev. A}}\ }\textbf {\bibinfo {volume} {95}},\
  \bibinfo {pages} {042302--} (\bibinfo {year} {2017})}\BibitemShut {NoStop}%
\bibitem [{\citenamefont {Salakhutdinov}\ and\ \citenamefont
  {Hinton}(2012)}]{Salakhutdinov:2012}%
  \BibitemOpen
  \bibfield  {author} {\bibinfo {author} {\bibfnamefont {Ruslan}\ \bibnamefont
  {Salakhutdinov}}\ and\ \bibinfo {author} {\bibfnamefont {Geoffrey}\
  \bibnamefont {Hinton}},\ }\bibfield  {title} {\enquote {\bibinfo {title} {An
  efficient learning procedure for deep boltzmann machines},}\ }\href {\doibase
  10.1162/NECO_a_00311} {\bibfield  {journal} {\bibinfo  {journal} {Neural
  Comput.}\ }\textbf {\bibinfo {volume} {24}},\ \bibinfo {pages} {1967--2006}
  (\bibinfo {year} {2012})}\BibitemShut {NoStop}%
\bibitem [{\citenamefont {Tang}\ \emph {et~al.}(2012)\citenamefont {Tang},
  \citenamefont {Salakhutdinov},\ and\ \citenamefont {Hinton}}]{Tang:2012}%
  \BibitemOpen
  \bibfield  {author} {\bibinfo {author} {\bibfnamefont {Y.}~\bibnamefont
  {Tang}}, \bibinfo {author} {\bibfnamefont {R.}~\bibnamefont {Salakhutdinov}},
  \ and\ \bibinfo {author} {\bibfnamefont {G.}~\bibnamefont {Hinton}},\
  }\bibfield  {title} {\enquote {\bibinfo {title} {Robust boltzmann machines
  for recognition and denoising},}\ }in\ \href {\doibase
  10.1109/CVPR.2012.6247936} {\emph {\bibinfo {booktitle} {2012 IEEE Conference
  on Computer Vision and Pattern Recognition}}}\ (\bibinfo {year} {2012})\ pp.\
  \bibinfo {pages} {2264--2271}\BibitemShut {NoStop}%
\bibitem [{\citenamefont {Vinci}\ \emph {et~al.}(2016)\citenamefont {Vinci},
  \citenamefont {Albash},\ and\ \citenamefont {Lidar}}]{vinci2015nested}%
  \BibitemOpen
  \bibfield  {author} {\bibinfo {author} {\bibfnamefont {Walter}\ \bibnamefont
  {Vinci}}, \bibinfo {author} {\bibfnamefont {Tameem}\ \bibnamefont {Albash}},
  \ and\ \bibinfo {author} {\bibfnamefont {Daniel~A}\ \bibnamefont {Lidar}},\
  }\bibfield  {title} {\enquote {\bibinfo {title} {Nested quantum annealing
  correction},}\ }\href {http://dx.doi.org/10.1038/npjqi.2016.17} {\bibfield
  {journal} {\bibinfo  {journal} {npj Quant. Inf.}\ }\textbf {\bibinfo {volume}
  {2}},\ \bibinfo {pages} {16017} (\bibinfo {year} {2016})}\BibitemShut
  {NoStop}%
\bibitem [{\citenamefont {Pudenz}\ \emph {et~al.}(2014)\citenamefont {Pudenz},
  \citenamefont {Albash},\ and\ \citenamefont {Lidar}}]{PAL:13}%
  \BibitemOpen
  \bibfield  {author} {\bibinfo {author} {\bibfnamefont {Kristen~L}\
  \bibnamefont {Pudenz}}, \bibinfo {author} {\bibfnamefont {Tameem}\
  \bibnamefont {Albash}}, \ and\ \bibinfo {author} {\bibfnamefont {Daniel~A}\
  \bibnamefont {Lidar}},\ }\bibfield  {title} {\enquote {\bibinfo {title}
  {Error-corrected quantum annealing with hundreds of qubits},}\ }\href
  {\doibase 10.1038/ncomms4243} {\bibfield  {journal} {\bibinfo  {journal}
  {Nat. Commun.}\ }\textbf {\bibinfo {volume} {5}},\ \bibinfo {pages} {3243}
  (\bibinfo {year} {2014})}\BibitemShut {NoStop}%
\bibitem [{\citenamefont {Pudenz}\ \emph {et~al.}(2015)\citenamefont {Pudenz},
  \citenamefont {Albash},\ and\ \citenamefont {Lidar}}]{PAL:14}%
  \BibitemOpen
  \bibfield  {author} {\bibinfo {author} {\bibfnamefont {Kristen~L.}\
  \bibnamefont {Pudenz}}, \bibinfo {author} {\bibfnamefont {Tameem}\
  \bibnamefont {Albash}}, \ and\ \bibinfo {author} {\bibfnamefont {Daniel~A.}\
  \bibnamefont {Lidar}},\ }\bibfield  {title} {\enquote {\bibinfo {title}
  {{Quantum annealing correction for random Ising problems}},}\ }\href
  {http://link.aps.org/doi/10.1103/PhysRevA.91.042302} {\bibfield  {journal}
  {\bibinfo  {journal} {{Phys. Rev. A}}\ }\textbf {\bibinfo {volume} {91}},\
  \bibinfo {pages} {042302} (\bibinfo {year} {2015})}\BibitemShut {NoStop}%
\bibitem [{\citenamefont {Matsuura}\ \emph {et~al.}(2016)\citenamefont
  {Matsuura}, \citenamefont {Nishimori}, \citenamefont {Albash},\ and\
  \citenamefont {Lidar}}]{MNAL:15}%
  \BibitemOpen
  \bibfield  {author} {\bibinfo {author} {\bibfnamefont {Shunji}\ \bibnamefont
  {Matsuura}}, \bibinfo {author} {\bibfnamefont {Hidetoshi}\ \bibnamefont
  {Nishimori}}, \bibinfo {author} {\bibfnamefont {Tameem}\ \bibnamefont
  {Albash}}, \ and\ \bibinfo {author} {\bibfnamefont {Daniel~A.}\ \bibnamefont
  {Lidar}},\ }\bibfield  {title} {\enquote {\bibinfo {title} {Mean field
  analysis of quantum annealing correction},}\ }\href
  {http://link.aps.org/doi/10.1103/PhysRevLett.116.220501} {\bibfield
  {journal} {\bibinfo  {journal} {Physical Review Letters}\ }\textbf {\bibinfo
  {volume} {116}},\ \bibinfo {pages} {220501--} (\bibinfo {year}
  {2016})}\BibitemShut {NoStop}%
\bibitem [{\citenamefont {Mishra}\ \emph {et~al.}(2015)\citenamefont {Mishra},
  \citenamefont {Albash},\ and\ \citenamefont {Lidar}}]{Mishra:2015}%
  \BibitemOpen
  \bibfield  {author} {\bibinfo {author} {\bibfnamefont {Anurag}\ \bibnamefont
  {Mishra}}, \bibinfo {author} {\bibfnamefont {Tameem}\ \bibnamefont {Albash}},
  \ and\ \bibinfo {author} {\bibfnamefont {Daniel~A.}\ \bibnamefont {Lidar}},\
  }\bibfield  {title} {\enquote {\bibinfo {title} {Performance of two different
  quantum annealing correction codes},}\ }\href {\doibase
  10.1007/s11128-015-1201-z} {\bibfield  {journal} {\bibinfo  {journal} {Quant.
  Inf. Proc.}\ }\textbf {\bibinfo {volume} {15}},\ \bibinfo {pages} {609--636}
  (\bibinfo {year} {2015})}\BibitemShut {NoStop}%
\bibitem [{\citenamefont {Vinci}\ and\ \citenamefont
  {Lidar}(2018)}]{Vinci:2017ab}%
  \BibitemOpen
  \bibfield  {author} {\bibinfo {author} {\bibfnamefont {Walter}\ \bibnamefont
  {Vinci}}\ and\ \bibinfo {author} {\bibfnamefont {Daniel~A.}\ \bibnamefont
  {Lidar}},\ }\bibfield  {title} {\enquote {\bibinfo {title} {Scalable
  effective-temperature reduction for quantum annealers via nested quantum
  annealing correction},}\ }\href
  {https://link.aps.org/doi/10.1103/PhysRevA.97.022308} {\bibfield  {journal}
  {\bibinfo  {journal} {Physical Review A}\ }\textbf {\bibinfo {volume} {97}},\
  \bibinfo {pages} {022308--} (\bibinfo {year} {2018})}\BibitemShut {NoStop}%
\bibitem [{\citenamefont {Matsuura}\ \emph {et~al.}(2019)\citenamefont
  {Matsuura}, \citenamefont {Nishimori}, \citenamefont {Vinci},\ and\
  \citenamefont {Lidar}}]{Matsuura:2018}%
  \BibitemOpen
  \bibfield  {author} {\bibinfo {author} {\bibfnamefont {Shunji}\ \bibnamefont
  {Matsuura}}, \bibinfo {author} {\bibfnamefont {Hidetoshi}\ \bibnamefont
  {Nishimori}}, \bibinfo {author} {\bibfnamefont {Walter}\ \bibnamefont
  {Vinci}}, \ and\ \bibinfo {author} {\bibfnamefont {Daniel~A.}\ \bibnamefont
  {Lidar}},\ }\bibfield  {title} {\enquote {\bibinfo {title} {Nested quantum
  annealing correction at finite temperature: $p$-spin models},}\ }\href
  {\doibase 10.1103/PhysRevA.99.062307} {\bibfield  {journal} {\bibinfo
  {journal} {Physical Review A}\ }\textbf {\bibinfo {volume} {99}},\ \bibinfo
  {pages} {062307--} (\bibinfo {year} {2019})}\BibitemShut {NoStop}%
\bibitem [{\citenamefont {Shin}\ \emph {et~al.}(2014)\citenamefont {Shin},
  \citenamefont {Smith}, \citenamefont {Smolin},\ and\ \citenamefont
  {Vazirani}}]{SSSV}%
  \BibitemOpen
  \bibfield  {author} {\bibinfo {author} {\bibfnamefont {Seung~Woo}\
  \bibnamefont {Shin}}, \bibinfo {author} {\bibfnamefont {Graeme}\ \bibnamefont
  {Smith}}, \bibinfo {author} {\bibfnamefont {John~A.}\ \bibnamefont {Smolin}},
  \ and\ \bibinfo {author} {\bibfnamefont {Umesh}\ \bibnamefont {Vazirani}},\
  }\bibfield  {title} {\enquote {\bibinfo {title} {How ``quantum" is the
  {D-Wave} machine?}}\ }\href {http://arXiv.org/abs/1401.7087} {\bibfield
  {journal} {\bibinfo  {journal} {arXiv:1401.7087}\ } (\bibinfo {year}
  {2014})}\BibitemShut {NoStop}%
\bibitem [{\citenamefont {Santoro}\ \emph {et~al.}(2002)\citenamefont
  {Santoro}, \citenamefont {Marto\v{n}\'{a}k}, \citenamefont {Tosatti},\ and\
  \citenamefont {Car}}]{Santoro}%
  \BibitemOpen
  \bibfield  {author} {\bibinfo {author} {\bibfnamefont {Giuseppe~E.}\
  \bibnamefont {Santoro}}, \bibinfo {author} {\bibfnamefont {Roman}\
  \bibnamefont {Marto\v{n}\'{a}k}}, \bibinfo {author} {\bibfnamefont {Erio}\
  \bibnamefont {Tosatti}}, \ and\ \bibinfo {author} {\bibfnamefont {Roberto}\
  \bibnamefont {Car}},\ }\bibfield  {title} {\enquote {\bibinfo {title} {Theory
  of quantum annealing of an {I}sing spin glass},}\ }\href
  {http://science.sciencemag.org/content/295/5564/2427} {\bibfield  {journal}
  {\bibinfo  {journal} {Science}\ }\textbf {\bibinfo {volume} {295}},\ \bibinfo
  {pages} {2427--2430} (\bibinfo {year} {2002})}\BibitemShut {NoStop}%
\bibitem [{\citenamefont {Kadowaki}\ and\ \citenamefont
  {Nishimori}(1998)}]{kadowaki_quantum_1998}%
  \BibitemOpen
  \bibfield  {author} {\bibinfo {author} {\bibfnamefont {Tadashi}\ \bibnamefont
  {Kadowaki}}\ and\ \bibinfo {author} {\bibfnamefont {Hidetoshi}\ \bibnamefont
  {Nishimori}},\ }\bibfield  {title} {\enquote {\bibinfo {title} {Quantum
  annealing in the transverse \uppercase{I}sing model},}\ }\href
  {http://journals.aps.org/pre/abstract/10.1103/PhysRevE.58.5355} {\bibfield
  {journal} {\bibinfo  {journal} {Phys. Rev. E}\ }\textbf {\bibinfo {volume}
  {58}},\ \bibinfo {pages} {5355} (\bibinfo {year} {1998})}\BibitemShut
  {NoStop}%
\bibitem [{\citenamefont {Vinci}\ \emph {et~al.}(2015)\citenamefont {Vinci},
  \citenamefont {Albash}, \citenamefont {Paz-Silva}, \citenamefont {Hen},\ and\
  \citenamefont {Lidar}}]{Vinci:2015jt}%
  \BibitemOpen
  \bibfield  {author} {\bibinfo {author} {\bibfnamefont {Walter}\ \bibnamefont
  {Vinci}}, \bibinfo {author} {\bibfnamefont {Tameem}\ \bibnamefont {Albash}},
  \bibinfo {author} {\bibfnamefont {Gerardo}\ \bibnamefont {Paz-Silva}},
  \bibinfo {author} {\bibfnamefont {Itay}\ \bibnamefont {Hen}}, \ and\ \bibinfo
  {author} {\bibfnamefont {Daniel~A.}\ \bibnamefont {Lidar}},\ }\bibfield
  {title} {\enquote {\bibinfo {title} {Quantum annealing correction with minor
  embedding},}\ }\href {http://link.aps.org/doi/10.1103/PhysRevA.92.042310}
  {\bibfield  {journal} {\bibinfo  {journal} {{Phys. Rev. A}}\ }\textbf
  {\bibinfo {volume} {92}},\ \bibinfo {pages} {042310--} (\bibinfo {year}
  {2015})}\BibitemShut {NoStop}%
\bibitem [{\citenamefont {Gottesman}(1997)}]{Gottesman:1997ub}%
  \BibitemOpen
  \bibfield  {author} {\bibinfo {author} {\bibfnamefont {Daniel}\ \bibnamefont
  {Gottesman}},\ }\bibfield  {title} {\enquote {\bibinfo {title} {{Stabilizer
  Codes and Quantum Error Correction}},}\ }\href@noop {} {\bibfield  {journal}
  {\bibinfo  {journal} {arXiv.org}\ } (\bibinfo {year} {1997})},\ \Eprint
  {http://arxiv.org/abs/quant-ph/9705052} {quant-ph/9705052} \BibitemShut
  {NoStop}%
\bibitem [{\citenamefont {Jordan}\ \emph {et~al.}(2006)\citenamefont {Jordan},
  \citenamefont {Farhi},\ and\ \citenamefont {Shor}}]{jordan2006error}%
  \BibitemOpen
  \bibfield  {author} {\bibinfo {author} {\bibfnamefont {S.~P.}\ \bibnamefont
  {Jordan}}, \bibinfo {author} {\bibfnamefont {E.}~\bibnamefont {Farhi}}, \
  and\ \bibinfo {author} {\bibfnamefont {P.~W.}\ \bibnamefont {Shor}},\
  }\bibfield  {title} {\enquote {\bibinfo {title} {Error-correcting codes for
  adiabatic quantum computation},}\ }\href
  {http://link.aps.org/doi/10.1103/PhysRevA.74.052322} {\bibfield  {journal}
  {\bibinfo  {journal} {{Phys. Rev. A}}\ }\textbf {\bibinfo {volume} {74}},\
  \bibinfo {pages} {052322} (\bibinfo {year} {2006})}\BibitemShut {NoStop}%
\bibitem [{\citenamefont {Bookatz}\ \emph {et~al.}(2015)\citenamefont
  {Bookatz}, \citenamefont {Farhi},\ and\ \citenamefont
  {Zhou}}]{Bookatz:2014uq}%
  \BibitemOpen
  \bibfield  {author} {\bibinfo {author} {\bibfnamefont {Adam~D.}\ \bibnamefont
  {Bookatz}}, \bibinfo {author} {\bibfnamefont {Edward}\ \bibnamefont {Farhi}},
  \ and\ \bibinfo {author} {\bibfnamefont {Leo}\ \bibnamefont {Zhou}},\
  }\bibfield  {title} {\enquote {\bibinfo {title} {Error suppression in
  hamiltonian-based quantum computation using energy penalties},}\ }\href
  {http://link.aps.org/doi/10.1103/PhysRevA.92.022317} {\bibfield  {journal}
  {\bibinfo  {journal} {{Phys. Rev. A}}\ }\textbf {\bibinfo {volume} {92}},\
  \bibinfo {pages} {022317--} (\bibinfo {year} {2015})}\BibitemShut {NoStop}%
\bibitem [{\citenamefont {Jiang}\ and\ \citenamefont
  {Rieffel}(2017)}]{Jiang:2015kx}%
  \BibitemOpen
  \bibfield  {author} {\bibinfo {author} {\bibfnamefont {Zhang}\ \bibnamefont
  {Jiang}}\ and\ \bibinfo {author} {\bibfnamefont {Eleanor~G.}\ \bibnamefont
  {Rieffel}},\ }\bibfield  {title} {\enquote {\bibinfo {title} {Non-commuting
  two-local hamiltonians for quantum error suppression},}\ }\href {\doibase
  10.1007/s11128-017-1527-9} {\bibfield  {journal} {\bibinfo  {journal}
  {{Quant. Inf. Proc.}}\ }\textbf {\bibinfo {volume} {16}},\ \bibinfo {pages}
  {89} (\bibinfo {year} {2017})}\BibitemShut {NoStop}%
\bibitem [{\citenamefont {Marvian}\ and\ \citenamefont
  {Lidar}(2017{\natexlab{a}})}]{Marvian-Lidar:16}%
  \BibitemOpen
  \bibfield  {author} {\bibinfo {author} {\bibfnamefont {Milad}\ \bibnamefont
  {Marvian}}\ and\ \bibinfo {author} {\bibfnamefont {Daniel~A.}\ \bibnamefont
  {Lidar}},\ }\bibfield  {title} {\enquote {\bibinfo {title} {{Error
  Suppression for Hamiltonian-Based Quantum Computation Using Subsystem
  Codes}},}\ }\href {https://link.aps.org/doi/10.1103/PhysRevLett.118.030504}
  {\bibfield  {journal} {\bibinfo  {journal} {Phys. Rev. Lett.}\ }\textbf
  {\bibinfo {volume} {118}},\ \bibinfo {pages} {030504--} (\bibinfo {year}
  {2017}{\natexlab{a}})}\BibitemShut {NoStop}%
\bibitem [{\citenamefont {Marvian}\ and\ \citenamefont
  {Lidar}(2017{\natexlab{b}})}]{Marvian:2017aa}%
  \BibitemOpen
  \bibfield  {author} {\bibinfo {author} {\bibfnamefont {Milad}\ \bibnamefont
  {Marvian}}\ and\ \bibinfo {author} {\bibfnamefont {Daniel~A.}\ \bibnamefont
  {Lidar}},\ }\bibfield  {title} {\enquote {\bibinfo {title} {Error suppression
  for hamiltonian quantum computing in markovian environments},}\ }\href
  {http://link.aps.org/doi/10.1103/PhysRevA.95.032302} {\bibfield  {journal}
  {\bibinfo  {journal} {Physical Review A}\ }\textbf {\bibinfo {volume} {95}},\
  \bibinfo {pages} {032302--} (\bibinfo {year}
  {2017}{\natexlab{b}})}\BibitemShut {NoStop}%
\bibitem [{\citenamefont {Lidar}(2019)}]{Lidar:2019ab}%
  \BibitemOpen
  \bibfield  {author} {\bibinfo {author} {\bibfnamefont {Daniel~A.}\
  \bibnamefont {Lidar}},\ }\bibfield  {title} {\enquote {\bibinfo {title}
  {Arbitrary-time error suppression for markovian adiabatic quantum computing
  using stabilizer subspace codes},}\ }\href {\doibase
  10.1103/PhysRevA.100.022326} {\bibfield  {journal} {\bibinfo  {journal}
  {Phys. Rev. A}\ }\textbf {\bibinfo {volume} {100}},\ \bibinfo {pages}
  {022326} (\bibinfo {year} {2019})}\BibitemShut {NoStop}%
\bibitem [{\citenamefont {Choi}(2011)}]{Choi2}%
  \BibitemOpen
  \bibfield  {author} {\bibinfo {author} {\bibfnamefont {V.}~\bibnamefont
  {Choi}},\ }\bibfield  {title} {\enquote {\bibinfo {title} {Minor-embedding in
  adiabatic quantum computation: Ii. minor-universal graph design},}\ }\href
  {https://link.springer.com/article/10.1007%2Fs11128-010-0200-3} {\bibfield
  {journal} {\bibinfo  {journal} {{Quant. Inf. Proc.}}\ }\textbf {\bibinfo
  {volume} {10}},\ \bibinfo {pages} {343--353} (\bibinfo {year}
  {2011})}\BibitemShut {NoStop}%
\bibitem [{\citenamefont {Klymko}\ \emph {et~al.}(2014)\citenamefont {Klymko},
  \citenamefont {Sullivan},\ and\ \citenamefont
  {Humble}}]{klymko_adiabatic_2012}%
  \BibitemOpen
  \bibfield  {author} {\bibinfo {author} {\bibfnamefont {Christine}\
  \bibnamefont {Klymko}}, \bibinfo {author} {\bibfnamefont {Blair~D.}\
  \bibnamefont {Sullivan}}, \ and\ \bibinfo {author} {\bibfnamefont
  {Travis~S.}\ \bibnamefont {Humble}},\ }\bibfield  {title} {\enquote {\bibinfo
  {title} {Adiabatic quantum programming: minor embedding with hard faults},}\
  }\href {\doibase 10.1007/s11128-013-0683-9} {\bibfield  {journal} {\bibinfo
  {journal} {Quant. Inf. Proc.}\ }\textbf {\bibinfo {volume} {13}},\ \bibinfo
  {pages} {709--729} (\bibinfo {year} {2014})}\BibitemShut {NoStop}%
\bibitem [{\citenamefont {Cai}\ \emph {et~al.}(2014)\citenamefont {Cai},
  \citenamefont {Macready},\ and\ \citenamefont {Roy}}]{Cai:2014nx}%
  \BibitemOpen
  \bibfield  {author} {\bibinfo {author} {\bibfnamefont {Jun}\ \bibnamefont
  {Cai}}, \bibinfo {author} {\bibfnamefont {William~G.}\ \bibnamefont
  {Macready}}, \ and\ \bibinfo {author} {\bibfnamefont {Aidan}\ \bibnamefont
  {Roy}},\ }\bibfield  {title} {\enquote {\bibinfo {title} {A practical
  heuristic for finding graph minors},}\ }\href
  {http://arXiv.org/abs/1406.2741} {\bibfield  {journal} {\bibinfo  {journal}
  {arXiv:1406.2741}\ } (\bibinfo {year} {2014})}\BibitemShut {NoStop}%
\bibitem [{\citenamefont {Hinton}\ \emph {et~al.}(2006)\citenamefont {Hinton},
  \citenamefont {Osindero},\ and\ \citenamefont {Teh}}]{Hinton:2006fv}%
  \BibitemOpen
  \bibfield  {author} {\bibinfo {author} {\bibfnamefont {Geoffrey~E.}\
  \bibnamefont {Hinton}}, \bibinfo {author} {\bibfnamefont {Simon}\
  \bibnamefont {Osindero}}, \ and\ \bibinfo {author} {\bibfnamefont {Yee-Whye}\
  \bibnamefont {Teh}},\ }\bibfield  {title} {\enquote {\bibinfo {title} {A fast
  learning algorithm for deep belief nets},}\ }\bibfield  {booktitle} {\emph
  {\bibinfo {booktitle} {Neural Computation}},\ }\href {\doibase
  10.1162/neco.2006.18.7.1527} {\bibfield  {journal} {\bibinfo  {journal}
  {Neural Computation}\ }\textbf {\bibinfo {volume} {18}},\ \bibinfo {pages}
  {1527--1554} (\bibinfo {year} {2006})}\BibitemShut {NoStop}%
\bibitem [{\citenamefont {Salakhutdinov}\ and\ \citenamefont
  {Larochelle}(2010)}]{Salakhutdinov:10a}%
  \BibitemOpen
  \bibfield  {author} {\bibinfo {author} {\bibfnamefont {Ruslan}\ \bibnamefont
  {Salakhutdinov}}\ and\ \bibinfo {author} {\bibfnamefont {Hugo}\ \bibnamefont
  {Larochelle}},\ }\bibfield  {title} {\enquote {\bibinfo {title} {Efficient
  learning of deep boltzmann machines},}\ }in\ \href@noop {} {\emph {\bibinfo
  {booktitle} {Proceedings of the Thirteenth International Conference on
  Artificial Intelligence and Statistics}}},\ \bibinfo {series} {Proceedings of
  Machine Learning Research}, Vol.~\bibinfo {volume} {9},\ \bibinfo {editor}
  {edited by\ \bibinfo {editor} {\bibfnamefont {Yee~Whye}\ \bibnamefont {Teh}}\
  and\ \bibinfo {editor} {\bibfnamefont {Mike}\ \bibnamefont {Titterington}}}\
  (\bibinfo  {publisher} {PMLR},\ \bibinfo {address} {Chia Laguna Resort,
  Sardinia, Italy},\ \bibinfo {year} {2010})\ pp.\ \bibinfo {pages}
  {693--700}\BibitemShut {NoStop}%
\bibitem [{\citenamefont {MacKay}(2003)}]{Mackay:2003}%
  \BibitemOpen
  \bibfield  {author} {\bibinfo {author} {\bibfnamefont {David~J.C.}\
  \bibnamefont {MacKay}},\ }\href@noop {} {\emph {\bibinfo {title} {Information
  Theory, Inference, and Learning Algorithms}}}\ (\bibinfo  {publisher}
  {Cambridge University Press},\ \bibinfo {year} {2003})\BibitemShut {NoStop}%
\bibitem [{\citenamefont {Hinton}(2002)}]{Hinton:2002aa}%
  \BibitemOpen
  \bibfield  {author} {\bibinfo {author} {\bibfnamefont {Geoffrey~E.}\
  \bibnamefont {Hinton}},\ }\bibfield  {title} {\enquote {\bibinfo {title}
  {Training products of experts by minimizing contrastive divergence},}\
  }\bibfield  {booktitle} {\emph {\bibinfo {booktitle} {Neural Computation}},\
  }\href {\doibase 10.1162/089976602760128018} {\bibfield  {journal} {\bibinfo
  {journal} {Neural Computation}\ }\textbf {\bibinfo {volume} {14}},\ \bibinfo
  {pages} {1771--1800} (\bibinfo {year} {2002})}\BibitemShut {NoStop}%
\bibitem [{\citenamefont {Amin}\ \emph {et~al.}(2018)\citenamefont {Amin},
  \citenamefont {Andriyash}, \citenamefont {Rolfe}, \citenamefont
  {Kulchytskyy},\ and\ \citenamefont {Melko}}]{Amin:2016}%
  \BibitemOpen
  \bibfield  {author} {\bibinfo {author} {\bibfnamefont {Mohammad~H.}\
  \bibnamefont {Amin}}, \bibinfo {author} {\bibfnamefont {Evgeny}\ \bibnamefont
  {Andriyash}}, \bibinfo {author} {\bibfnamefont {Jason}\ \bibnamefont
  {Rolfe}}, \bibinfo {author} {\bibfnamefont {Bohdan}\ \bibnamefont
  {Kulchytskyy}}, \ and\ \bibinfo {author} {\bibfnamefont {Roger}\ \bibnamefont
  {Melko}},\ }\bibfield  {title} {\enquote {\bibinfo {title} {Quantum boltzmann
  machine},}\ }\href {\doibase 10.1103/PhysRevX.8.021050} {\bibfield  {journal}
  {\bibinfo  {journal} {Physical Review X}\ }\textbf {\bibinfo {volume} {8}},\
  \bibinfo {pages} {021050--} (\bibinfo {year} {2018})}\BibitemShut {NoStop}%
\bibitem [{\citenamefont {Benedetti}\ \emph {et~al.}(2017)\citenamefont
  {Benedetti}, \citenamefont {Realpe-G{\'o}mez}, \citenamefont {Biswas},\ and\
  \citenamefont {Perdomo-Ortiz}}]{Benedetti:2016oz}%
  \BibitemOpen
  \bibfield  {author} {\bibinfo {author} {\bibfnamefont {Marcello}\
  \bibnamefont {Benedetti}}, \bibinfo {author} {\bibfnamefont {John}\
  \bibnamefont {Realpe-G{\'o}mez}}, \bibinfo {author} {\bibfnamefont {Rupak}\
  \bibnamefont {Biswas}}, \ and\ \bibinfo {author} {\bibfnamefont {Alejandro}\
  \bibnamefont {Perdomo-Ortiz}},\ }\bibfield  {title} {\enquote {\bibinfo
  {title} {Quantum-assisted learning of hardware-embedded probabilistic
  graphical models},}\ }\href {\doibase 10.1103/PhysRevX.7.041052} {\bibfield
  {journal} {\bibinfo  {journal} {Physical Review X}\ }\textbf {\bibinfo
  {volume} {7}},\ \bibinfo {pages} {041052--} (\bibinfo {year}
  {2017})}\BibitemShut {NoStop}%
\bibitem [{\citenamefont {Adachi}\ and\ \citenamefont
  {Henderson}(2015)}]{2012arXiv1204.2821S}%
  \BibitemOpen
  \bibfield  {author} {\bibinfo {author} {\bibfnamefont {Steven~H.}\
  \bibnamefont {Adachi}}\ and\ \bibinfo {author} {\bibfnamefont {Maxwell~P.}\
  \bibnamefont {Henderson}},\ }\bibfield  {title} {\enquote {\bibinfo {title}
  {Application of quantum annealing to training of deep neural networks},}\
  }\href {http://arXiv.org/abs/1510.06356} {\bibfield  {journal} {\bibinfo
  {journal} {arXiv:1510.06356}\ } (\bibinfo {year} {2015})}\BibitemShut
  {NoStop}%
\bibitem [{\citenamefont {{Lecun}}\ \emph {et~al.}(1998)\citenamefont
  {{Lecun}}, \citenamefont {{Bottou}}, \citenamefont {{Bengio}},\ and\
  \citenamefont {{Haffner}}}]{MNIST}%
  \BibitemOpen
  \bibfield  {author} {\bibinfo {author} {\bibfnamefont {Y.}~\bibnamefont
  {{Lecun}}}, \bibinfo {author} {\bibfnamefont {L.}~\bibnamefont {{Bottou}}},
  \bibinfo {author} {\bibfnamefont {Y.}~\bibnamefont {{Bengio}}}, \ and\
  \bibinfo {author} {\bibfnamefont {P.}~\bibnamefont {{Haffner}}},\ }\bibfield
  {title} {\enquote {\bibinfo {title} {Gradient-based learning applied to
  document recognition},}\ }\href {\doibase 10.1109/5.726791} {\bibfield
  {journal} {\bibinfo  {journal} {Proceedings of the IEEE}\ }\textbf {\bibinfo
  {volume} {86}},\ \bibinfo {pages} {2278--2324} (\bibinfo {year}
  {1998})}\BibitemShut {NoStop}%
\bibitem [{\citenamefont {Jaynes}(1957)}]{Jaynes:1957aa}%
  \BibitemOpen
  \bibfield  {author} {\bibinfo {author} {\bibfnamefont {E.~T.}\ \bibnamefont
  {Jaynes}},\ }\bibfield  {title} {\enquote {\bibinfo {title} {Information
  theory and statistical mechanics},}\ }\href {\doibase
  10.1103/PhysRev.106.620} {\bibfield  {journal} {\bibinfo  {journal} {Physical
  Review}\ }\textbf {\bibinfo {volume} {106}},\ \bibinfo {pages} {620--630}
  (\bibinfo {year} {1957})}\BibitemShut {NoStop}%
\bibitem [{\citenamefont {Nishimori}\ \emph {et~al.}(2015)\citenamefont
  {Nishimori}, \citenamefont {Tsuda},\ and\ \citenamefont
  {Knysh}}]{Nishimori:2015dp}%
  \BibitemOpen
  \bibfield  {author} {\bibinfo {author} {\bibfnamefont {Hidetoshi}\
  \bibnamefont {Nishimori}}, \bibinfo {author} {\bibfnamefont {Junichi}\
  \bibnamefont {Tsuda}}, \ and\ \bibinfo {author} {\bibfnamefont {Sergey}\
  \bibnamefont {Knysh}},\ }\bibfield  {title} {\enquote {\bibinfo {title}
  {Comparative study of the performance of quantum annealing and simulated
  annealing},}\ }\href {http://link.aps.org/doi/10.1103/PhysRevE.91.012104}
  {\bibfield  {journal} {\bibinfo  {journal} {Phys. Rev. E}\ }\textbf {\bibinfo
  {volume} {91}},\ \bibinfo {pages} {012104--} (\bibinfo {year}
  {2015})}\BibitemShut {NoStop}%
\bibitem [{\citenamefont {Albash}\ \emph {et~al.}(2019)\citenamefont {Albash},
  \citenamefont {Martin-Mayor},\ and\ \citenamefont
  {Hen}}]{albash_2019_analogerrors}%
  \BibitemOpen
  \bibfield  {author} {\bibinfo {author} {\bibfnamefont {T.}~\bibnamefont
  {Albash}}, \bibinfo {author} {\bibfnamefont {V.}~\bibnamefont
  {Martin-Mayor}}, \ and\ \bibinfo {author} {\bibfnamefont {I.}~\bibnamefont
  {Hen}},\ }\bibfield  {title} {\enquote {\bibinfo {title} {Analog errors in
  ising machines},}\ }\href@noop {} {\bibfield  {journal} {\bibinfo  {journal}
  {Quantum Sci. Technol.}\ }\textbf {\bibinfo {volume} {4}},\ \bibinfo {pages}
  {02LT03} (\bibinfo {year} {2019})}\BibitemShut {NoStop}%
\bibitem [{\citenamefont {Inc.}(2018)}]{DW2KQ}%
  \BibitemOpen
  \bibfield  {author} {\bibinfo {author} {\bibfnamefont {D-Wave~Systems}\
  \bibnamefont {Inc.}},\ }\href
  {https://www.dwavesys.com/sites/default/files/D-Wave%202000Q%20Tech%20Collateral_1029F.pdf}
  {\enquote {\bibinfo {title} {{The D-Wave 2000Q Quantum Computer Technology
  Overview}},}\ } (\bibinfo {year} {2018})\BibitemShut {NoStop}%
\bibitem [{\citenamefont {Albash}\ \emph {et~al.}(2015)\citenamefont {Albash},
  \citenamefont {R{\o}nnow}, \citenamefont {Troyer},\ and\ \citenamefont
  {Lidar}}]{Albash:2014if}%
  \BibitemOpen
  \bibfield  {author} {\bibinfo {author} {\bibfnamefont {T.}~\bibnamefont
  {Albash}}, \bibinfo {author} {\bibfnamefont {T.~F.}\ \bibnamefont
  {R{\o}nnow}}, \bibinfo {author} {\bibfnamefont {M.}~\bibnamefont {Troyer}}, \
  and\ \bibinfo {author} {\bibfnamefont {D.~A.}\ \bibnamefont {Lidar}},\
  }\bibfield  {title} {\enquote {\bibinfo {title} {{Reexamining classical and
  quantum models for the D-Wave One processor}},}\ }\href
  {https://link.springer.com/article/10.1140%2Fepjst%2Fe2015-02346-0}
  {\bibfield  {journal} {\bibinfo  {journal} {Eur. Phys. J. Spec. Top.}\
  }\textbf {\bibinfo {volume} {224}},\ \bibinfo {pages} {111--129} (\bibinfo
  {year} {2015})}\BibitemShut {NoStop}%
\bibitem [{\citenamefont {Crowley}\ and\ \citenamefont
  {Green}(2016)}]{Crowley:2016aa}%
  \BibitemOpen
  \bibfield  {author} {\bibinfo {author} {\bibfnamefont {Philip J.~D.}\
  \bibnamefont {Crowley}}\ and\ \bibinfo {author} {\bibfnamefont {A.~G.}\
  \bibnamefont {Green}},\ }\bibfield  {title} {\enquote {\bibinfo {title}
  {Anisotropic landau-lifshitz-gilbert models of dissipation in qubits},}\
  }\href {https://link.aps.org/doi/10.1103/PhysRevA.94.062106} {\bibfield
  {journal} {\bibinfo  {journal} {Physical Review A}\ }\textbf {\bibinfo
  {volume} {94}},\ \bibinfo {pages} {062106--} (\bibinfo {year}
  {2016})}\BibitemShut {NoStop}%
\bibitem [{\citenamefont {Pedregosa}\ \emph {et~al.}(2011)\citenamefont
  {Pedregosa}, \citenamefont {Varoquaux}, \citenamefont {Gramfort},
  \citenamefont {Michel}, \citenamefont {Thirion}, \citenamefont {Grisel},
  \citenamefont {Blondel}, \citenamefont {Prettenhofer}, \citenamefont {Weiss},
  \citenamefont {Dubourg}, \citenamefont {Vanderplas}, \citenamefont {Passos},
  \citenamefont {Cournapeau}, \citenamefont {Brucher}, \citenamefont {Perrot},\
  and\ \citenamefont {{{\'E}}douard Duchesnay}}]{scikit-learn}%
  \BibitemOpen
  \bibfield  {author} {\bibinfo {author} {\bibfnamefont {Fabian}\ \bibnamefont
  {Pedregosa}}, \bibinfo {author} {\bibfnamefont {Ga{{\"e}}l}\ \bibnamefont
  {Varoquaux}}, \bibinfo {author} {\bibfnamefont {Alexandre}\ \bibnamefont
  {Gramfort}}, \bibinfo {author} {\bibfnamefont {Vincent}\ \bibnamefont
  {Michel}}, \bibinfo {author} {\bibfnamefont {Bertrand}\ \bibnamefont
  {Thirion}}, \bibinfo {author} {\bibfnamefont {Olivier}\ \bibnamefont
  {Grisel}}, \bibinfo {author} {\bibfnamefont {Mathieu}\ \bibnamefont
  {Blondel}}, \bibinfo {author} {\bibfnamefont {Peter}\ \bibnamefont
  {Prettenhofer}}, \bibinfo {author} {\bibfnamefont {Ron}\ \bibnamefont
  {Weiss}}, \bibinfo {author} {\bibfnamefont {Vincent}\ \bibnamefont
  {Dubourg}}, \bibinfo {author} {\bibfnamefont {Jake}\ \bibnamefont
  {Vanderplas}}, \bibinfo {author} {\bibfnamefont {Alexandre}\ \bibnamefont
  {Passos}}, \bibinfo {author} {\bibfnamefont {David}\ \bibnamefont
  {Cournapeau}}, \bibinfo {author} {\bibfnamefont {Matthieu}\ \bibnamefont
  {Brucher}}, \bibinfo {author} {\bibfnamefont {Matthieu}\ \bibnamefont
  {Perrot}}, \ and\ \bibinfo {author} {\bibnamefont {{{\'E}}douard
  Duchesnay}},\ }\bibfield  {title} {\enquote {\bibinfo {title} {Scikit-learn:
  Machine learning in python},}\ }\href
  {http://jmlr.org/papers/v12/pedregosa11a.html} {\bibfield  {journal}
  {\bibinfo  {journal} {Journal of Machine Learning Research}\ }\textbf
  {\bibinfo {volume} {12}},\ \bibinfo {pages} {2825--2830} (\bibinfo {year}
  {2011})}\BibitemShut {NoStop}%
\bibitem [{\citenamefont {Boothby}\ \emph {et~al.}(2019)\citenamefont
  {Boothby}, \citenamefont {Bunyk}, \citenamefont {Raymond},\ and\
  \citenamefont {Roy}}]{DWave-Pegasus-techreport}%
  \BibitemOpen
  \bibfield  {author} {\bibinfo {author} {\bibfnamefont {Kelly}\ \bibnamefont
  {Boothby}}, \bibinfo {author} {\bibfnamefont {Paul}\ \bibnamefont {Bunyk}},
  \bibinfo {author} {\bibfnamefont {Jack}\ \bibnamefont {Raymond}}, \ and\
  \bibinfo {author} {\bibfnamefont {Aidan}\ \bibnamefont {Roy}},\ }\href
  {https://www.dwavesys.com/sites/default/files/14-1026A-C_Next-Generation-Topology-of-DW-Quantum-Processors.pdf}
  {\emph {\bibinfo {title} {{Next-Generation Topology of D-Wave Quantum
  Processors}}}},\ \bibinfo {type} {Tech. Rep.}\ (\bibinfo  {institution}
  {D-Wave Systems Inc.},\ \bibinfo {year} {2019})\BibitemShut {NoStop}%
\bibitem [{\citenamefont {Oreshkov}\ and\ \citenamefont
  {Calsamiglia}(2010)}]{oreshkov_adiabatic_2010}%
  \BibitemOpen
  \bibfield  {author} {\bibinfo {author} {\bibfnamefont {Ognyan}\ \bibnamefont
  {Oreshkov}}\ and\ \bibinfo {author} {\bibfnamefont {John}\ \bibnamefont
  {Calsamiglia}},\ }\bibfield  {title} {\enquote {\bibinfo {title} {Adiabatic
  {Markovian} {Dynamics}},}\ }\href {\doibase 10.1103/PhysRevLett.105.050503}
  {\bibfield  {journal} {\bibinfo  {journal} {Phys. Rev. Lett.}\ }\textbf
  {\bibinfo {volume} {105}},\ \bibinfo {pages} {050503} (\bibinfo {year}
  {2010})}\BibitemShut {NoStop}%
\bibitem [{\citenamefont {Avron}\ \emph {et~al.}(2012)\citenamefont {Avron},
  \citenamefont {Fraas}, \citenamefont {Graf},\ and\ \citenamefont
  {Grech}}]{Avron:2012tv}%
  \BibitemOpen
  \bibfield  {author} {\bibinfo {author} {\bibfnamefont {J.~E.}\ \bibnamefont
  {Avron}}, \bibinfo {author} {\bibfnamefont {M.}~\bibnamefont {Fraas}},
  \bibinfo {author} {\bibfnamefont {G.~M.}\ \bibnamefont {Graf}}, \ and\
  \bibinfo {author} {\bibfnamefont {P.}~\bibnamefont {Grech}},\ }\bibfield
  {title} {\enquote {\bibinfo {title} {Adiabatic theorems for generators of
  contracting evolutions},}\ }\href {\doibase 10.1007/s00220-012-1504-1}
  {\bibfield  {journal} {\bibinfo  {journal} {Comm. Math. Phys.}\ }\textbf
  {\bibinfo {volume} {314}},\ \bibinfo {pages} {163--191} (\bibinfo {year}
  {2012})}\BibitemShut {NoStop}%
\bibitem [{\citenamefont {Choi}(2020)}]{Choi:19}%
  \BibitemOpen
  \bibfield  {author} {\bibinfo {author} {\bibfnamefont {Vicky}\ \bibnamefont
  {Choi}},\ }\bibfield  {title} {\enquote {\bibinfo {title} {The effects of the
  problem hamiltonian parameters on the minimum spectral gap in adiabatic
  quantum optimization},}\ }\href {\doibase 10.1007/s11128-020-2582-1}
  {\bibfield  {journal} {\bibinfo  {journal} {Quantum Information Processing}\
  }\textbf {\bibinfo {volume} {19}},\ \bibinfo {pages} {90} (\bibinfo {year}
  {2020})}\BibitemShut {NoStop}%
\bibitem [{\citenamefont {Albash}\ and\ \citenamefont
  {Lidar}(2018{\natexlab{b}})}]{Albash:2017aa}%
  \BibitemOpen
  \bibfield  {author} {\bibinfo {author} {\bibfnamefont {Tameem}\ \bibnamefont
  {Albash}}\ and\ \bibinfo {author} {\bibfnamefont {Daniel~A.}\ \bibnamefont
  {Lidar}},\ }\bibfield  {title} {\enquote {\bibinfo {title} {Demonstration of
  a scaling advantage for a quantum annealer over simulated annealing},}\
  }\href {\doibase 10.1103/PhysRevX.8.031016} {\bibfield  {journal} {\bibinfo
  {journal} {Physical Review X}\ }\textbf {\bibinfo {volume} {8}},\ \bibinfo
  {pages} {031016--} (\bibinfo {year} {2018}{\natexlab{b}})}\BibitemShut
  {NoStop}%
\bibitem [{\citenamefont {Pearson}\ \emph {et~al.}(2019)\citenamefont
  {Pearson}, \citenamefont {Mishra}, \citenamefont {Hen},\ and\ \citenamefont
  {Lidar}}]{Pearson:2019aa}%
  \BibitemOpen
  \bibfield  {author} {\bibinfo {author} {\bibfnamefont {Adam}\ \bibnamefont
  {Pearson}}, \bibinfo {author} {\bibfnamefont {Anurag}\ \bibnamefont
  {Mishra}}, \bibinfo {author} {\bibfnamefont {Itay}\ \bibnamefont {Hen}}, \
  and\ \bibinfo {author} {\bibfnamefont {Daniel~A.}\ \bibnamefont {Lidar}},\
  }\bibfield  {title} {\enquote {\bibinfo {title} {Analog errors in quantum
  annealing: doom and hope},}\ }\href {\doibase 10.1038/s41534-019-0210-7}
  {\bibfield  {journal} {\bibinfo  {journal} {npj Quantum Information}\
  }\textbf {\bibinfo {volume} {5}},\ \bibinfo {pages} {107} (\bibinfo {year}
  {2019})}\BibitemShut {NoStop}%
\bibitem [{\citenamefont {Marshall}\ \emph {et~al.}(2020)\citenamefont
  {Marshall}, \citenamefont {Di~Gioacchino},\ and\ \citenamefont
  {Rieffel}}]{1909.12184}%
  \BibitemOpen
  \bibfield  {author} {\bibinfo {author} {\bibfnamefont {Jeffrey}\ \bibnamefont
  {Marshall}}, \bibinfo {author} {\bibfnamefont {Andrea}\ \bibnamefont
  {Di~Gioacchino}}, \ and\ \bibinfo {author} {\bibfnamefont {Eleanor~G.}\
  \bibnamefont {Rieffel}},\ }\bibfield  {title} {\enquote {\bibinfo {title}
  {Perils of embedding for sampling problems},}\ }\href {\doibase
  10.1103/PhysRevResearch.2.023020} {\bibfield  {journal} {\bibinfo  {journal}
  {Phys. Rev. Research}\ }\textbf {\bibinfo {volume} {2}},\ \bibinfo {pages}
  {023020} (\bibinfo {year} {2020})}\BibitemShut {NoStop}%
\bibitem [{DW-(2019)}]{DW-white-paper-1overf}%
  \BibitemOpen
  \href
  {https://www.dwavesys.com/sites/default/files/14-1037A-A_Improved_coherence_leads_to_gains_QA_performance.pdf}
  {\enquote {\bibinfo {title} {{D-Wave White Paper: Improved coherence leads to
  gains in quantum annealing performance}},}\ } (\bibinfo {year}
  {2019})\BibitemShut {NoStop}%
\end{thebibliography}%

\appendix

\section{DW processor used in this work}
\label{app:DW}

\begin{figure}[b]
\begin{center}
 \includegraphics[width=1.3\columnwidth,angle=0]{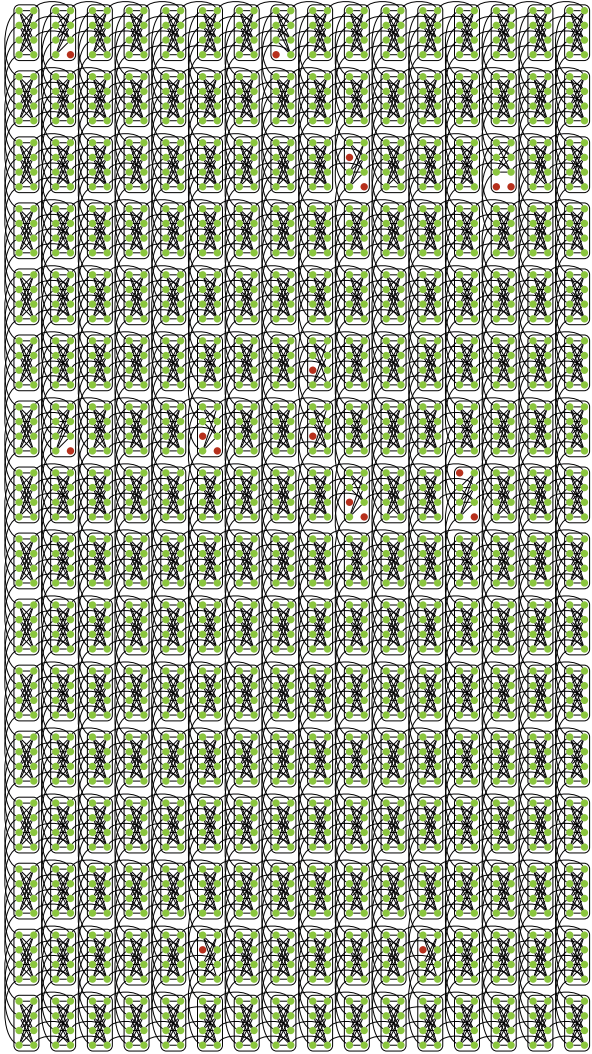}
 \end{center}
 \caption{Chimera graph of the NASA DW 2000Q processor. Green circles indicate active qubits, red circles are inactive qubits, and active physical couplers are indicated by black lines.}
 \label{fig:chimera}
\end{figure}

In this work we used the DW 2000Q processor located at Moffett Field and managed by NASA-Ames. The hardware graph consists of $2031$ qubits and is depicted in Fig.~\ref{fig:chimera}.

\section{Optimal penalty values}
\label{app:opt-gamma}

\begin{figure}[t]
\includegraphics[width=\columnwidth]{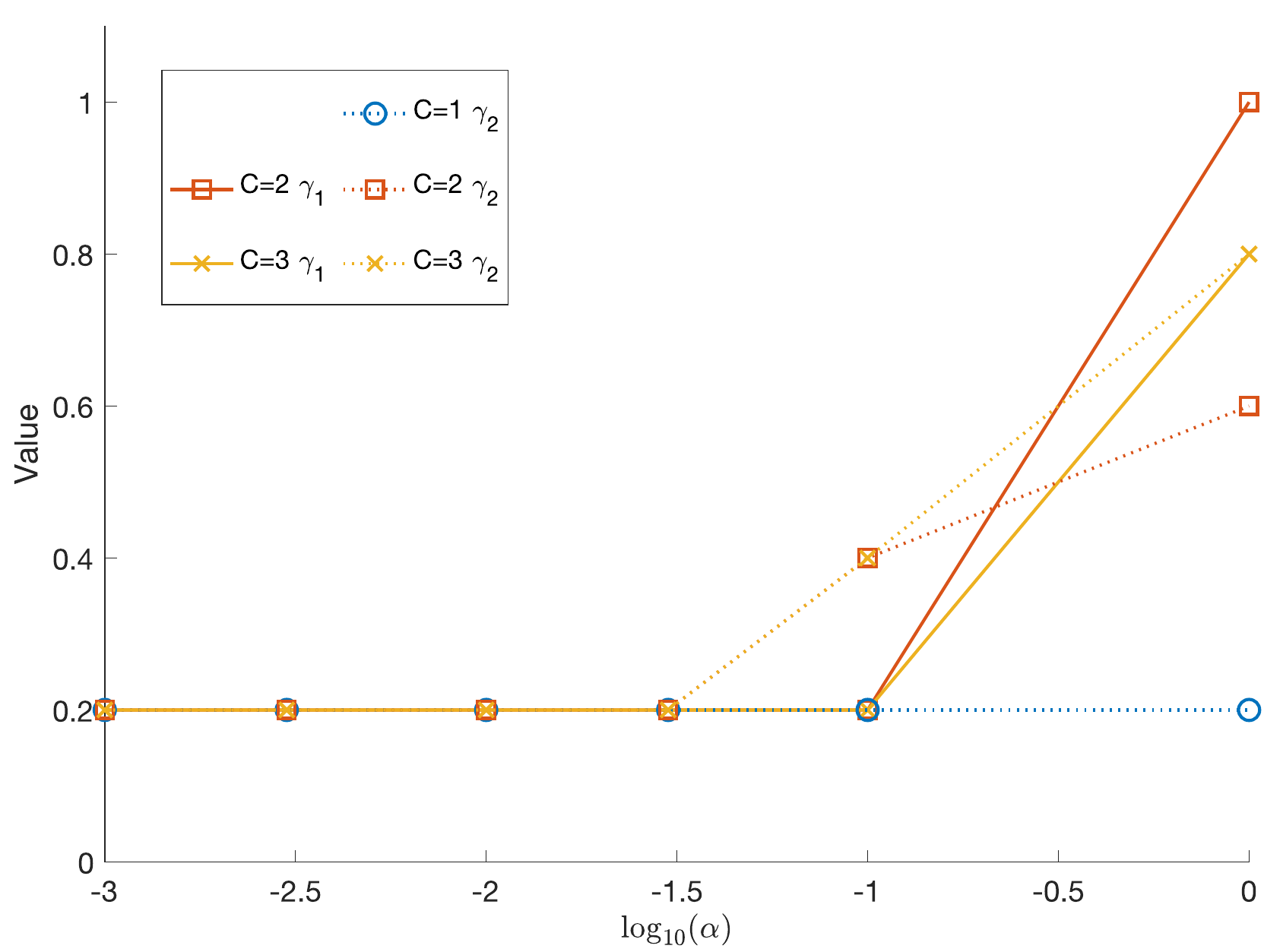}
\caption{Optimal penalty values for the MNIST dataset as a function of the scaling parameter $\alpha$, for different nesting levels $C$.}
\label{fig:MNIST_opt_gammas}
\end{figure}

Figure~\ref{fig:MNIST_opt_gammas} shows the optimal penalty values we obtained for training DW subject to the MNIST dataset. The key observation is that the penalty remains below the hardware limit of $1$, with the one exception of $C=2$, where $\gamma_1=1$ at $\alpha=1$. This might explain why $C=2$ is overtaken by $C=1$ in terms of the performance seen in Fig.~\ref{fig:MNISTResultsvsAlpha} in the main text. Also note that the chain penalty strength $\gamma_2$ generally grows with $C$, which means that the penalty plays a larger role in suppressing errors due to broken chains.

\begin{figure}[t]
\includegraphics[width=\columnwidth]{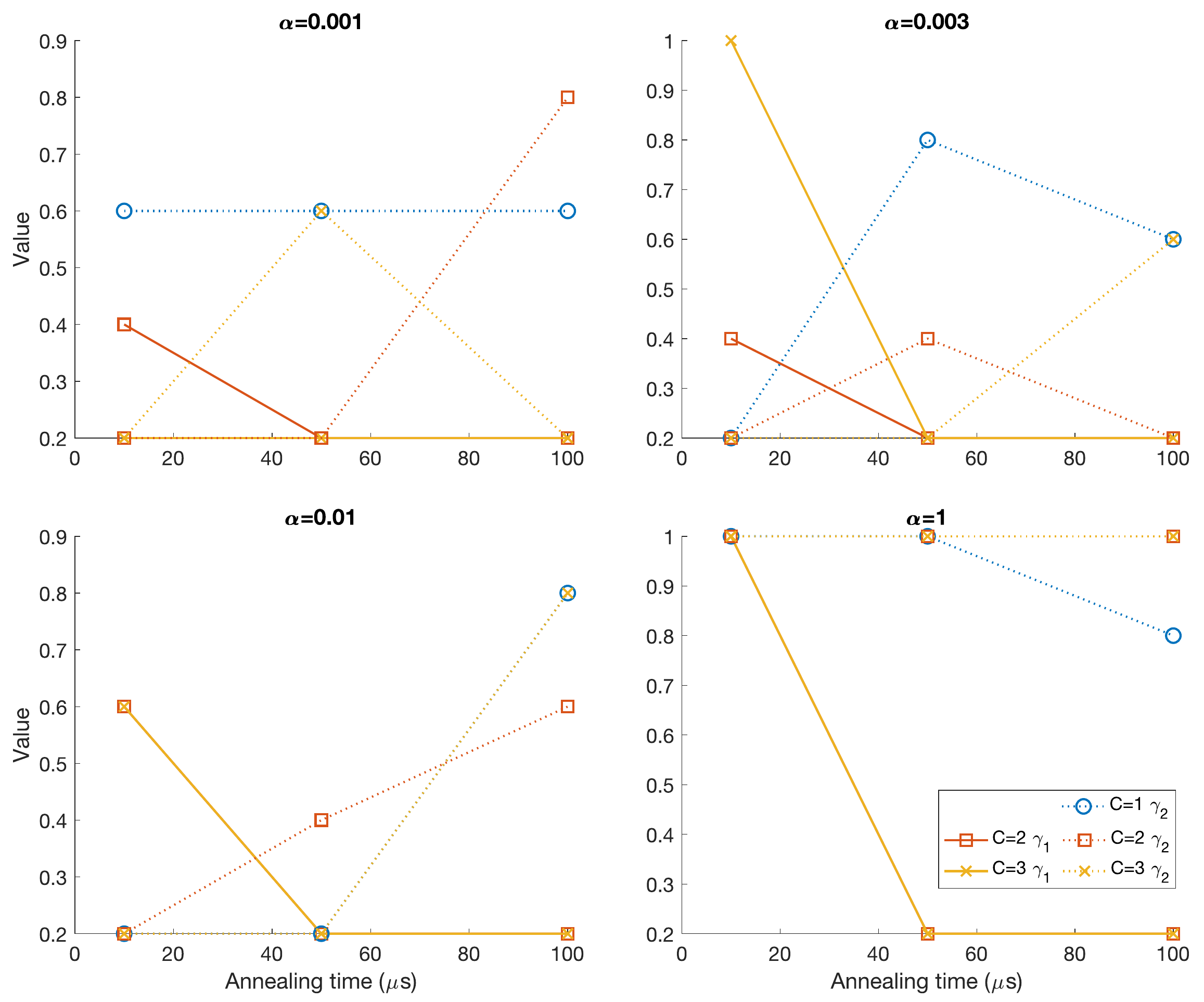}
\caption{Optimal penalty values for the BAS dataset, as a function of the anneal time $t_f$, for different values of the scaling parameter $\alpha$, and for different nesting levels $C$. This figure complements Fig.~\ref{fig:optimalgamma} in the main text.}
\label{fig:BAS_opt_gammas}
\end{figure}

Figure~\ref{fig:BAS_opt_gammas} shows the optimal penalty values we obtained for training DW subject to the BAS dataset, for additional values not shown in 
Fig.~\ref{fig:optimalgamma}. For $\alpha \geq 0.03$ the minor embedding penalty $\gamma_2$ becomes hardware-limited for $C=3$, except at a few anneal time values. For $\alpha=1$ the minor embedding penalty is  hardware-limited for all $C$ values, except for $C=1$ at the highest anneal time.

\begin{figure*}[ht]
\includegraphics[width=\textwidth]{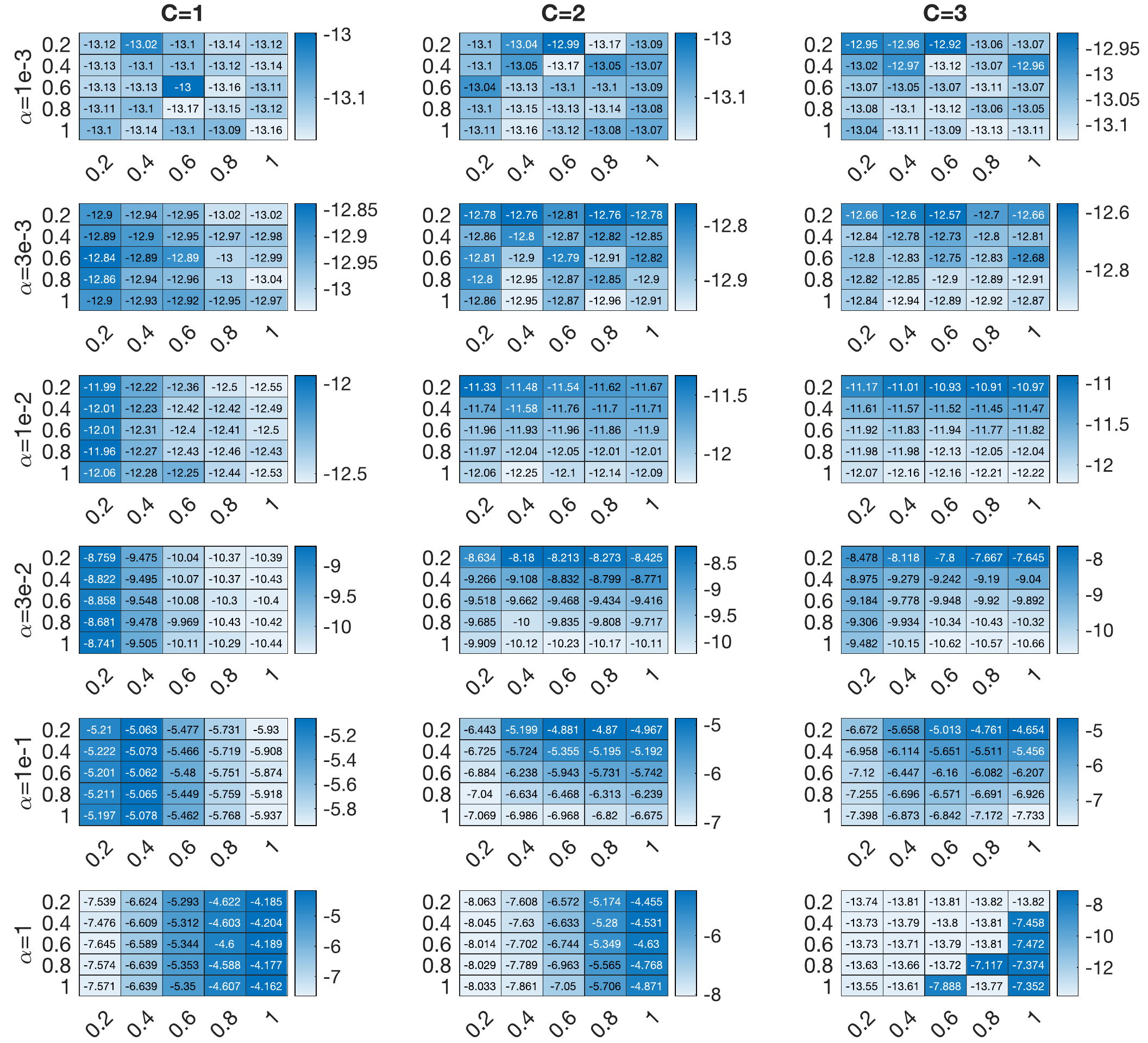}
\caption{Heat map of the empirical log-likelihood. $\gamma_1$ is on the vertical axis and $\gamma_2$ on the horizontal axis. The value shown is the average empirical log-likelihood after training for 5 epochs.}
\label{fig:BAS_ll_vs_gammas}
\end{figure*}

Figure~\ref{fig:BAS_ll_vs_gammas} shows a heatmap of the empirical log-likelihood after tuning the hyper-parameters. For the two smallest values of $\alpha$ the exact values of $\gamma_1$ and $\gamma_2$ do not have much of an effect; the empirical log-likelihood after training for 5 epochs (see Sec.~\ref{sec:BAS-ta}) varies by around 0.5 at most. In other words, as might be expected, noise is dominant and learning is difficult. This makes it difficult to draw conclusions about the optimal values of the $\gamma$'s. At $\alpha = 10^{-2}$ the value of $\gamma_1$ and $\gamma_2$ do not seem to be penalty-limited for $C=2$ and $C=3$, with a clear preference for $\gamma_1=0.2$. However as $\alpha$ continues to increase, the optimal value of at least one of the $\gamma$'s begins to increase: $\gamma_2$ quickly shifts towards $1$. There seems to be a very limited range of $\alpha$ where the $\gamma$'s are not penalty-limited for higher values of $C$; $C=3$ appears to be penalty-limited by $\alpha = 3\times10^{-2}$, $C=2$ at $\alpha = 10^{-1}$ and $C=1$ by $\alpha=1$. 

\section{Discarding broken chains}
\label{app:decoding}
For the results in the main text, we employed a simple majority vote to decode broken chains. If prior to this `correction' step the states obeyed a classical Boltzmann distribution, then the correction step can result in a distortion such that the resulting distribution is no longer Boltzmann \cite{1909.12184}. Here we focus on the subset of the states without broken chains and discard all the rest.  The reason is that if the states were originally Boltzmann distributed, then so will be the subset of unbroken chain states, although a decoding step is nevertheless still needed to go from the physical problem to the original logical problem. To understand the effect of the penalties without decoding, we study the case where broken chains were discarded at $\alpha=0.03$. We repeat the same training procedure on the BAS dataset as in the main text except that we discard any samples with broken chains: we first find the optimal values of $\gamma_1$ and $\gamma_2$, then use these values to train for 10 epochs. Based on the excellent agreement between D-Wave and SQA found in the main text, we use SQA with a temperature of 12 mK and $8 \times 10^4$ sweeps and $10^4$ repetitions to generate these results. 

\begin{figure*}[t]
\includegraphics[width=\textwidth]{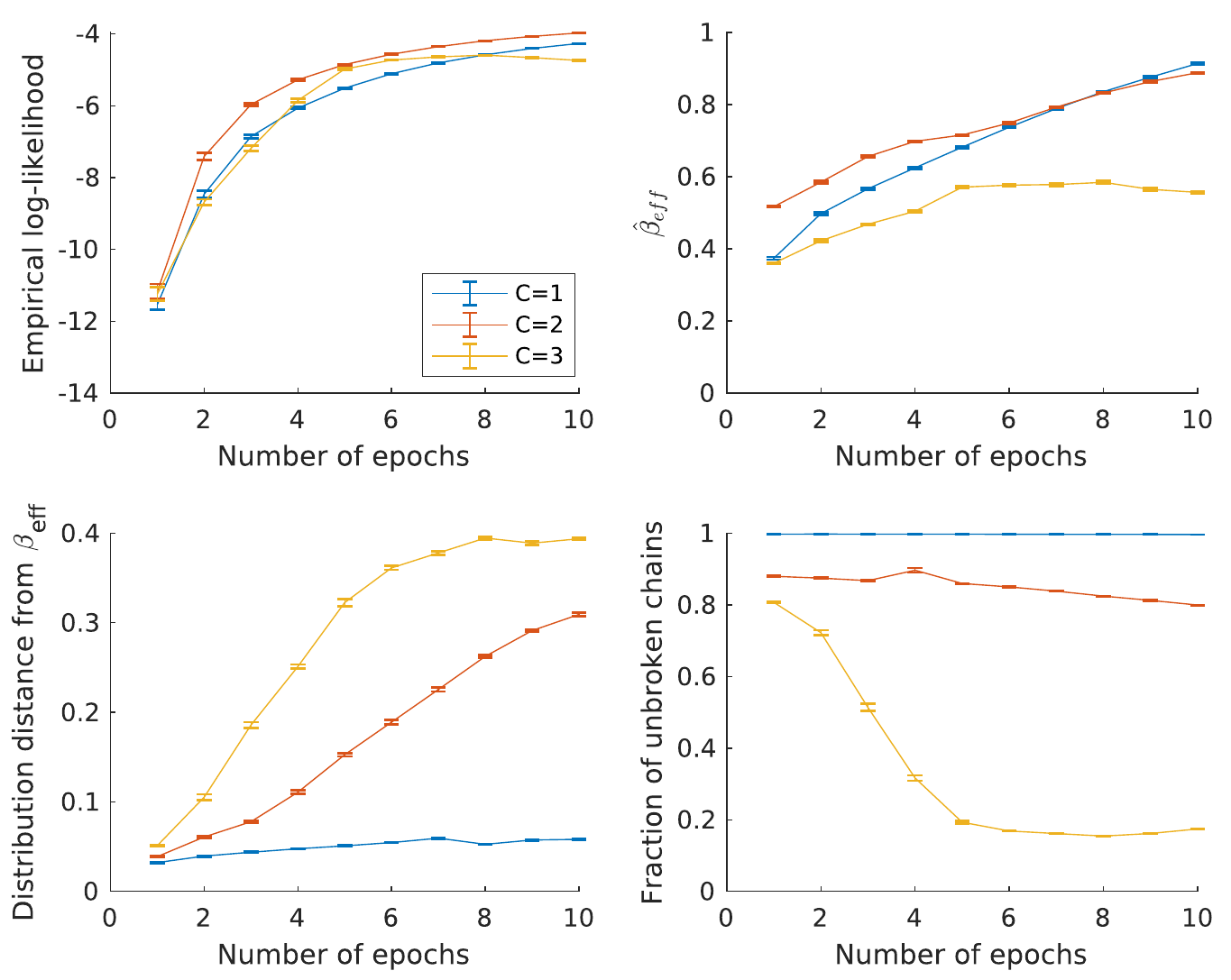}
\caption{NQAC training results for BAS dataset at $\alpha=0.03$ when discarding broken chains using SQA. The optimal values of $\gamma_1$ and $\gamma_2$ are both 0.2 for all nesting levels.}\label{fig:sqa_alpha-0.03}
\end{figure*}

Figure~\ref{fig:sqa_alpha-0.03} shows the results for training. When discarding broken chains, the optimal value of both $\gamma_1$ and $\gamma_2$ are 0.2 for all nesting levels. For $C=1$ and $C=2$ the majority of the samples do not have broken chains, but for $C=3$, the fraction of samples with unbroken chains decreases significantly with the number of epochs before plateauing. We observe the same plateauing in our performance metrics, suggesting that the small fraction of available samples with unbroken chains is hindering learning. It is possible that calling for more samples until a minimum number of unbroken chains is reached may improve performance. 

\begin{figure*}[h]
\includegraphics[width=\textwidth]{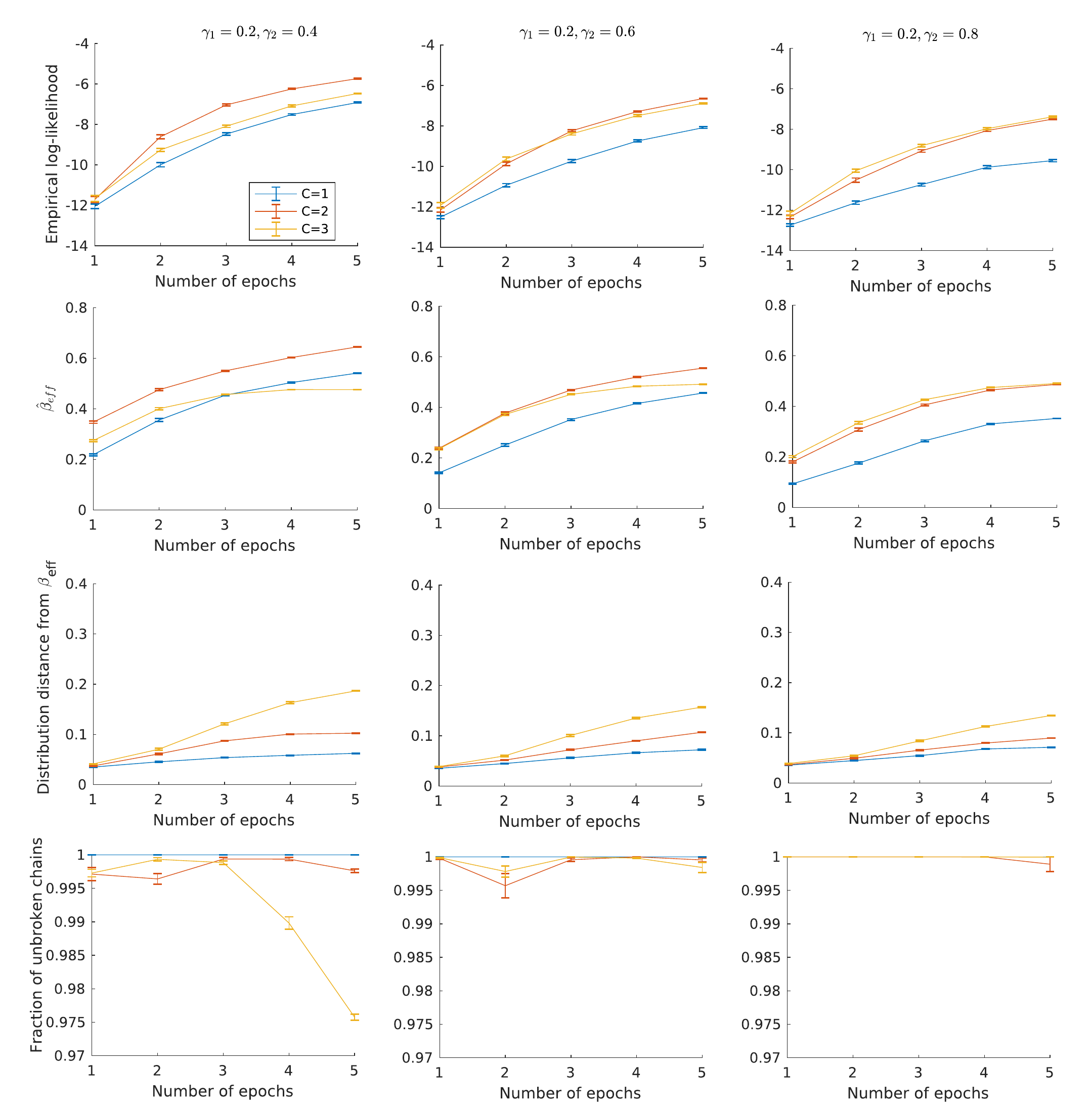}
\caption{NQAC training results for BAS dataset at $\alpha=0.03$ with (suboptimal) values of the penalty strengths $\gamma_1$ and $\gamma_2$ when discarding broken chains using SQA.}\label{fig:sqa_tune_gammas}
\end{figure*}

Figure~\ref{fig:sqa_tune_gammas} shows the parameter tuning results for three different values of the minor-embedding penalty $\gamma_2$. As can be seen, when $\gamma_2$ increases, we find fewer broken chains; for $C=1$ and $C=2$, the chains are short enough that the vast majority of chains are unbroken with a small $\gamma_2$ , but for $C=3$, $\gamma_2=0.2$ is too small to keep the chains intact, and a higher $\gamma_2$ helps alleviate this problem.  However, we also find that the performance as measured by the empirical log-likelihood decreases for all $C$ values.  We observe the largest drops in performance for $C=1$, where we believe that the strongly coupled chains, corresponding to higher $\gamma_2$ values, dominate over the problem implementation and effectively hinder learning. The energy boost associated with the repetition code help alleviate this problem in the case of $C=2$ and $C=3$, so we observe less dramatic drops in performance.  Nevertheless, learning suffers generally in the case of more strongly coupled chains, and we believe this is because such chains are harder to collectively flip thermally, so learning suffers because it is harder to explore the configuration space even if the fraction of unbroken chains is higher. Hence, our results indicate that the optimal strategy is to use values of $\gamma_1$ and $\gamma_2$ that are relatively weak and to employ a decoding strategy to ``fix" the broken chains. 

\section{SQA and SVMC sweeps}
\label{app:sweeps}

\begin{figure*}[t]
\includegraphics[width=\textwidth]{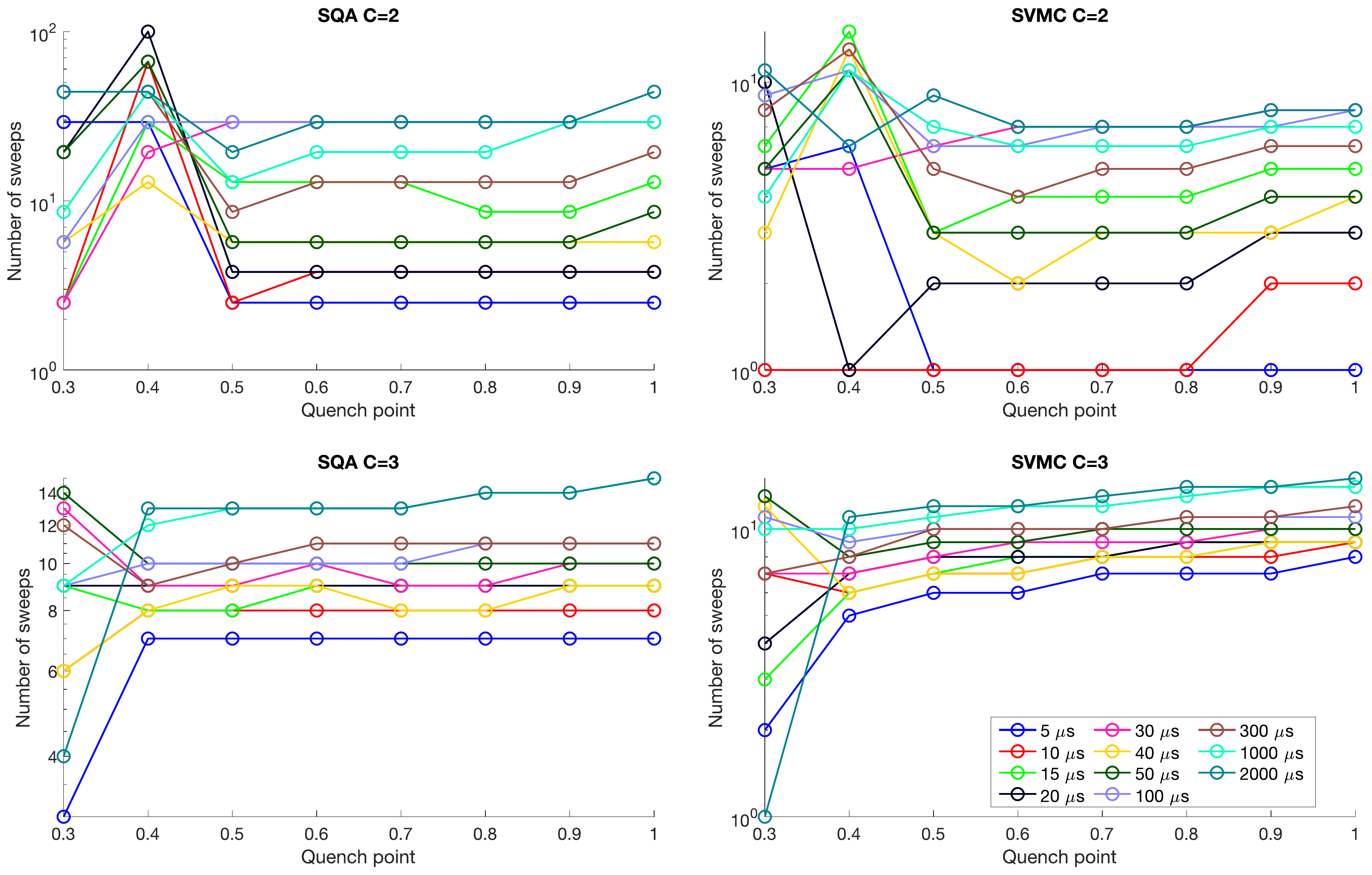}
\caption{Number of SQA and SVMC sweeps (logarithmic scale) as a function of quench point. Top row: $C=2$, bottom row: $C=3$. Left: SQA, right: SVMC. The legend is the corresponding DW anneal time in $\mu$s. What is plotted is the mode, i.e., the number of sweeps that appeared the most often. We used the mode because SQA and SVMC were compared to $20$ different noisy realizations from DW, so each realization of DW had a distribution with a slightly different temperature.
}
\label{fig:sweeps}
\end{figure*}

To recap Sec.~\ref{sec:SQAandSVMC}, for each Hamiltonian at each quench point, we ran both SQA and SVMC on the encoded physical problem Hamiltonian sent to DW, and selected the number of sweeps that gave an effective inverse temperature on the logical problem distribution that was closest to DW's. We compared SQA and SVMC at each quench point for each instance to $20$ noisy realizations of DW and thus the closest $\beta_{\mathrm{eff}}$ could vary slightly from realization to realization.

Figure~\ref{fig:sweeps} shows (the mode of) the number of sweeps at each quench point for $C=2,3$, for both SQA and SVMC. 
Before $s=0.5$ the number of sweeps for both SQA and SVMC varies significantly, with no discernible trend for the longer anneal times. This is unsurprising as freezing and the minimum gap likely occur before this point, and the contribution from the transverse field is still large. The $1\mu$s DW quench time is probably not  fast enough, so the trends in the number of sweeps in the early stages of the anneal are not consistent, as can be seen in Fig.~\ref{fig:dw_sqa_svmc_closest_to_boltzmann}. 
At later points in the anneal, trends for the number of sweeps are stable. In particular, as expected shorter anneal times require fewer sweeps.

\section{More details on the comparison with SQA and SVMC}
\label{app:D}

\begin{figure*}[t]
\includegraphics[width=.91\textwidth]{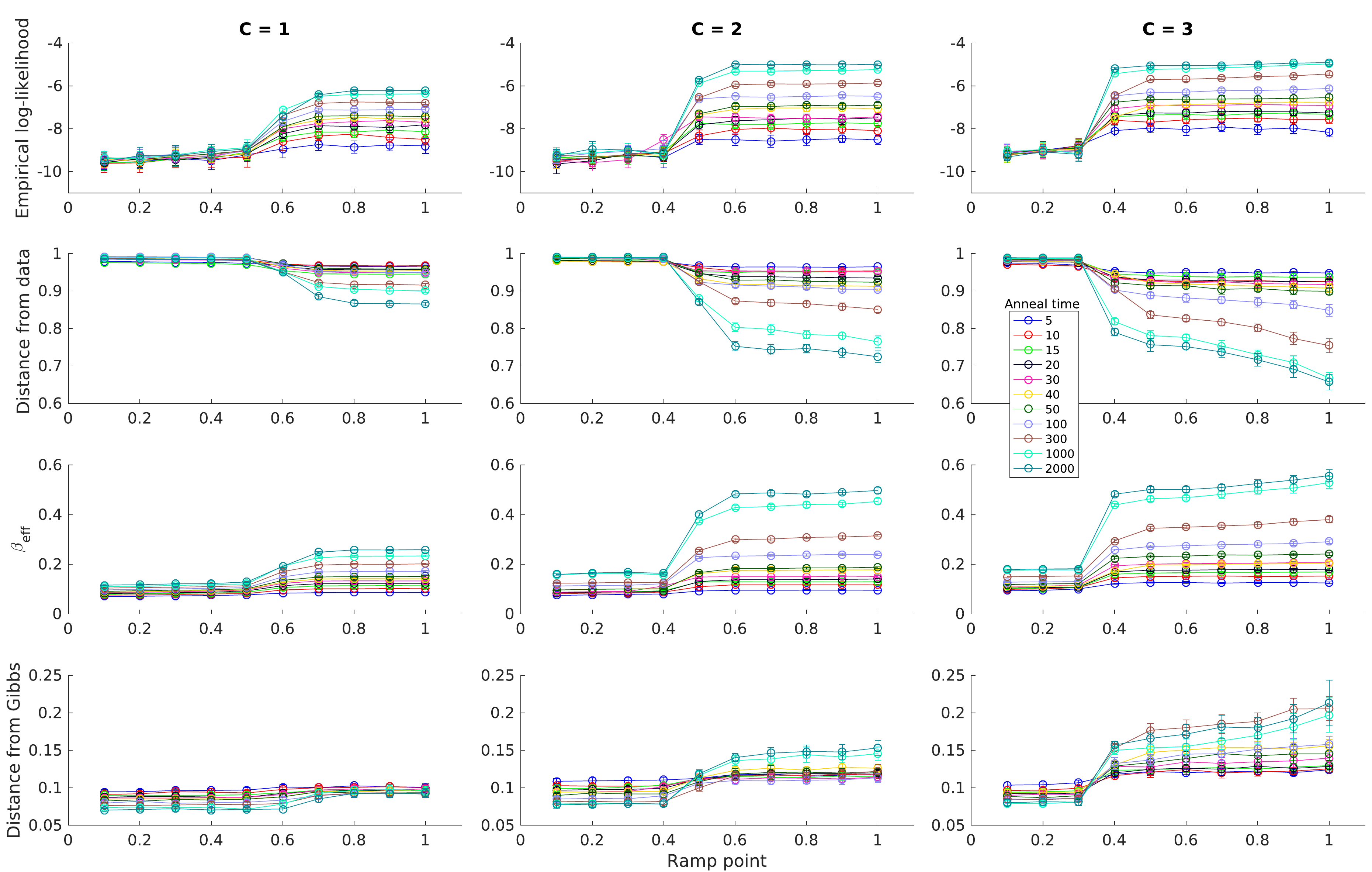}
\caption{Comparison of DW results for nesting levels $C=1,2,3$ at different anneal times with quench at various intermediate points in the anneal, for the BAS dataset. A $1\mu$s quench was used throughout; $\alpha$ was fixed at $0.03$. Here error bars represent only one standard deviation to improve readability. For all $C$ values there is a transition that becomes sharper with increasing anneal time. The transition point shifts to earlier in the anneal with increasing $C$; after the transition all quantities freeze, except at the largest anneal times. The trend with increasing anneal time is consistent throughout: the empirical log-likelihood, fidelity with data ($1-$distance) and inverse temperature all increase. In contrast, the distance from the Gibbs state at $t_f$ is at first smaller as a function of quench point for larger $t_f$, then becomes larger for larger $t_f$ past the freeze-out point.}
\label{fig:dw_quenchpoint}
\end{figure*}

\begin{figure*}[t]
\includegraphics[width=.91\textwidth]{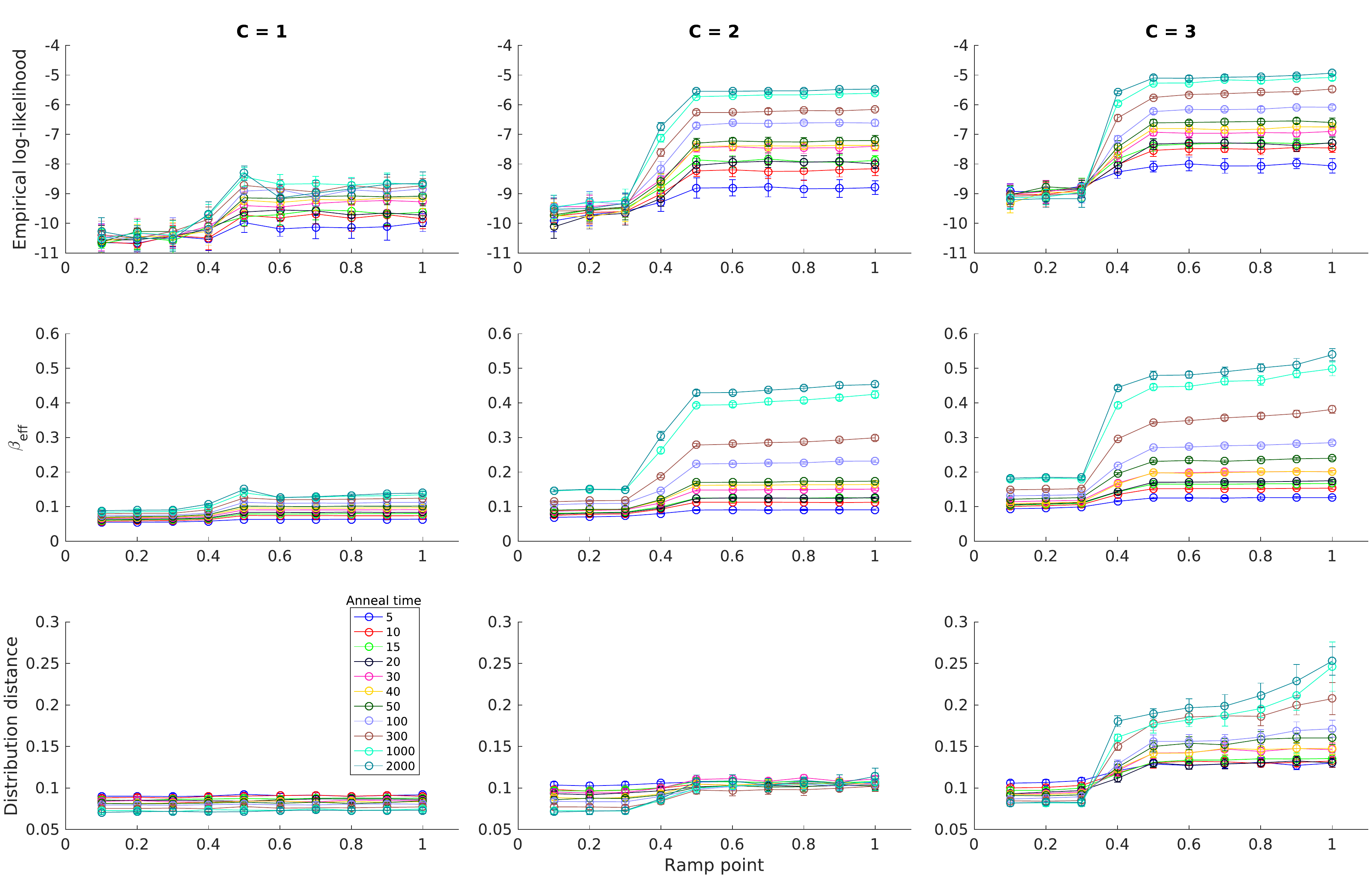}
\caption{As in Fig.~\ref{fig:dw_quenchpoint}, but with a fixed logical problem Hamiltonian, as in Ref.~\cite{Vinci:2017ab}. The trends are similar to Fig.~\ref{fig:dw_quenchpoint}, where the Hamiltonian was updated for each anneal time. Freezing occurs earlier with higher nesting level. For $C=3$, the distribution distance from Gibbs increases with longer anneal times. In addition, at longer anneal times, the distance from the Gibbs distribution of the final Hamiltonian increases at later quench points. }
\label{fig:dw_quenchpoint_fixed_w}
\end{figure*}

\begin{figure*}[t]
\includegraphics[width=\textwidth]{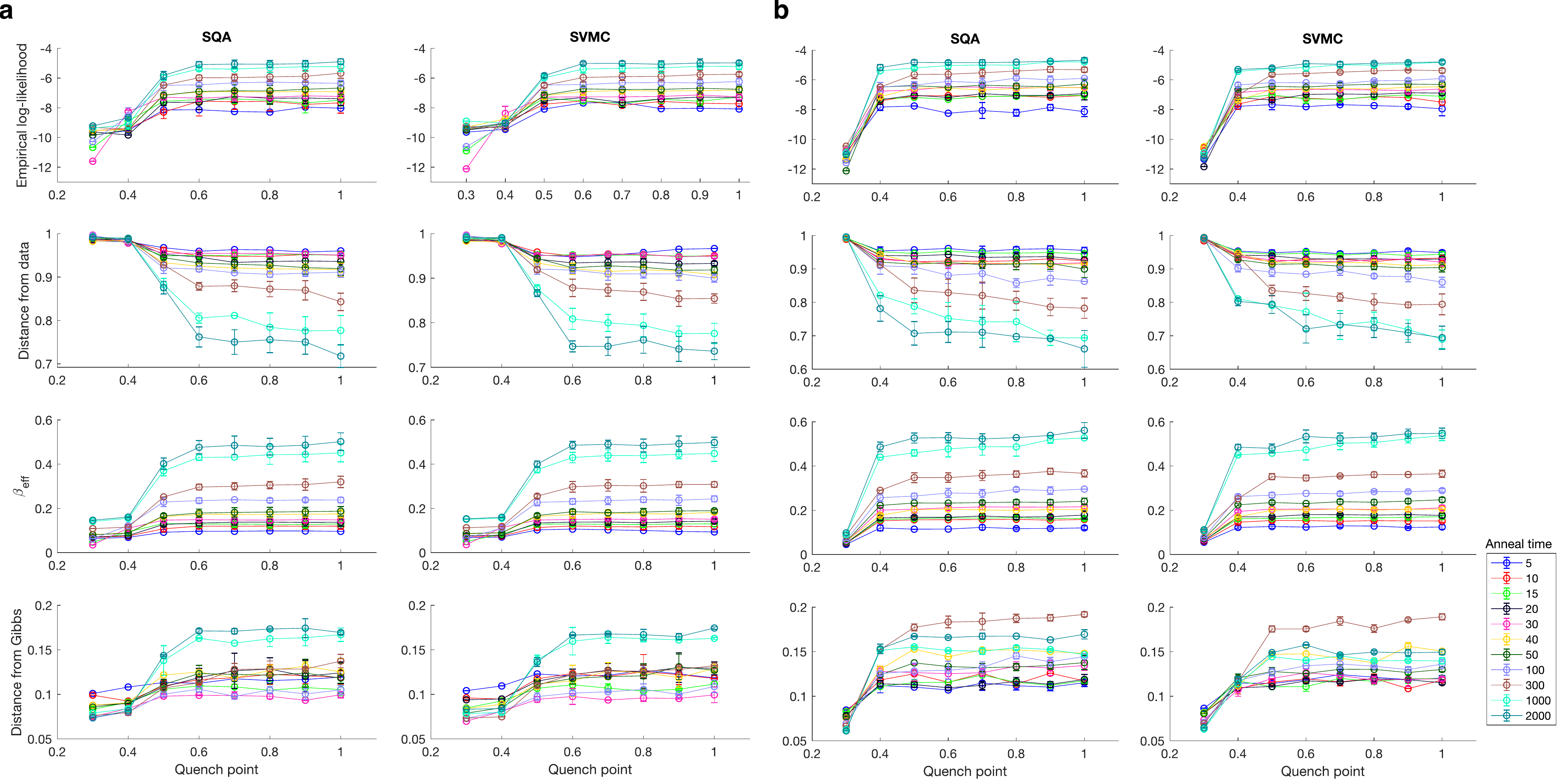}
 \caption{SQA and SVMC simulations with quench point for $C=2$ (left) and $C=3$ (right). Error bars are one standard deviation. The ``anneal time'' in the legend corresponds to the final physical problem Hamiltonian trained by DW at the specified anneal time, as in Fig.~\ref{fig:dw_quenchpoint}. The corresponding number of sweeps for SQA and SVMC was selected such that the effective inverse temperature of the distribution was closest to that found by DW at the specified anneal time.}
 \label{fig:C_2and3_sqa_svmc_quenchpoint}
\end{figure*}
\begin{figure*}[t]
\subfigure[\ $C=2$]{\includegraphics[width=.91\textwidth]{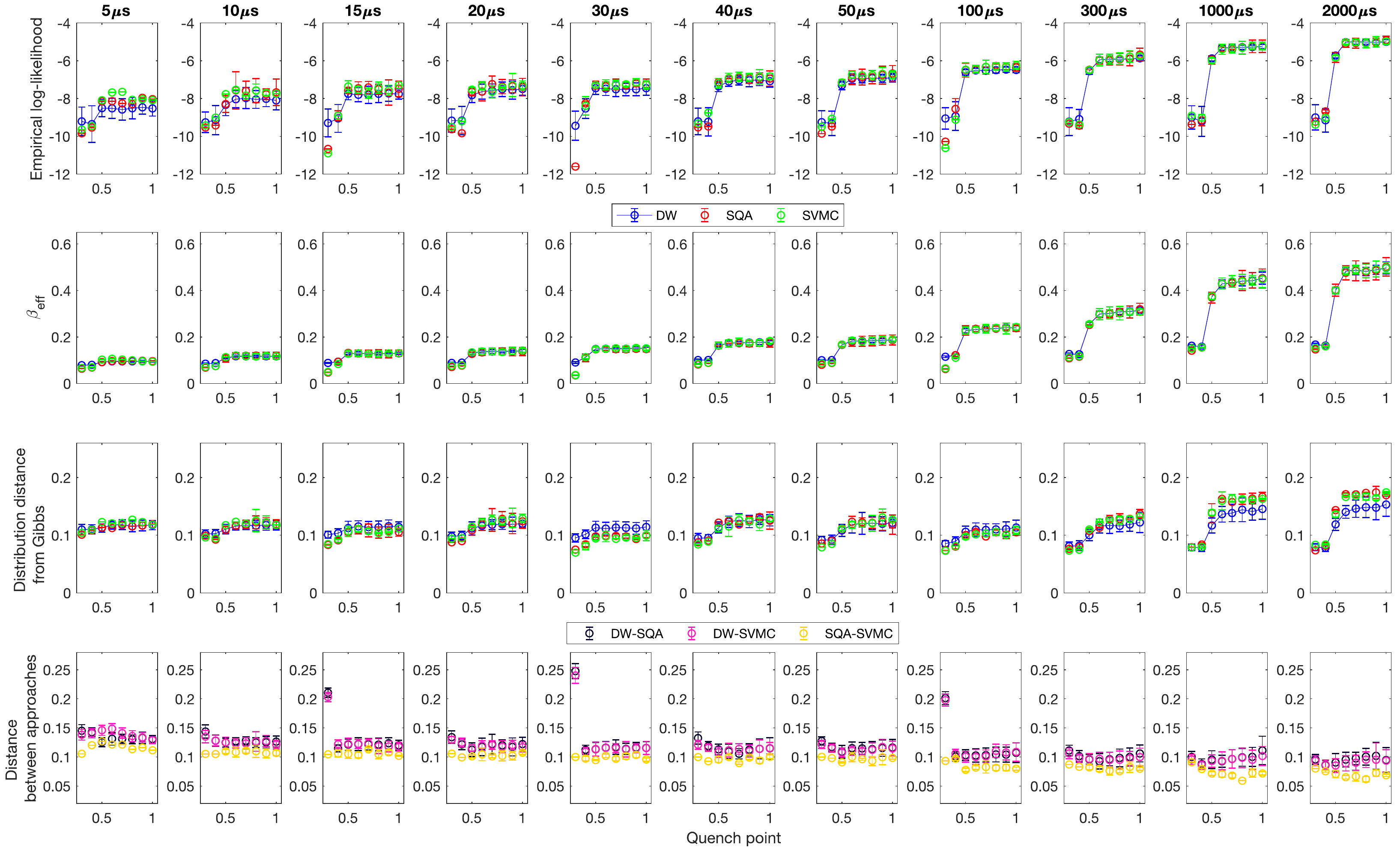}\label{fig:dw_sqa_svmc_closest_to_boltzmann_C=2}}
\subfigure[\ $C=3$]{\includegraphics[width=.91\textwidth]{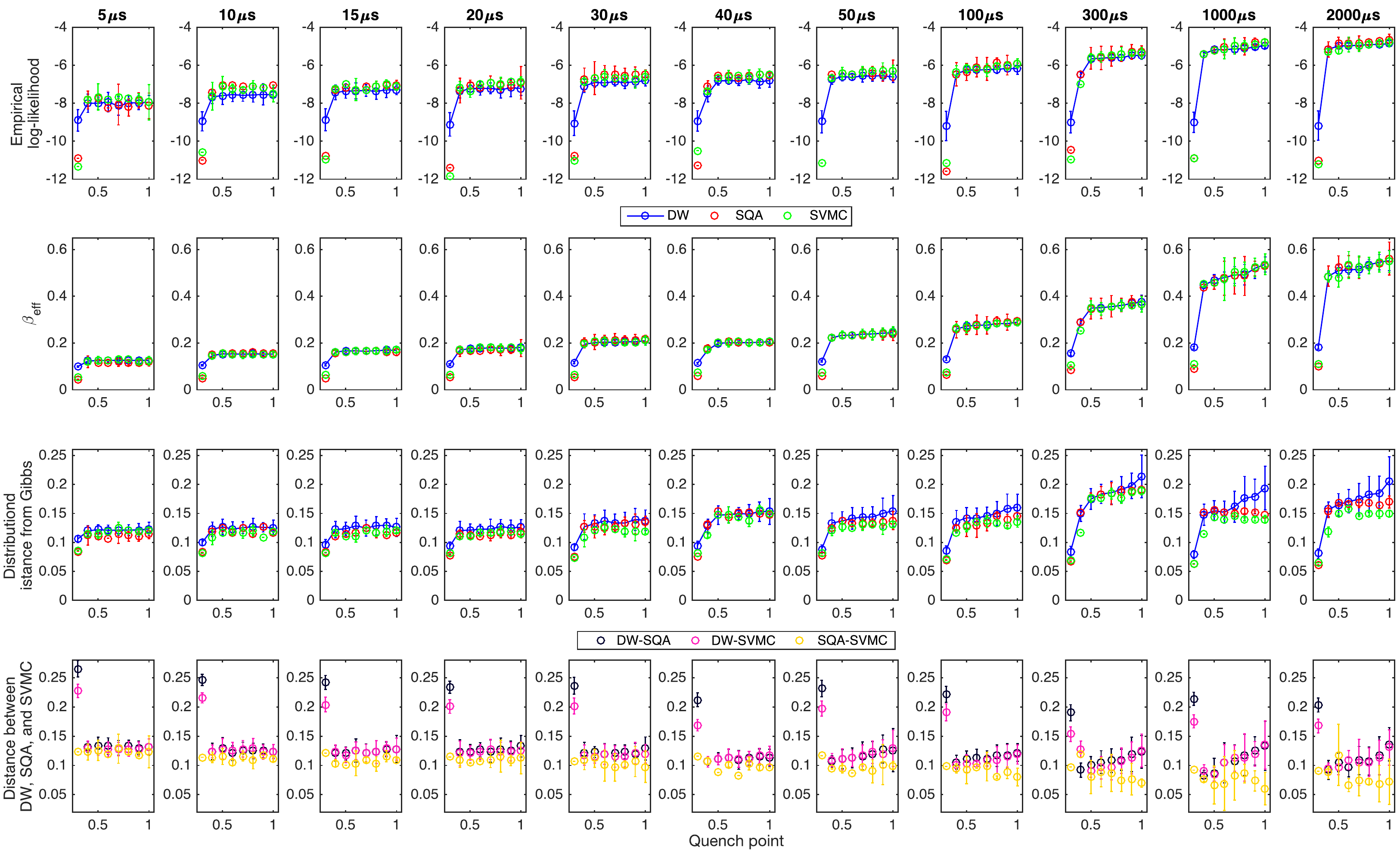}\label{fig:dw_sqa_svmc_closest_to_boltzmann_C=3}}
\caption{Direct comparison between DW, SQA, and SVMC for  (a) $C=2$ and (b) $C=3$. Each column represents the final logical Hamiltonian after training with a particular anneal time, i.e., each column is a different color curve in Fig.~\ref{fig:dw_quenchpoint}. The SQA and SVMC data is the same as in Fig.~\ref{fig:C_2and3_sqa_svmc_quenchpoint} (left). In the first three rows, blue, red and green represent DW, SQA and SVMC, respectively. In the last row, black, pink and yellow represent the distribution distances between DW and SQA, DW and SVMC, and SQA and SVMC, respectively. The close agreement in the second row is by design: we chose the number of sweeps of SQA and SVMC to match DW's $\beta_{\mathrm{eff}}$ at each anneal time and quench point (see Sec.~\ref{app:sweeps}).  For $C=3$ the SQA and SVMC data is the same as in Fig.~\ref{fig:C_2and3_sqa_svmc_quenchpoint} (right).}
\label{fig:dw_sqa_svmc_closest_to_boltzmann2}
\end{figure*}

To complement Sec.~\ref{sec:comp_SQA-SVMC}, we present results obtained using the quench feature of the DW2000Q devices, using the final trained Hamiltonians at $\alpha=0.03$ [see Fig.~\ref{fig:ResultsvsAT_alpha=0.03}].
Figure~\ref{fig:dw_quenchpoint} shows the effect of quenching at intervals of $\delta s = 0.1$. For these results we did not repeat the machine learning procedure from Sec.~\ref{sec:BAS-ta} (i.e., starting from a random set of weights and biases and training for $10$ epochs), but instead we used the final problem Hamiltonian after training for $10$ epochs with a fixed anneal time $t_f$. I.e., each colored curve in Fig.~\ref{fig:dw_quenchpoint} at a different $t_f$ corresponds to a different logical problem Hamiltonian. The results with a fixed Hamiltonian~\cite{Vinci:2017ab} are qualitatively very similar: see Fig.~\ref{fig:dw_quenchpoint_fixed_w}.
Statistics were then obtained by sampling $20$ times from each trained Hamiltonian at its corresponding $t_f$. 

Similarly to Fig.~\ref{fig:dw_sqa_svmc_closest_to_boltzmann}, which used the average energy over the logical problem as a metric, at all nesting levels shown in Fig.~\ref{fig:dw_quenchpoint} there is a clear transition point, after which all quantities plotted remain nearly constant. Hence this is a ``freeze-out" point, signifying the effective reezing of the dynamics. Beyond this point, the empirical log-likelihood [Eq.~\eqref{eq:emp-ll}], the distance from data [Eq.~\eqref{eq:dist_from_data}], the distance from the final Gibbs distribution [Eq.~\eqref{eq:tvd}], and the effective inverse temperature [Eq.~\eqref{eq:beta}] do not change significantly for most anneal times. Note that this freeze-out point happens earlier in the anneal with increasing $C$. This is expected because we are only able to encode the final Hamiltonian, $H_P$, which corresponds to increasing the scale of $B(s)$ relative to $A(s)$ [see Eq.~\eqref{eq:encoded_H}] \cite{PAL:13,vinci2015nested}. 

Note, though, that while the freeze-out point on the \textit{physical} problem happens earlier with increasing nesting level, the effective temperature of the \textit{logical} problem is still decreasing.
 In this case, the final distribution will depend on the dynamics in this interval and may not correspond to a Gibbs distribution at the end or at any particular point~\cite{Amin:2016}.

When the transverse field is very strong, the spins are mostly random and there is very poor agreement with the target data distribution (second row of Fig.~\ref{fig:dw_quenchpoint}). However, the sharp change in distribution distances at the freezing point is consistent with a large reorganization of the state, in which spins flip such that the distribution more closely resembles the data distribution. This trend is consistent across nesting levels. While the behavior after the freeze-out point is suggestive of a tunneling event, the tunneling interpretation is not tenable due to agreement of the DW data with the classical SVMC model, as discussed in Sec.~\ref{sec:comp_SQA-SVMC}, and in more detail below.

Figure~\ref{fig:dw_quenchpoint} shows that in addition to the distribution distance increasing with anneal time, the distribution distance also increases slightly even after the freeze-out point, most noticeably at $C=3$ and for the largest $t_f$ values. In order to gain more insight into the behavior of DW and to explore whether tunneling plays a role in DW's performance, we show the results of SQA and SVMC in Fig.~\ref{fig:C_2and3_sqa_svmc_quenchpoint} for $C=2$ and $C=3$. 

For the most part, both SQA and SVMC exhibit remarkably similar behavior to DW at $C=2$ and $C=3$. SQA and SVMC both exhibit a transition in the middle of the anneal, and trends with anneal time of the empirical log-likelihood, distance from data, and effective inverse temperature all echo the DW trends. The distance from the Gibbs distribution generally increases with the final annealed Hamiltonian, but not as consistently as  the DW results. This is probably because we did not fine-tune the number of SQA and SVMC sweeps to match DW's results. 

A direct comparison of the DW, SQA, and SVMC at $C=2$ and $C=3$ is presented in Fig.~\ref{fig:dw_sqa_svmc_closest_to_boltzmann2}. There is very clear agreement between all three in almost every respect, except at early points in the anneal and at late points in the anneal for the Hamiltonians trained at longer anneal times (the rightmost columns in Fig.~\ref{fig:dw_sqa_svmc_closest_to_boltzmann}). At $s=0.3$ (the first data point in each column) and $C=3$, DW differs significantly from SQA and SVMC, in that SQA and SVMC tend to have much lower empirical log-likelihoods and much smaller $\beta_{\mathrm{eff}}$. The reason for this may be an artifact in the length of the quench. For DW, the quench used is 1 $\mu$s, and for SQA and SVMC we set the quench at $500$ sweeps. For the trained Hamiltonian at $5\mu$s, quenching at $s=0.3$ means that the total time the system has been allowed to evolve is $1.5\mu$s, comparable to the $1\mu$s quench. For the Hamiltonian trained with a total anneal time of $2000\mu$s, quenching at $s=0.3$ means that the system is allowed to evolve for $600\mu$s before a $1\mu$s quench. In other words, if the quench is relatively long, we may not have an accurate representation of the state of the system in the middle of the anneal. The quench may still be too long for DW, such that even at $s=0.3$ across all anneal times, the system is able to thermalize as the quench is taking place. 

For long anneal times at later points in the anneal, the distance from Gibbs does not increase as noticeably with quench point for SQA and SVMC as it does with DW. For both SQA and SVMC, the distance from a Gibbs distribution stays approximately the same after the ``freeze-out'' point, whereas DW's distance from Gibbs increases. This may be attributable to the presence of $1/f$ noise, whose low frequency components impact performance at long anneal times~\cite{DW-white-paper-1overf}. 

The fact that SVMC and SQA yield nearly the same trends as DW suggests that for the most part DW is performing similarly to both SQA and SVMC, including in having larger distribution distances from Gibbs with longer anneal times. To the extent that we trust SQA and SVMC as realistic simulations of the underlying physics, this increase in distribution distance is not an artifact of DW but consistent with other physicals models. We offer two possible, complementary explanations for the increase in distance from Gibbs. The first relates primarily to the increase of distance from Gibbs with the anneal time: longer anneal times allow better thermalization, but to the physical problem. The embedding procedure preserves the structure of low-lying states but may not guarantee the ordering of higher-energy states. In the process of decoding the physical problem states to the encoded problem states and then the logical problem states, the lowest part of the energy spectrum may be preserved, but upon decoding the order of the higher-energy states may not be preserved, and as such $\beta_{\mathrm{eff}}$ increases from more probability on the lower energy states, but the distribution in terms of the logical problem appears less Gibbsian. The fact that the increase in distribution distance increases for larger $C$ supports this idea; with larger $C$ there are more physical qubits and hence more energy levels of the physical problem, but the same number of states for the logical problem. The second explanation relates primarily to the increase of distance from Gibbs with the quench point: it is that the system may be thermalizing to a Gibbs state at some intermediate point in the anneal, where the dynamics freezes. The distance from Gibbs is calculated with reference to the final classical Hamiltonian, but if DW, SQA, and SVMC are thermalizing to a different (but close) Gibbs distribution, the distance may increase as a function of $s$.

\section{Anneal times at different values of $\alpha$}
\label{app:alpha_at}

\begin{figure*}[t]
\centering
\includegraphics[width=\textwidth]{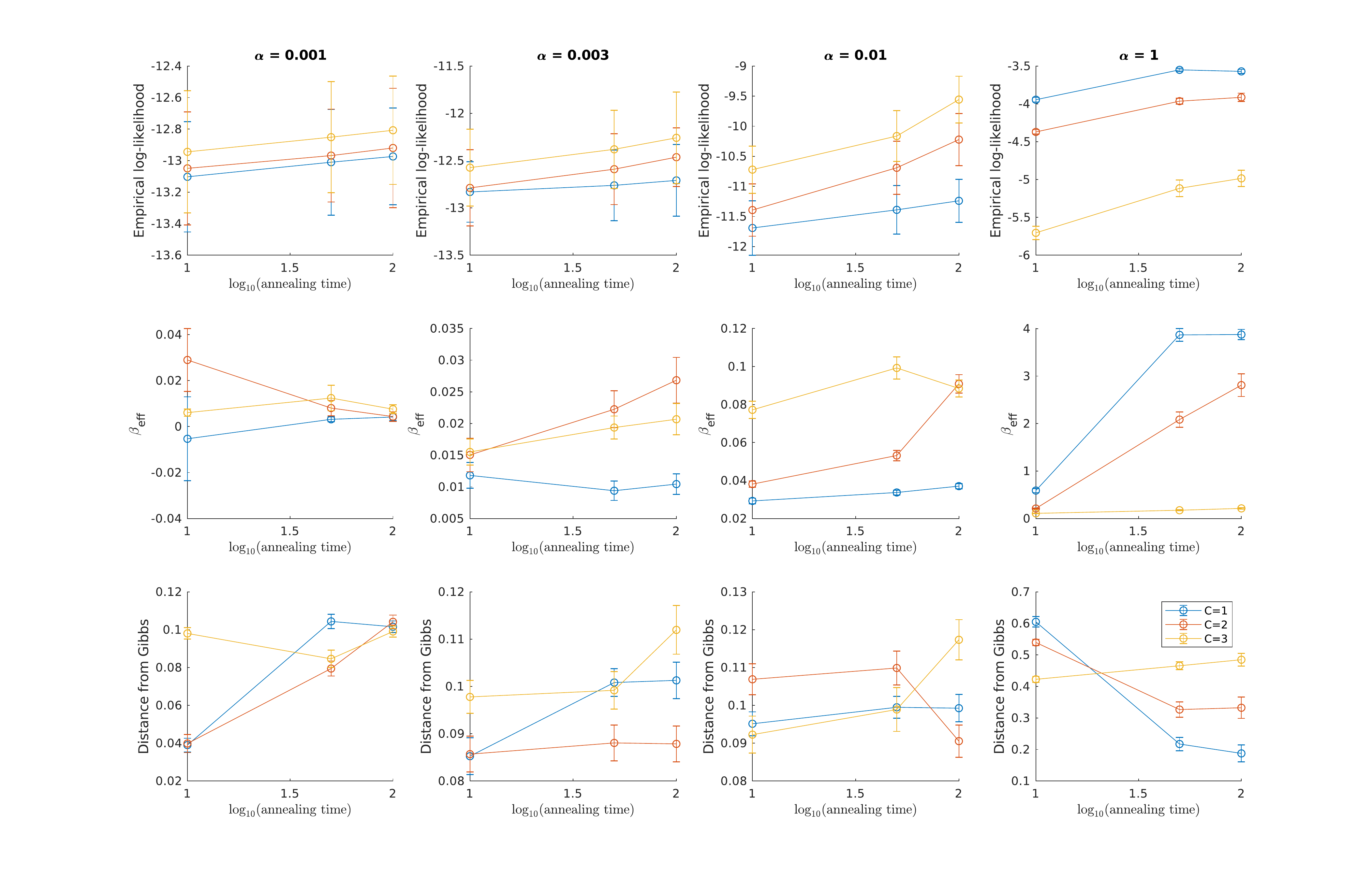}
\caption{Effect of using NQAC with different anneal times for the BAS dataset, measured in microseconds, for several values of $\alpha$. The maximum possible value of the empirical log-likelihood in this case is $-3.41$. Error bars are two standard deviations.}
\label{fig:alpha_at}
\end{figure*}

To complement the results in Fig.~\ref{fig:ResultsvsAT}, we explored other values of $\alpha$ in Fig.~\ref{fig:alpha_at}. Data was collected for a smaller set of anneal times in this case (data collection is very time consuming) to get a sense of the trends.
 
The overall trend is one of improvement with increasing anneal times, and of higher $C$ leading to higher (but not statistically significant) empirical log-likelihood, except for $\alpha=1$, where the ordering is reversed. This might be expected because for $\alpha=1$ both $C=2,3$ are in the penalty-limited regime, meaning that $\gamma_2$ is pegged at $1$. Except for $\alpha=1$ the effective inverse temperature and distance from a Gibbs distribution on the logical problem do not exhibit trends, and it is difficult to interpret their behavior. However, note that for $\alpha < 1$ the maximum value of the empirical log-likelihood attained is very far from the maximum possible value of $-3.41$, so that the inconsistent trends observed may be attributable to the anneal time being too short for the results to have converged.

%

\end{document}